\documentclass[nofootinbib,a4paper,12pt]{article}
\usepackage{jheppub}
\usepackage[T1]{fontenc}
\usepackage[utf8]{inputenc}
\usepackage{float}
\usepackage{slashed}
\usepackage{multicol,multirow}
\usepackage{amssymb}
\usepackage{amsmath}
\usepackage{bbm, slashed}
\usepackage{subcaption}
\usepackage{array}
\usepackage{xcolor}
\usepackage{hyperref}
\hypersetup{colorlinks=true}
\usepackage{longtable}
\usepackage{graphics,graphicx,epsfig,ulem,booktabs}
\numberwithin{equation}{section}
\usepackage{url}
\usepackage{physics}
\usepackage{graphicx}
\usepackage{simpler-wick}
\usepackage{feynmp}
\usepackage{tikz-feynman} 
\usepackage{bm}
\usepackage[fleqn]{nccmath}
\usepackage{bbold}
\usepackage{kantlipsum}
\usepackage{braket}
\usepackage{empheq}
\usepackage[most]{tcolorbox}
\usepackage{caption}
\usepackage{subcaption}
\usepackage{fancyhdr} 
\usepackage{pgfplots}

\definecolor{green}{rgb}{0.1,0.8,0.2}
\definecolor{orange}{rgb}{1.0,0.5,0.0}
\definecolor{cyan}{rgb}{0.0,0.75,0.8}
\definecolor{brown}{rgb}{0.7,0.35,0.05}
\definecolor{am}{rgb}{0,0,1.0}

\newcolumntype{C}[1]{>{\centering\let\newline\\\arraybackslash\hspace{0pt}}m{#1}}

\newcommand{\GeV}{\, {\rm GeV}}

\title{\begin{center}
\boldmath 
Are Subleading Effects Really Subleading? $B$-Meson Decays in Mesogenesis
\end{center}
}

\preprint{\small P3H-25-089, SI-HEP-2025-24}

\author[]{Ali Mohamed}

\affiliation[]{Physik Department, Universit\"{a}t Siegen, Walter-Flex-Str. 3, 57068 Siegen, Germany}

\emailAdd{ali.mohamed@uni-siegen.de}

\abstract{
We calculate inclusive $B$-meson decay rates in the Mesogenesis framework—a model explaining baryogenesis and the existence of dark matter—using the Heavy Quark Expansion (HQE), up to the dimension-six two-quark Darwin term.
By systematically studying the power-suppressed contributions, we identify regions of parameter space where subleading terms exceed the leading contribution, i.e., the free $b$-quark decay, highlighting the limits of the HQE in this BSM scenario. This behavior is reminiscent of the Standard Model only under artificially heavy charm masses, and can be used to study the HQE close to its breakdown. We further update the lower bounds on the exclusive decay mode $B^+ \to p^+ \psi$ by incorporating the fully HQE-corrected inclusive width in the ratio $\Gamma_{\mathrm{excl}}/\Gamma_{\mathrm{incl}}$. Extending the analysis from total decay rates to the lifetime ratio $\tau(B_s)/\tau(B_d)$, we find no additional constraints on the couplings beyond existing collider bounds, consistent with analogous results for $\tau(B^+)/\tau(B_d)$. We further compare the sensitivity of both lifetime ratios. 
}

\begin{document}
\maketitle
\section{Introduction}
The \textit{B}-Mesogenesis model~\cite{Elor:2018twp,Alonso-Alvarez:2021qfd,Elahi:2021jia,Nelson:2019fln} addresses two of the most compelling puzzles in modern physics: the origin of the baryon asymmetry of the universe~\cite{Planck:2015fie,Planck:2018vyg,Cyburt:2015mya} and the nature of dark matter~\cite{Trimble:1987ee}. In this scenario, new decay channels of $B$-mesons connect these phenomena by producing visible baryons alongside a dark-sector antibaryon $\psi$, resulting in final states with missing energy.\footnote{We do not consider here the recent modifications of the original \textit{B}-Mesogenesis setup proposed in Ref.~\cite{Elor:2025fcp}.} These novel decay modes also modify the total decay widths of $B$-mesons, impacting precision observables such as meson lifetimes and their ratios. Studying these effects provides a valuable window into the underlying new physics responsible for baryogenesis and dark matter. 

Exclusive $B$-meson decay channels into visible light baryons and a dark antibaryon $\psi$ 
have been studied in the framework of Light-Cone Sum Rules (LCSR) 
\cite{Khodjamirian:2022vta,Boushmelev:2023huu,Elor:2022jxy}, using the proton distribution amplitude~\cite{Braun:2001tj,Lenz:2003tq,Braun:2006hz,Lenz:2009ar}, and 
including higher-twist corrections and extensions to a variety of baryonic final states. 
The framework was also applied to the decays of heavy $\Lambda_b$ and $\Xi_b$ baryons into a pseudoscalar meson and a dark baryon in Ref.~\cite{Shi:2024uqs}. The analysis employed the $\Lambda_b$ distribution amplitudes from Refs.~\cite{Ball:2008fw,Duan:2022uzm}. On the experimental side, dedicated searches for $B$-meson decays into baryons plus missing energy 
have been carried out at BaBar, Belle, and LHCb~\cite{BaBar:2023dtq,Belle:2021gmc,Rodriguez:2021dm}, with the Belle II collaboration (using the dataset from Belle experiment) now providing new preliminary constraints~\cite{Belle-II:2026tyb}.

In our previous work~\cite{Lenz:2024rwi}, we analyzed the impact of these new decay channels on $B$-meson lifetimes within the framework of the Heavy Quark Expansion (HQE), see e.g. the reviews~\cite{Lenz:2014jha,Albrecht:2024oyn}. In particular, we focused on the implications of the new operators for the lifetime ratio $\tau(B^+)/\tau(B_d)$, which is mainly sensitive to the dimension-six four-quark operator due to isospin symmetry. Moreover, by studying the ratio $\Gamma_{\text{excl}} / \Gamma_{\text{incl}}$, we obtained lower limits on the exclusive decay $B^+ \to p^+ \psi$, where only the leading dimension-three term was included in the inclusive contribution.

In this paper, we extend that analysis by systematically computing the sub-leading corrections in the HQE, incorporating the dimension-5 chromomagnetic and dimension-6 two-quark Darwin operators. These contributions enter at $\mathcal{O}(1/m_b^2)$ and $\mathcal{O}(1/m_b^3)$, respectively. Their inclusion is not only a matter of formal completeness: in the Standard Model (SM) the Darwin term alone is known to induce a sizeable correction of about $-4\%$ to the non-leptonic $B$-meson decay width, representing the dominant subleading effect in $B_{d,s}$ mesons and the second-largest in $B^+$ mesons, after the dimension-six four-quark contribution (arising from Pauli interference diagram (PI))~\cite{Mannel:2020fts,Lenz:2020oce}. Given the percent-level size of these effects and the per-mille precision of experimental lifetime data~\cite{PDG:2024,HeavyFlavorAveragingGroupHFLAV:2024ctg}, it is essential to account for them when assessing the impact of new invisible decay channels on inclusive observables.

We therefore revisit the total lifetimes of $B$-mesons in this extended framework and analyze the impact of subleading terms on lifetime observables, comparing them with the corresponding effects in the SM. In addition, we investigate their implications for the lifetime ratio $\tau(B_s)/\tau(B_d)$. While $\tau(B^+)/\tau(B_d)$ remains insensitive to two-quark contributions due to isospin symmetry, the ratio $\tau(B_s)/\tau(B_d)$ becomes sensitive at this level as a result of $SU(3)_F$ breaking effects, which can be sizable, see e.g.~\cite{King:2021jsq,Lenz:2022rbq}. Finally, we improve the lower bounds on the exclusive decay $B^+ \to p^+ \psi$ by incorporating the updated HQE contributions.

The structure of this paper is as follows: In Sec.~\ref{model}, we introduce the model and its Lagrangian. The theoretical framework is presented in Sec.~\ref{theorytools}, where we discuss the HQE, lifetime ratios, and the exclusive predictions from LCSR. Sec.~\ref{numerics} contains the numerical analysis and phenomenological discussion. Finally, in Sec.~\ref{summary}, we conclude with a summary of the results and an outlook for future developments. 
\section{\textit{B}-Mesogenesis}
\label{model}
The Mesogenesis framework~\cite{Elor:2018twp,Alonso-Alvarez:2021qfd,Elahi:2021jia,Nelson:2019fln} proposes a mechanism for baryogenesis and dark matter based on $B$-meson production in the early Universe, known as “$B$-Mesogenesis.” The scenario assumes the early Universe is dominated by radiation and a weakly coupled scalar field $\Phi$, which decays into $b\bar{b}$ pairs. These $b$-quarks hadronize into $B$ and $\bar{B}$ mesons, which subsequently decay not only via SM channels but also through new decay modes into visible baryons $\mathcal{B}$ and dark sector antibaryons $\psi$, the latter appearing as missing energy.\footnote{While \cite{Nelson:2019fln} generates the baryon asymmetry via the same mechanism, it does not include a dark matter candidate.} CP violation in the mixing of the neutral $B_d$ and $B_s$ mesons causes a difference in the decay rates of $B_q$ and $\bar{B}_q$, producing the observed baryon asymmetry. This links the matter-antimatter imbalance directly to $B$-meson CP violation and simultaneously provides a dark matter candidate originating from the $B$-meson decays. The decay $B \to \psi \mathcal{B}$ is mediated by a heavy color-triplet scalar $Y$, carrying baryon number $-2/3$ and hypercharge $-1/3$ or $+2/3$. The most general renormalizable Lagrangian describing the interaction of $Y$ with SM quarks and the singlet baryon $\psi$ is given by:
\begin{equation}
\begin{aligned}
\mathcal{L}_{-1/3} = -\sum_{\alpha,\beta}  y_{u_{\alpha}d_{\beta}} \epsilon_{{i}{j}{k}} \kern0.18emY^{*{i}} \kern0.18em \bar{u}^{{j}}_{\alpha R}\kern0.18em d^{c,{k}}_{\beta R} - \sum_{\gamma} y_{\psi d_{\gamma}}  Y_{{i}} \kern0.18em \bar{\psi}\kern0.18em d^{c,{i}}_{\gamma R} + \text{h.c.}\kern0.18em,\\
\mathcal{L}_{+2/3} = -\sum_{\alpha,\beta}  y_{d_{\alpha}d_{\beta}} \epsilon_{{i}{j}{k}} \kern0.18emY^{*{i}} \kern0.18em \bar{d}^{{j}}_{\alpha R}\kern0.18em d^{c,{k}}_{\beta R} - \sum_{\gamma} y_{\psi u_{\gamma}}  Y_{{i}} \kern0.18em \bar{\psi}\kern0.18em u^{c,{i}}_{\gamma R} + \text{h.c.}\kern0.18em,
\label{eq:ModelLagrangians}
\end{aligned}
\end{equation}
where $y$ denotes coupling constants, ${i},{j},{k}$ are color indices, $\alpha,\beta,\gamma$ label quark flavors, and the superscript $c$ denotes charge conjugation. It has been pointed out from different perspectives, see 
Refs.~\cite{Lenz:2024rwi,Miro:2024fid}, that the effects of the Lagrangian $\mathcal{L}_{+2/3}$ are expected to be smaller than those of $\mathcal{L}_{-1/3}$. We also anticipate this behaviour, therefore, in the following, 
we restrict our analysis to the contributions stemming from $\mathcal{L}_{-1/3}$.
Integrating out the heavy color-triplet $Y$, the effective Hamiltonian mediating the $B$-meson decays can be written as 

\begin{align}
\mathcal H_{-1/3} \;=\; \sum_{u,d}\Bigg\{&
\underbrace{\epsilon_{{i}{j}{k}}\,
   C_{\psi}^{(ud)}\,(\bar u_R^{\,{i}} b_R^{c,{j}})\,(\bar\psi\, d_R^{c,{k}})
   + \epsilon_{{i}{j}{k}}\,
   \widetilde C_{\psi}^{(ud)}\,(\bar u_R^{\,{i}} d_R^{c,{j}})\,(\bar\psi\, b_R^{c,{k}})}_{\text{(i) $\psi$-type operators}}
\notag\\[4pt]
&+\,\underbrace{\epsilon_{{i}{j}{k}}\epsilon_{{i}{m}{n}}\,
   C_{uu}^{(d)}\,(\bar u_R^{\,{j}} b_R^{c,{k}})\,(\bar d_R^{c,{m}} u_R^{\,{n}})
   + C_{\psi\psi}^{(d)}\,(\bar\psi\, d_R^{c,{i}})\,(\bar b_R^{c,{i}}\,\psi)}_{\text{(ii) hadronic and $\psi\psi$ operators}}
\Bigg\} \;+\; \text{h.c.},
\label{eq:Heff}
\end{align}
where $u \in \{u,c\},\ d \in \{d,s\},$ and the Wilson coefficients, including the heavy scale suppression, are
\begin{align}
C_{\psi}^{(ud)} &= -\frac{y_{ub}\,y_{\psi d}}{M_Y^2}, 
\ 
\widetilde C_{\psi}^{(ud)} = -\frac{y_{ud}\,y_{\psi b}}{M_Y^2}, \
C_{uu}^{(d)} = \frac{y_{ub}\,y_{ud}^{*}}{M_Y^2}, 
\ 
C_{\psi\psi}^{(d)} = -\frac{y_{\psi d}\,y_{\psi b}^{*}}{M_Y^2}.
\label{eq:Cpsipsi_map}
\end{align}
In the following, we focus the detailed analysis on the operators proportional to $C_{\psi}^{(ud)}$ and $\widetilde C_{\psi}^{(ud)}$. While previous works~\cite{Khodjamirian:2022vta,Boushmelev:2023huu,Lenz:2024rwi} referred to these two as ``model (d)'' and ``model (b),'' respectively, we use them here as representative examples for the main discussion. The calculations for the remaining operators in Eq.~\eqref{eq:Heff} have been fully performed and their results are provided in the Appendix, with references to them given at the relevant points in the main text.

\section{Theoretical Framework for $B$ decays}
\label{theorytools}
\subsection{HQE within the $B$-Mesogenesis Model}
We perform our calculations within the framework of the HQE, a well-established tool for computing inclusive decay rates of heavy hadrons. Using the optical theorem, the total decay width of a $B$-meson can be written in terms of the time-ordered product of two insertions of the effective Hamiltonian:
\begin{equation}
    \Gamma (B) = \frac{1}{2 m_B}{\rm Im} \langle B|  {\cal T} | B \rangle \,,
    \label{eq:GammaB}
\end{equation}
with the transition operator given by
\begin{equation}
    {\cal T} = {i} \int d^4 x \, T\, \{ {\cal H}_{\rm eff}(x), {\cal H}_{\rm eff}(0)\}\,.
    \label{eq:T-operator}
\end{equation}
\begin{figure}[t!]
  \centering
  \setlength{\unitlength}{1pt}

  \begin{tabular}{ccccc}
    \begin{picture}(150,120)
      \put(0,0){\includegraphics[width=0.25\textwidth]{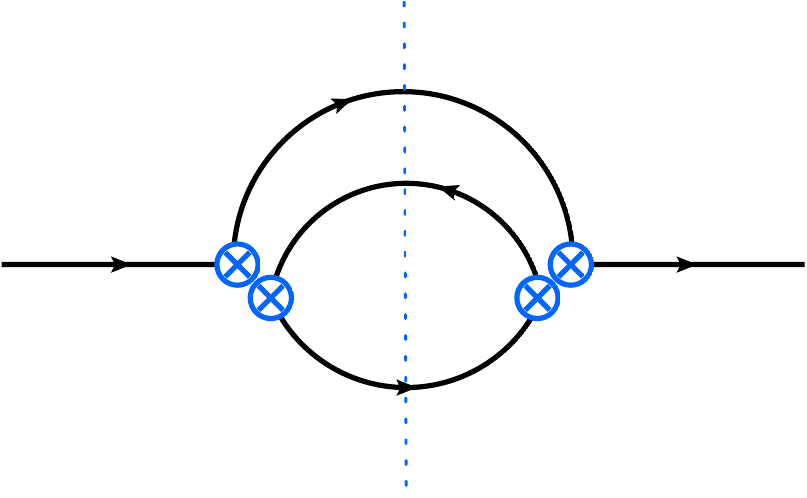}}
      \put(50,68){$f_1$}
      \put(50,35){$f_2$}
      \put(70,10){$f_3$}
      \put(15,40){$b$}
      \put(100,40){$b$}
       \put(133,33){$+$}
    \end{picture}
    \begin{picture}(150,120)
      \put(0,0){\includegraphics[width=0.25\textwidth]{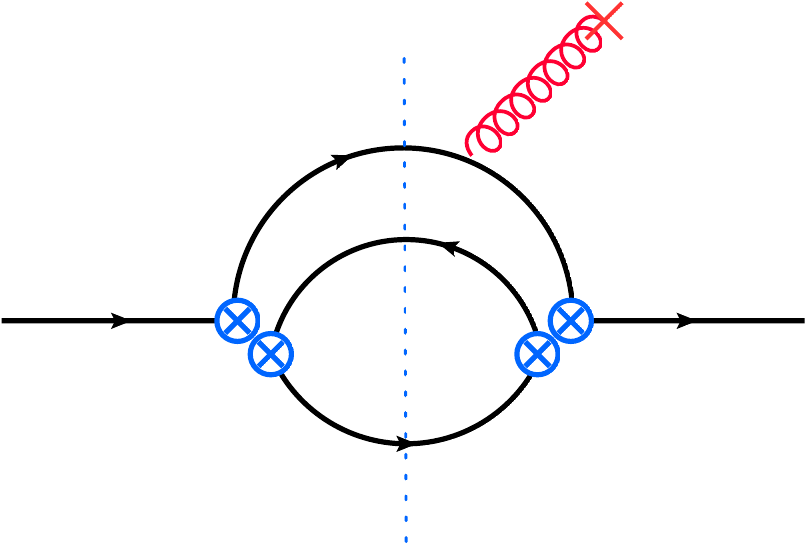}}
      \put(50,68){$f_1$}
      \put(50,35){$f_2$}
      \put(70,10){$f_3$}
      \put(15,40){$b$}
      \put(100,40){$b$}
      \put(133,33){$+ \cdots +$}
    \end{picture}
\hspace{0.3cm}
    \begin{picture}(150,120)
      \put(0,0){\includegraphics[width=0.25\textwidth]{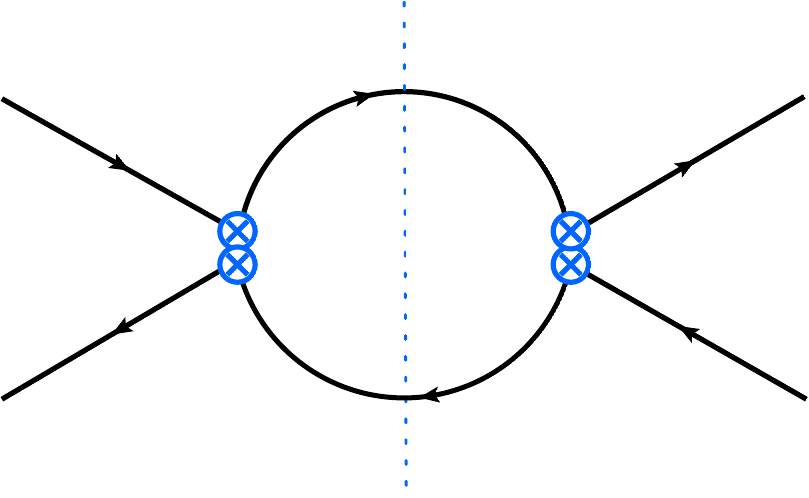}}
      \put(45,66){$f_1$}
      \put(70,7){$f_2$}
      \put(20,55){$b$}
      \put(100,55){$b$}
      \put(20,18){$q$}
      \put(95,15){$q$}
         \put(120,33){$+ \cdots $}
    \end{picture}
  \end{tabular}

 \caption{Schematic representation of the HQE of the $B$-meson. 
The two-quark contribution arises from two loops, whereas the four-quark 
contribution involves a single loop. Crossed vertices indicate insertions of operators from the effective $\Delta B=1$ Hamiltonian. The vertical dotted line indicates 
taking the imaginary part of the diagrams, and $f_1$, $f_2$, $f_3$ 
represent general fermions into which the $b$ quark can decay.}
  \label{fig:HQEDiags}
\end{figure}
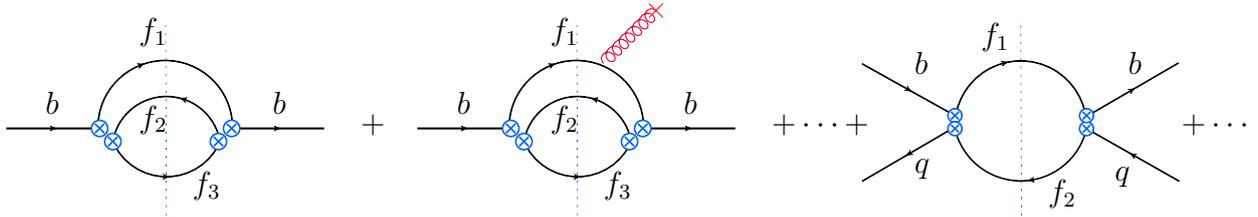
The effective Hamiltonian ${\cal H}_{\rm eff}$ is given by the sum of the SM part and the BSM part
\begin{eqnarray}
{\cal H}_{\rm eff} & = &
{\cal H}_{\rm eff}^{\rm SM} + 
{\cal H}_{-1/3} \, .
\end{eqnarray}
Exploiting the hierarchy $m_b \gg \Lambda_{\mathrm{QCD}}$, the non-local operator in Eq.~\eqref{eq:T-operator} can be systematically expanded in local $\Delta B = 0$ operators, ordered by increasing dimension. This heavy-quark expansion~\cite{Shifman:1986mx} takes the form, see Fig.~\ref{fig:HQEDiags},
\begin{equation}
\Gamma(B) = 
\Gamma_3  
+ \frac{\langle \mathcal{O}_5 \rangle}{m_b^2} \, \Gamma_5
+ \frac{\langle \mathcal{O}_6 \rangle}{m_b^3} \, \Gamma_6
+ \ldots
+ 16\pi^2 \left(
  \frac{\langle \tilde{\mathcal{O}}_6 \rangle}{m_b^3} \, \tilde{\Gamma}_6
  + \frac{\langle \tilde{\mathcal{O}}_7 \rangle}{m_b^4} \, \tilde{\Gamma}_7
  + \ldots
\right) ,
\label{eq:HQE}
\end{equation}
where the short-distance coefficients $\Gamma_i$ are calculable in perturbative QCD, and $\langle \mathcal{O}_i \rangle \equiv \frac{\langle B | \mathcal{O}_i | B \rangle}{2 m_B}$
denotes the hadronic matrix elements.
The most recent calculations are the NLO corrections to $\Gamma_5$ in non-leptonic decays, both for $b \to c \bar{c} s$~\cite{Mannel:2025fvj} and $b \to c \bar{u} d$~\cite{Mannel:2024uar}, as well as the NNLO determination of the leading term $\Gamma_3$~\cite{Egner:2024azu}. For a comprehensive overview of the calculations of the various terms in the SM, see   Ref~\cite{Albrecht:2024oyn}. In the context of $B$-Mesogenesis, the contributions to both $\Gamma_3$ and $\tilde{\Gamma}_6$ have been computed for the Lagrangians $\mathcal{L}_{-1/3}$ and $\mathcal{L}_{+2/3}$ in Ref.~\cite{Lenz:2024rwi}.

Here, we compute the contributions of the coefficients $\Gamma_5$ and $\Gamma_6$, which correspond to the nonperturbative parameters $\mu_\pi^2$ (kinetic energy), $\mu_G^2$ (chromomagnetic interaction), $\rho_D^3$ (Darwin term), and $\rho_{LS}^3$ (spin–orbit interaction) in the HQE. These parameters encode the matrix elements of the dimension-five and dimension-six two-quark operators. They are defined as, see e.g.~\cite{Dassinger:2006md}
\begin{equation}
\begin{aligned}
2 m_B\,\mu_\pi^2(B) &= -\langle B(p_B)\,|\,\bar b_v \,(iD_\mu)(iD^\mu)\,b_v\,|\,B(p_B)\rangle , \\[4pt]
2 m_B\,\mu_G^2(B) &= \langle B(p_B)\,|\,\bar b_v \,(iD_\mu)(iD_\nu)(-i\sigma^{\mu\nu})\,b_v\,|\,B(p_B)\rangle , \\[6pt]
2 m_B\,\rho_D^3(B) &= \langle B(p_B)\,|\,\bar b_v \,(iD_\mu)(i v\!\cdot\!D)(iD^\mu)\,b_v\,|\,B(p_B)\rangle , \\[6pt]
2 m_B\,\rho_{LS}^3(B) &= \langle B(p_B)\,|\,\bar b_v \,(iD_\mu)(i v\!\cdot\!D)(iD_\nu)(-i\sigma^{\mu\nu})\,b_v\,|\,B(p_B)\rangle ,
\end{aligned}
\label{eq:hadronic_parameters}
\end{equation}
where $\sigma^{\mu\nu} = \frac{i}{2}[\gamma^\mu,\gamma^\nu]$, $b_v$ denotes the Heavy Quark Effective Theory (HQET) $b$-quark field with four-velocity $v^\mu = p_B^\mu/m_B$, 
\begin{equation}
    b(x) = e^{-i m_b v\cdot x}\,b_v(x),
\end{equation}
and $D_\mu$ is the QCD covariant derivative $D_\mu = \partial_\mu - i g_s t^a A^a_\mu$, where $g_s$ is the strong coupling constant, $t^a$ are the 
${SU}(3)$ color generators in the fundamental representation, and 
$A^a_\mu$ denotes the gluon field.

To obtain the contribution up to dimension-six operators, the propagator must be expanded up to a single covariant derivative acting on one gluon field strength tensor. The latter is defined as $G_{\mu \nu} = -i  \left[iD_\mu, iD_\nu \right].$ We use the result derived in~\cite{Lenz:2020oce} by expanding the propagator in the soft background gluon field using the Fock-Schwinger gauge:
\begin{equation}
S(x,0) =  \int \frac{d^4 p}{(2\pi)^4}\, {\cal S}(p)  \, e^{-i p \cdot x} \,,
\label{eq:Sx0}
\end{equation}\\
with
\begin{align}
{\cal S}(p) & = \frac{\slashed p + m}{p^2 - m^2 } -\frac{m}{2} \frac{G_{\rho \mu } \,\sigma^{\rho \mu}}{(p^2 -m^2 )^2}  +  \frac{\tilde G_{\sigma \eta} \ p^\sigma \gamma^\eta \gamma^5}{(p^2 -m^2 )^2} - \frac23 \frac{p^\alpha D_\alpha  G_{\rho \mu}}{(p^2 -m^2)^3} \gamma^\mu p^\rho 
\nonumber \\[3mm]
&+ \frac23 \frac{D_\alpha G_{\alpha \mu}}{(p^2 - m^2)^3} \Big[ \gamma^\mu  (p^2 -m^2) - p^\mu (\slashed p + 2 m) \Big]
 + 2 i \frac{D_\alpha \tilde G_{\tau \eta}}{(p^2 -m^2)^3} p^\alpha p^\tau \gamma^\eta \gamma^5 
\nonumber \\[3mm]
& + \frac23 m \frac{D_\alpha G_{\rho \mu}}{(p^2 -m^2)^3} \Big( p^\alpha \gamma^\rho \gamma^\mu - p^\rho \gamma^\mu \gamma^\alpha \Big)
+ \, \ldots \,.
\label{eq:S1p}
\end{align}
where the ellipses denote higher-dimension terms with additional derivatives, while the dual field tensor is $\tilde G_{\mu \nu} = (1/2) \epsilon_{\mu \nu \rho \sigma} G^{\rho \sigma}$ with the Levi-Civita tensor $\epsilon_{\mu \nu \rho \sigma}$, and $ D_\rho G_{\mu \nu} = -\left[ iD_\rho,  \left[iD_\mu, iD_\nu \right] \right].$
The calculation proceeds by inserting the effective Hamiltonian $\mathcal{H}_{-1/3}$ in Eq.~\eqref{eq:Heff} into Eq.~\eqref{eq:T-operator}, taking the time-ordered product, and substituting the propagator from Eq.~\eqref{eq:S1p}. This corresponds to the calculation of the diagram with the soft-gluon emission in Fig.~\ref{fig:HQEDiags}.\footnote{In principle, diagrams involving the interference between a BSM and an SM vertex can contribute, corresponding to topologies without the $\psi$ particle. However, since the associated couplings $C_{uu}^{(d)}$ are extremely suppressed~\cite{Alonso-Alvarez:2021qfd}, these mixed contributions are numerically negligible and are therefore omitted in our analysis.} The soft gluon can be emitted from any of the quark propagators, but not from the dark fermion $\psi$, as it does not carry color charge. After rewriting $b(x)=e^{-ip_{p}x} \ b(0)$ \footnote{This follows from translation invariance of the free field and reflects that the external $b$-quark is described by a plane-wave state with momentum $p_b$.}
, where $p_b$ is the external $b$-quark momentum, and performing the $x-$ and momentum integrations, the resulting loop integrals take the generic form
\begin{equation}
\hspace*{-3.5mm}
 {\cal I}_{n_1 n_2 n_3}
\equiv 
\int \frac{d^4 l_1}{(2 \pi)^4} \, \int \frac{d^4 l_2}{(2 \pi)^4}  \frac{ f(p_b, l_1, l_2) }{ \big[ l_1^2 - m_1^2 + i \varepsilon \big]^{n_1} \big[ l_2^2 - m_2^2 + i \varepsilon \big]^{n_2} \big[ (l_1 + l_2 -p_b)^2 - m_3^2 + i \varepsilon \big]^{n_3}} \,,
\label{eq:MI-def}
\end{equation}
where $n_i \in {\mathbb N}_0$, and $f(p_b, l_1, l_2)$ is a function of all possible scalar products. Since these integrals are free of ultraviolet divergences, we work directly in $d = 4$ dimensions. They can 
be reduced to master integrals whose imaginary parts are known; see Ref.~\cite{Lenz:2020oce} and references therein for details.
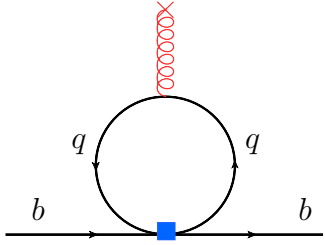
\begin{figure}[t!]
  \centering
  \setlength{\unitlength}{1pt}

  \begin{tabular}{ccccc}
    \begin{picture}(150,120)
      \put(0,0){\includegraphics[width=0.25\textwidth]{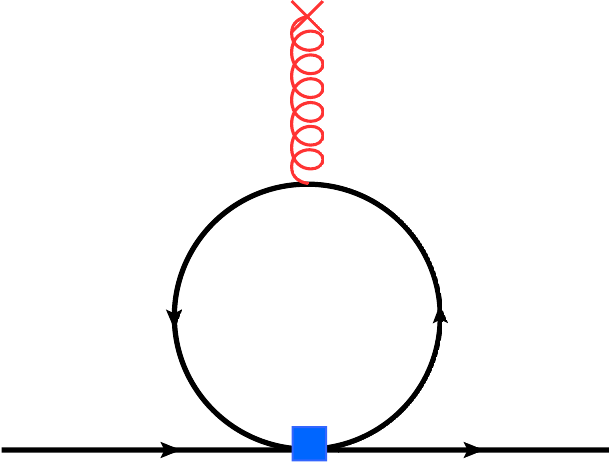}}
      \put(25,35){$q$}
      \put(90,35){$q$}
      \put(10,8){$b$}
      \put(110,8){$b$}
    \end{picture}
     \end{tabular}
     \caption{Diagram illustrating the mixing of the four-quark operator with the Darwin operator.}
\label{fig:MatrixMix}
    \end{figure}

For diagrams with a soft gluon emission from light quarks $q \in \{u,d,s\}$, the corresponding master integrals develop an infrared divergence in the $m_q \to 0$ limit. We regulate these divergences by keeping the light-quark masses finite, such that they appear as logarithmic terms of the form $\text{log}(m_q^2/{m_b^2})$ in the intermediate expressions. These infrared logarithms cancel against the renormalised one-loop matrix elements of the corresponding four-quark operators, see Fig.~\ref{fig:MatrixMix}, which are derived in our previous work~\cite{Lenz:2024rwi}. The resulting $\Delta B = 0$ four-quark operators $\tilde{O}_{n}$ are defined as \begin{align}
& \tilde{Q}^{(q)}_1 = 4 \,
( \bar{q}^i \gamma_\mu P_R \, b^i) 
(\bar b^j \gamma^\mu P_R \, q^j)\,,
\quad &
& \tilde{Q}^{(q)}_3 = 4 \,
(\bar q^i \gamma_\mu P_R t^{a}_{ij} \, b^j) 
(\bar b^r \gamma^\mu P_R t^{a}_{rm} \, q^m)\,,
\label{eq:Q1-Q2}
\\[2mm]
& \tilde{Q}^{(q)}_2 = 4 \,
(\bar q^i P_L \, b^i) 
(\bar b^j P_R \, q^j)\,,
\quad &
& \tilde{Q}^{(q)}_4 = 4 \,
(\bar q^i P_L t^{a}_{ij}\, b^j) 
(\bar b^r P_R t^{a}_{rm} \, q^m)\,. 
\label{eq:Q3-Q4}
\end{align}  
Notice that these operators coincide with the corresponding SM ones, up to the interchange $P_L \leftrightarrow P_R$ in all quark bilinears. The one-loop corrections to the above matrix elements are computed in dimensional regularization and the $\overline{\text{MS}}$ scheme, following the same cancellation mechanism as described in Sec.~2.2 of Ref.~\cite{Lenz:2020oce}.  
The renormalised matrix elements are found to be identical to the corresponding SM results, namely  
\begin{equation}
\begin{aligned}
\langle {\tilde{\cal O}}^{(q)}_{1,2} \rangle^{\text{ren}} &= 
 \frac{a_{1,2}}{12 \pi^2}  \left[ \log \left( \frac{m_b^2}{m_q^2}\right) + b_{1,2} \right] \langle {{\cal O}}_{\rho_D} \rangle + {\cal{O}}\left(\frac{1}{m_b}\right) , \\
 \langle {\tilde{\cal O}}^{(q)}_{3,4} \rangle^{\text{ren}} &= -\frac{1}{2N_c} \langle {\tilde{\cal O}}^{(q)}_{1,2} \rangle^{\text{ren}},
  \end{aligned}
\label{eq:crD-finite}
\end{equation}
where ${\cal O}_{\rho_D}$ is the Darwin operator,  
\[
{\cal O}_{\rho_D} = \bar b_v (i D_\mu) (i v \cdot D)(i D^\mu) b_v \,,
\]
see $\rho_D$ in Eq.~\eqref{eq:hadronic_parameters}, with $a_1=2$, $a_2=-1$, $b_1=-1$ and $b_2=0$.  
We note that Ref.~\cite{Lenz:2020oce} uses a different operator basis (the so-called colour-rearranged basis), and the constants above are basis-dependent.

After performing the loop integrations and removing the divergences, we obtain 
a matrix element of the generic form
\begin{equation}
\begin{aligned}
 \mathcal{F}(p_b) \ 
 \langle B(p_B)\,|\,\bar b_v(0) \, 
 \mathcal{M}_{\mu_1 \dots \mu_3} \,
 (iD_{\mu_1}) \dots (iD_{\mu_3}) \,
 b_v(0) \,|\, B(p_B) \rangle ,
\end{aligned}
\label{eq:MEGeneral}
\end{equation}
where $\mathcal{F}(p_b)$ is a function of the quark masses and the external 
$b$-quark momentum, and $\mathcal{M}_{\mu_1 \dots \mu_3}$ denotes a generic 
Dirac structure. For contributions of order $\mathcal{O}(1/m_b^3)$, 
there can be at most three covariant derivatives. By applying the HQET techniques, the $b-$momentum can be written as
\begin{equation}
p_b^\mu = m_b v^\mu + i D^\mu,
\end{equation}
where $v^\mu$ denotes the meson’s four-velocity and $D^\mu$ represents the residual momentum associated with soft gluon interactions with the spectator.  
Expanding in powers of $D^\mu/m_b$ generates local operators with increasing numbers of covariant derivatives.  
The ordering of Lorentz indices for multiple derivatives is fixed by symmetrization,
\begin{equation}
p_b^{\mu_1} p_b^{\mu_2} \dots p_b^{\mu_n} 
= \frac{1}{n!} \sum_{\sigma \in S_n}
p_b^{\sigma(\mu_1)} p_b^{\sigma(\mu_2)} \dots p_b^{\sigma(\mu_n)},
\end{equation}
where $S_n$ denotes the set of all permutations of $n$ indices.  
Finally, the matrix elements of the resulting local operators can be expressed in terms of the standard HQE basis involving the parameters $\mu_\pi^2$, $\mu_G^2$, $\rho_{LS}^3$, and $\rho_D^3$ in Eq.~\eqref{eq:hadronic_parameters} using the trace formalism of Refs.~\cite{Dassinger:2006md,Mannel:2023yqf}.

The final results of the Mesogenesis operators can be written as
\begin{equation}
\Gamma^{(i)} \;=\; 
\frac{\big|C^{(i)}\big|^2\, m_b^5}{3072\,\pi^3}\;
\left[
   f_{0}^{(i)}(\rho_f) \left(1- \frac{\mu_\pi^2}{2m_b^2}\right) 
 + f_{G}^{(i)}(\rho_f)\,\frac{\mu_G^2}{m_b^2}
 + f_{D}^{(i)}(\rho_f)\,\frac{\rho_D^3}{m_b^3}
\right],
\label{eq:MasterWidth}
\end{equation}
with \(\rho_f \equiv (m_f/m_b)^2\). Note that the spin–orbit parameter \(\rho_{LS}^3\) does not contribute to these decay widths, as its associated operator yields a vanishing matrix element for the considered Dirac structures. The Wilson coefficients \(C^{(i)}\) are defined in Eq.~\eqref{eq:Cpsipsi_map}, and \(f_{0}^{(i)}(\rho_f)\), \(f_{G}^{(i)}(\rho_f)\), and \(f_{D}^{(i)}(\rho_f)\) are dimensionless kinematic functions 
that depend on the operator structure. The results for the operator proportional to $C_\psi^{(ud)}$ read
\begin{flalign}
\nonumber f_{0}^{(\psi ud)}(\rho_\psi) &=\, 
1 - 8\,\rho_\psi + 8\,\rho_\psi^{3} - \rho_\psi^{4} - 12\,\rho_\psi^{2}\log\rho_\psi, & \\[4pt]
\nonumber f_{G}^{(\psi ud)}(\rho_\psi) &=\, 
\frac{5}{2} - 8\,\rho_\psi + 8\,\rho_\psi^{3} - \frac{5}{2}\,\rho_\psi^{4} - 6\,\rho_\psi^{2}\log\rho_\psi, & \\[6pt]
f_{D}^{(\psi ud)}(\rho_\psi) &=\, 
\frac{2}{3}\!\left(
22 - 31\,\rho_\psi + 3\,\rho_\psi^{2} + 11\,\rho_\psi^{3} - 5\,\rho_\psi^{4} - 12\,(1-\rho_\psi)^{2}(3+\rho_\psi)\log(1-\rho_\psi)\right. \nonumber \\
& \quad \left.
+ 6\,\rho_\psi^{2}(1+\rho_\psi)\log\rho_\psi
\right). &
\end{flalign}
and for $\widetilde{C}_\psi^{(ud)}$
\begin{flalign}
\nonumber f_{0}^{(\tilde\psi ud)}(\rho_\psi) &=\;
1 - 8\rho_\psi + 8\rho_\psi^{3} - \rho_\psi^{4} - 12\rho_\psi^{2}\log\rho_\psi, & \\[6pt]
\nonumber f_{G}^{(\tilde\psi ud)}(\rho_\psi) &=\;
-\tfrac{1}{2}\Big(3 - 8\rho_\psi + 24\rho_\psi^{2} - 24\rho_\psi^{3} + 5\rho_\psi^{4} + 12\rho_\psi^{2}\log\rho_\psi\Big), & \\[6pt]
f_{D}^{(\tilde\psi ud)}(\rho_\psi) &=\;
\tfrac{2}{3}\,(1-\rho_\psi )\Big(9 + 11\rho_\psi - 25\rho_\psi^{2} + 5\rho_\psi^{3} 
   - 24(1-\rho_\psi^{2})\log(1-\rho_\psi) - 12\rho_\psi^{2}\log\rho_\psi\Big). &
\end{flalign}
The results for the remaining operators are given in Appendix~\ref{app:C}.

As a nontrivial cross-check, we verified that the infrared divergences cancel as described above. Moreover, after properly accounting for the color factors and Dirac structures of the operators, we reproduce the known SM results~\cite{Mannel:2020fts,Lenz:2020oce}, with $\rho_\psi$ in our case corresponding to $\rho_c$ in the SM. Appendix~\ref{app:B} illustrates how these ingredients enter the comparison with the SM case.
\subsection{Lifetime Ratios}
Experimentally, the lifetime ratios of $B$ mesons are measured with high precision~\cite{HeavyFlavorAveragingGroupHFLAV:2024ctg}:
\begin{equation}
\left(\frac{\tau(B^+)}{\tau(B_d)}\right)^{\rm Exp.} = 1.076 \pm 0.004, \quad \left(\frac{\tau(B_s)}{\tau(B_d)}\right)^{\rm Exp.} = 0.993 \pm 0.004.
\label{eq:ratio_exp}
\end{equation}
The corresponding SM predictions from the HQE are in good agreement, despite the large uncertainties, see e.g.~\cite{Albrecht:2024oyn,Lenz:2022rbq, Egner:2024lay}:
\begin{equation}
\left.\left(\frac{\tau(B^+)}{\tau(B_d)}\right)^{\rm HQE}\right|_{\rm SM}
= 1.081^{+0.014}_{-0.016}, 
\qquad
\left.\left(\frac{\tau(B_s)}{\tau(B_d)}\right)^{\rm HQE}\right|_{\rm SM}
= 1.013^{+0.007}_{-0.007}.
\label{eq:ratio_HQE}
\end{equation}
We now extend the HQE expression for the lifetime ratios by including BSM contributions. The lifetime ratios are then expressed as
\begin{equation}
 \frac{\tau (B_{q_1})}{\tau (B_{q_2})}^{\rm HQE} \!\!\!\!\! = 1 + 
\left[\Gamma^{\rm SM} (B_{q_2}) - \Gamma^{\rm SM} (B_{q_1}) \right] 
\tau^{\rm Exp.} (B_{q_1}) 
+
\left[\Gamma^{\rm BSM} (B_{q_2}) - \Gamma^{\rm BSM} (B_{q_1}) \right]
\tau^{\rm Exp.} (B_{q_1}),
\label{eq:ratio_BSM}
\end{equation}
where $\Gamma^{\rm SM}$ denotes the SM contribution to the inclusive decay width, while $\Gamma^{\rm BSM}$ represents the additional contribution arising purely from BSM operators.

Because of isospin symmetry, the contributions from the two-quark operator terms $\Gamma_i$ in Eq.~\eqref{eq:HQE} cancel in the difference~\eqref{eq:ratio_BSM} for $\tau(B^+)/\tau(B_d)$. Consequently, this ratio is predominantly sensitive to the dimension-six four-quark operators, arising from spectator effects such as Weak Annihilation (WA) and Pauli Interference (PI), namely the terms $\tilde{\Gamma}_i$ in Eq.~\eqref{eq:HQE}.  In contrast, the ratio $\tau(B_s)/\tau(B_d)$ is largely protected from these spectator effects in the exact $SU(3)_F$ limit, and deviations from unity are driven by $SU(3)_F$-breaking contributions.  
These originate from subleading terms, including dimension-six two-quark operators, making $\tau(B_s)/\tau(B_d)$ a sensitive probe of small $SU(3)_F$-breaking effects.

Using our results for the contributions of the two-quark operators up to and including the Darwin term, together with the four-quark operator contributions determined in Ref.~\cite{Lenz:2024rwi}, and applying Eq.~\eqref{eq:ratio_BSM}, we can place constraints on the coupling appearing in the effective Hamiltonian~\eqref{eq:Heff}.
\subsection{Exclusive Decays}
The exclusive decay mode $B^+ \to p^{+}\psi$ has been investigated in Ref.~\cite{Boushmelev:2023huu} for the $\psi$-type operators within the framework of QCD LCSR. The calculation employed nucleon distribution amplitudes (DAs) up to twist six. In that work, the terms with Wilson coefficients $C_\psi^{ud}$ and $\widetilde{C}_\psi^{ud}$ from Eq.~\eqref{eq:Heff} were referred to as “model (d)” and “model (b),” respectively.
The corresponding decay width takes the form 
\begin{align}
	&\Gamma_{(d)}(B^+\to p\psi) =
	|G_{(d)}|^2 \Bigg\{\Bigg[\Big(F^{(d)}_{B\to p_R}(m_\psi^2)\Big)^2
	+ \frac{m_\psi^2}{m_p^2}\Big(\widetilde{F}^{(d)}_{B\to p_L}(m_\psi^2)\Big)^2\Bigg] \nonumber \\
	&\times \big(m_B^2-m_p^2-m_\psi^2\big)
	+ \;
	2 m_\psi^2F^{(d)}_{B\to p_R}(m_\psi^2)\widetilde{F}^{(d)}_{B\to p_L}(m_\psi^2)\Bigg\}
	\,\frac{\lambda^{1/2}(m_B^2,m_p^2,m_\psi^2)}{16\pi m_B^3}\, ,
	\label{eq:widthd}
\end{align}\noindent
where $  |G_{(d)}|^2=|C^{(ud)}_{\psi}|^{2}$ in our notation, and $F^{(d)}_{B \to p_R}(q^2)$ and $ \widetilde{F}^{(d)}_{B\to p_L} (q^2)$ are the form factors
\begin{align}
	F^{(d)}_{B \to p_R}(q^2) =& \; \frac{F^{(d)}_{B \to p_R}(0)}{1 - q^2/m_{\Lambda_b}^2} \Bigg[1 + b^{(d)}_{B \to p_R} \bigg(z(q^2) - z(0) + \frac{1}{2} \Big[z(q^2)^2 - z(0)^2\Big]\bigg)\Bigg]\,,  \label{eq:FFzExpFinal}
\end{align}
and $ \widetilde{F}^{(d)}_{B\to p_L} (q^2)$ is obtained by the replacement $F^{(d)}_{B \to p_R}(q^2) \to \, \widetilde{F}^{(d)}_{B\to p_L} (q^2)$. The conformal variable $z(q^2)$ is defined as
\begin{align}
z(q^2) = \frac{\sqrt{t_+ - q^2} - \sqrt{t_+ - t_0}}{\sqrt{t_+ - q^2} + \sqrt{t_+ - t_0}},
\end{align}
with $t_0 = (m_B+m_p)(\sqrt{m_B}-\sqrt{m_p})^2$ and $t_\pm = (m_B \pm m_p)^2$. Here $m_B$, $m_p$, and $m_{\Lambda_b}$ denote the masses of the $B$-meson, proton, and $\Lambda_b$ baryon, respectively. The function $\lambda(x,y,z) = x^2+y^2+z^2-2xy-2xz-2yz$ is the standard Källén function. The slope parameter $b^{(d)}_{B \to p_R}$ (and analogously for the other form factors) is determined from fits to the LCSR results in Ref.~\cite{Boushmelev:2023huu}. The expressions for model (b) follow from the replacement $(d) \to (b)$ ($ |G_{(b)}|^2=|\widetilde{C}^{ud}_{\psi}|^{2}$).

The numerical inputs employed for the form factors and slope parameters are summarized in Table~\ref{tab:LCSRinput}.
\begin{table}[h]
\centering
\scalebox{1}{
\begin{tabular}{|c|c|c|c|}
\hline
$F^{(d)}_{B\to p_R}(0)$ & $\widetilde{F}^{(d)}_{B\to p_L}(0)$ & $F^{(b)}_{B\to p_R}(0)$ & $\widetilde{F}^{(b)}_{B\to p_L}(0)$ \\
\hline
$0.022^{+0.013}_{-0.013}$ & $0.005^{+0.002}_{-0.001}$ & $-0.041^{+0.019}_{-0.018}$ & $-0.007^{+0.003}_{-0.002}$ \\
\hline
$b^{(d)}_{B\to p_R}$ & $b^{(d)}_{B\to p_L}$ & $b^{(b)}_{B\to p_R}$ & $b^{(b)}_{B\to p_L}$ \\
\hline
$4.46^{+0.97}_{-1.72}$ & $-2.27^{+0.10}_{-0.08}$ & $-2.00^{+1.58}_{-3.62}$ & $-2.85^{+0.17}_{-0.15}$ \\
\hline
\end{tabular}}
\caption{Input parameters in the LCSR~\cite{Boushmelev:2023huu}. All values are given in GeV$^2$.}
\label{tab:LCSRinput}
\end{table}

\section{Numerical Analysis and Phenomenology}
\label{numerics}
\subsection{Inclusive Decays}
\label{sec:InclusiveTotalLifetimes}
Since the HQE is derived under the assumption that the final-state masses are small compared to $m_b$, one expects the OPE to converge well in the regime $m_\psi \ll m_b$. However, in the Mesogenesis framework the dark-sector mass $m_\psi$ can range up to values comparable to $m_b$, where the validity of the OPE is not guaranteed. It is therefore essential to analyze systematically how the subleading terms behave as functions of $m_\psi$, in order to assess the range of applicability of the HQE and to identify the point at which the expansion begins to break down.

In the following, we analyze, for each operator in the effective Hamiltonian~\eqref{eq:Heff}, the individual contributions of the HQE terms in Eq.~\eqref{eq:HQE} to the total decay width of the $B^+$ meson. We focus on the $B^+$ meson because, as we will demonstrate, it exhibits unique and interesting topologies, and it allows us to derive a relevant experimental lower limit for the exclusive decay $B^+ \to p^+ \psi$ in Section.~(\ref{lower limits}). We note that, generally, the HQE contributions are nearly identical for two-quark operators across all $B$ mesons ($B^+$, $B^0$, $B_s^0$). However, the four-quark operator contributions are weaker for the $B^0$ and $B_s^0$ mesons. Furthermore, certain operators—notably those with two massive fermions besides the $b$-quark—do not contribute to the $B^+$ width. We derive the contributions to $B^0$ for these specific operators, see Appendix~\ref{app:D}.

Since in the scenarios of interest a single operator is expected to dominate~\cite{Alonso-Alvarez:2021qfd}, we adopt the strategy of switching on one operator at a time while setting all others to zero. For each case, we display the separate contributions to $\Gamma(B^+)$ from the leading dimension-three term $\Gamma_3$, the dimension-five term $\Gamma_5$, the Darwin term $\Gamma_6$, and the dimension-six four-quark operators $\tilde{\Gamma}_6$ as functions of $\rho_f = (m_f/m_b)^2$. We also define the ratio $R_d=|\Gamma_d/\Gamma_3|$, where the numerator corresponds to one of the subleading contributions listed above. The total decay width, obtained by summing all contributions, is likewise shown as a function of $\rho_f$. For clarity, the numerical inputs used throughout this analysis are collected in Appendix~\ref{app:A}.

We stress that, throughout this analysis of the total decay rate, we consider exclusively the new-physics (NP) contributions from the Mesogenesis operators, omitting the SM part of the width. This allows us to assess the relative importance of subleading terms directly within the BSM framework.\footnote{Using couplings values as constrained by current ATLAS/CMS collider limits~\cite{Alonso-Alvarez:2021qfd}, the new-physics contribution remains at most $\sim 3\%$ of the SM total width, i.e.\ negligible compared to present experimental and theoretical uncertainties.}

\begin{figure}[t]
    \centering
    \begin{subfigure}{0.49\textwidth}
        \centering
        \includegraphics[width=\textwidth]{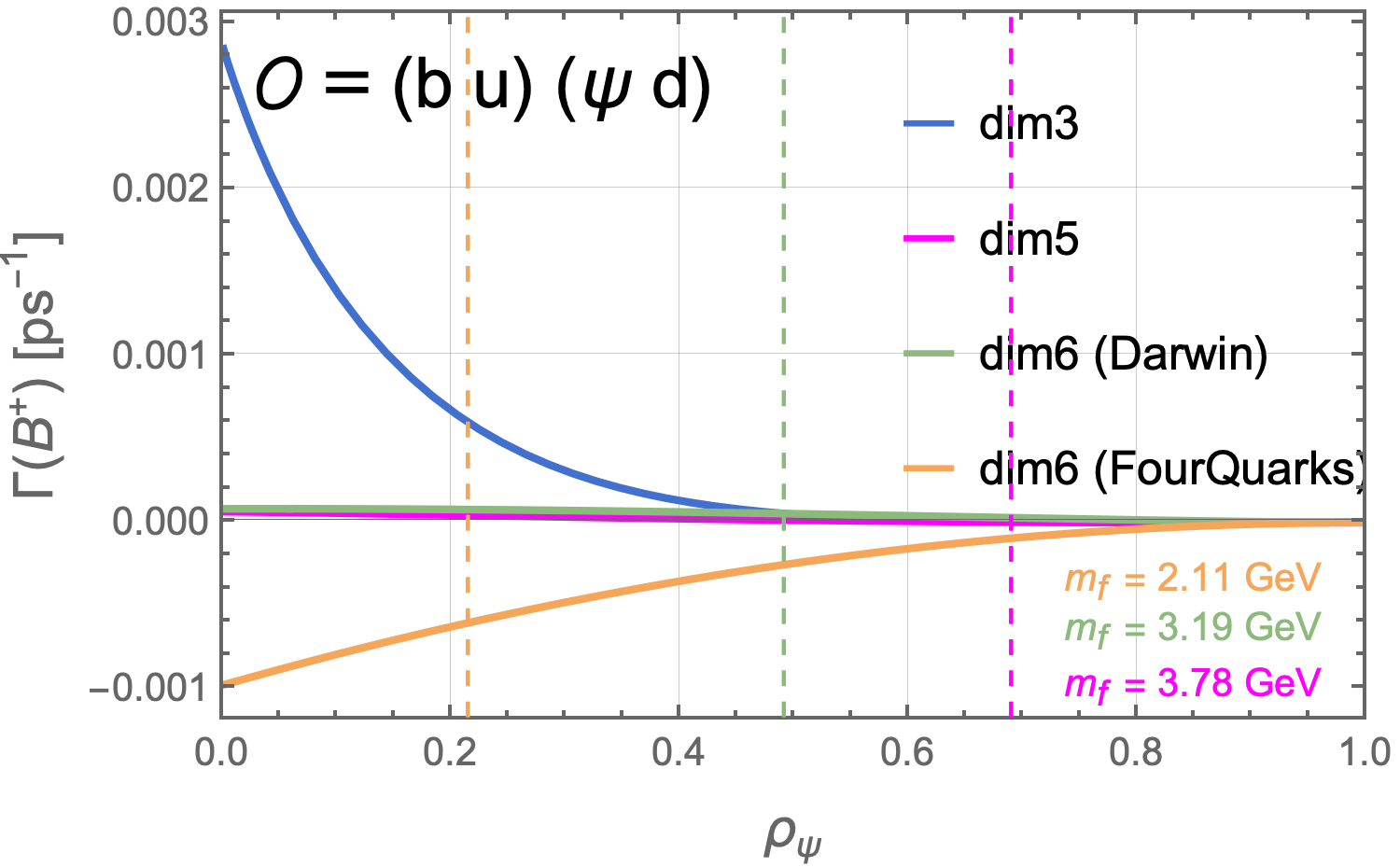}
    \end{subfigure}
    \hfill
    \begin{subfigure}{0.47\textwidth}
        \centering
        \includegraphics[width=\textwidth]{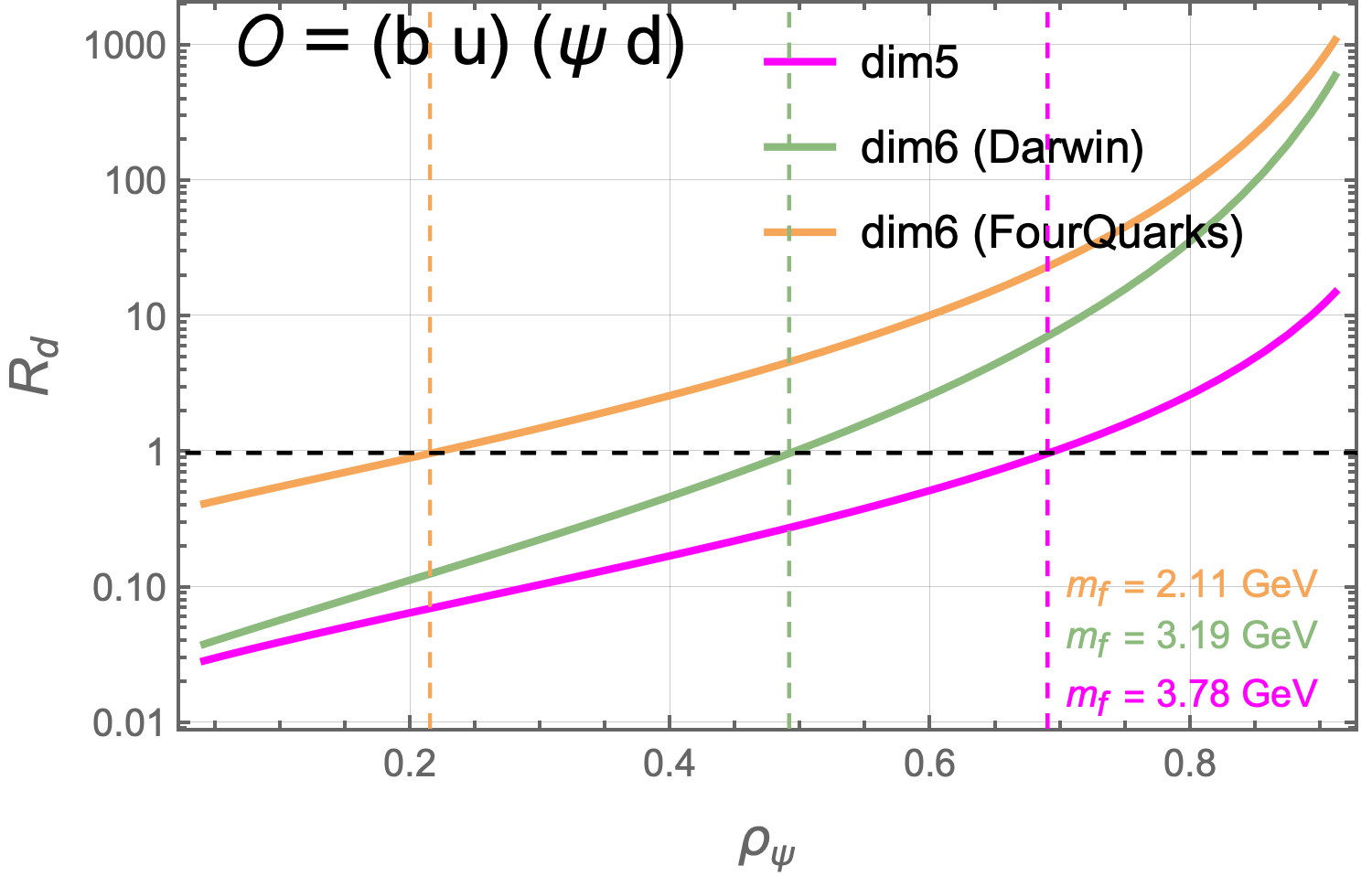}
    \end{subfigure}
    
    \medskip 
    
    \begin{subfigure}{0.49\textwidth}
        \centering
        \includegraphics[width=\textwidth]{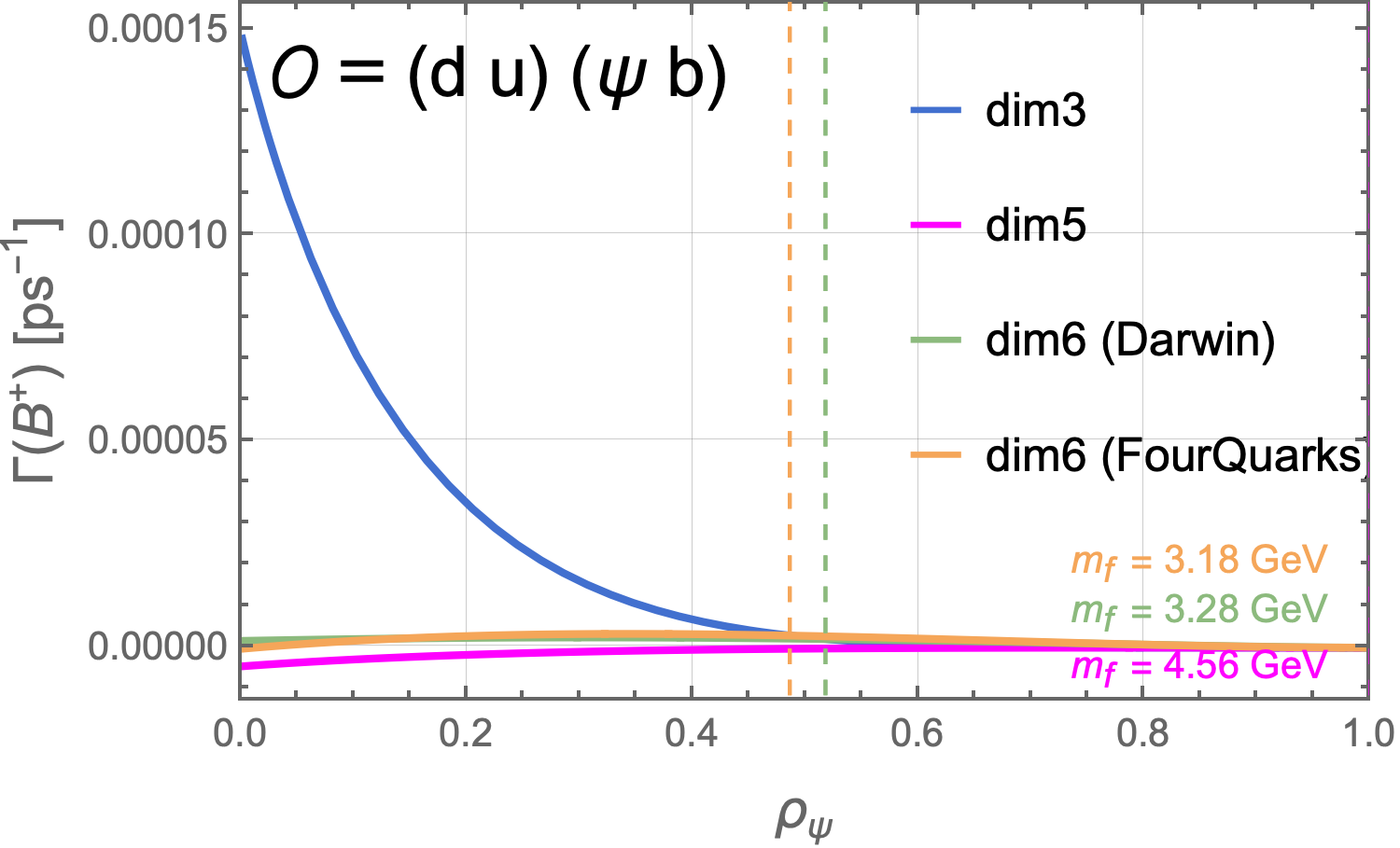}
    \end{subfigure}
    \hfill
    \begin{subfigure}{0.47\textwidth}
        \centering
        \includegraphics[width=\textwidth]{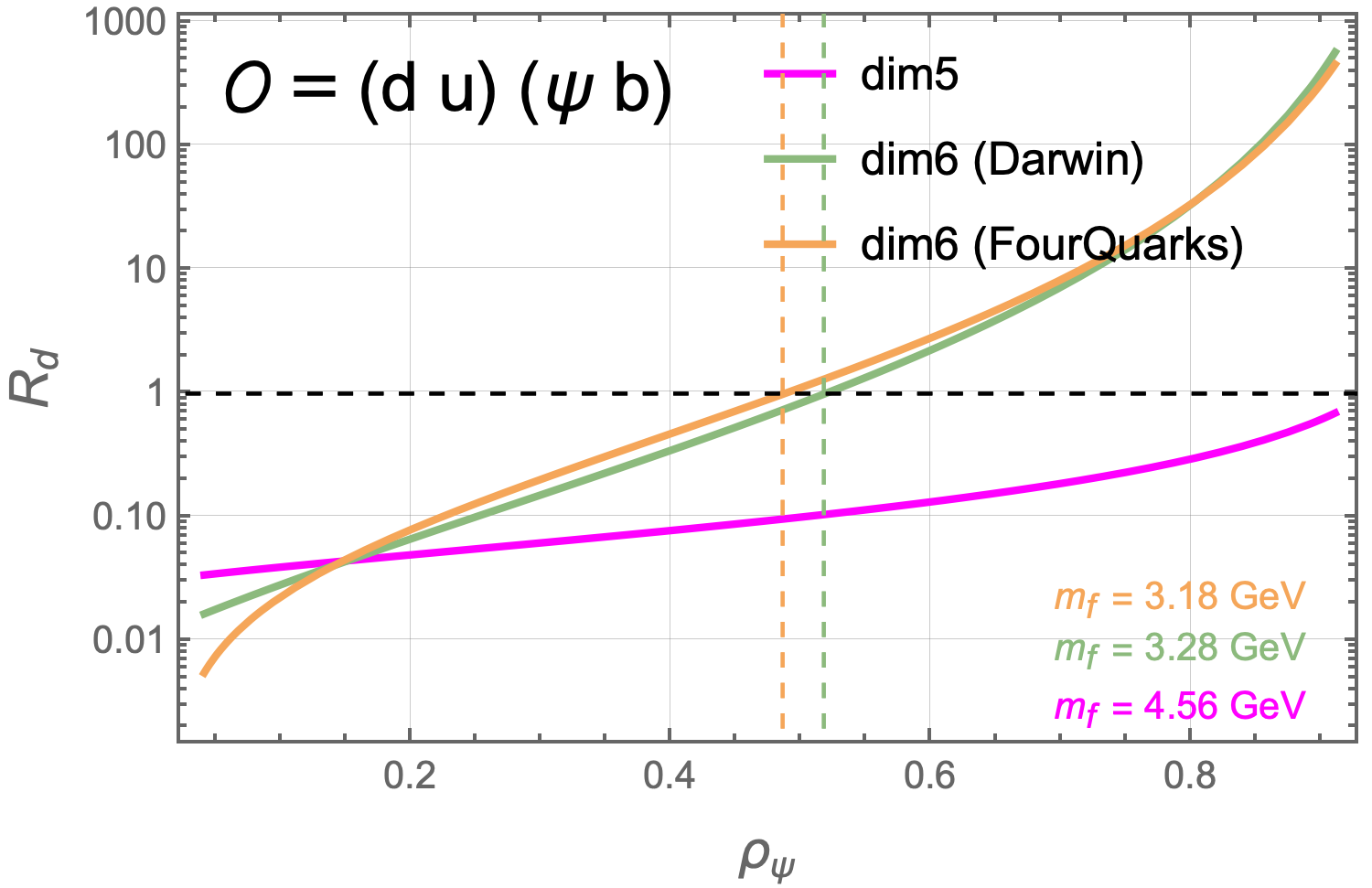}
    \end{subfigure}
    \caption{Total decay width $\Gamma(B^+)$ arising purely from the Mesogenesis (BSM) operators as a function of $\rho_f= (m_f/m_b)^2$. The left column displays individual HQE contributions, and the right column shows the ratio $R_d=|\Gamma_d/\Gamma_3|$, where $\Gamma_d$ is the subleading term indicated. Vertical lines mark the $\rho_f$ values where the subleading contribution exceeds the leading dimension-three term $\Gamma_3$ (the corresponding $m_f$ values are also indicated).}
    \label{fig:DecayRates}
\end{figure}

\begin{figure}[h]
    \centering
    \begin{subfigure}{0.49\textwidth}
        \centering
        \includegraphics[width=\textwidth]{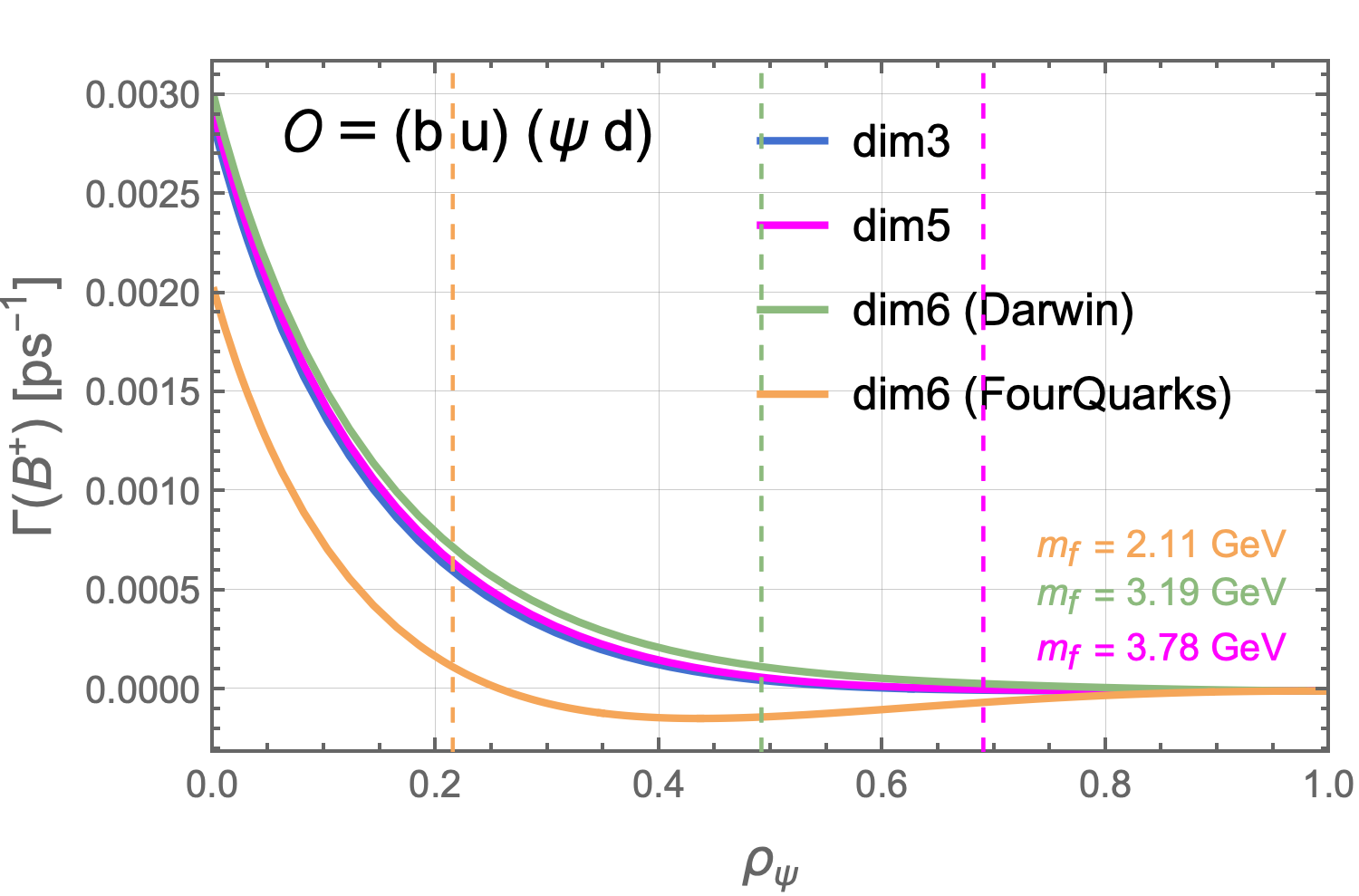}
    \end{subfigure}
    \hfill
    \begin{subfigure}{0.49\textwidth}
        \centering
        \includegraphics[width=\textwidth]{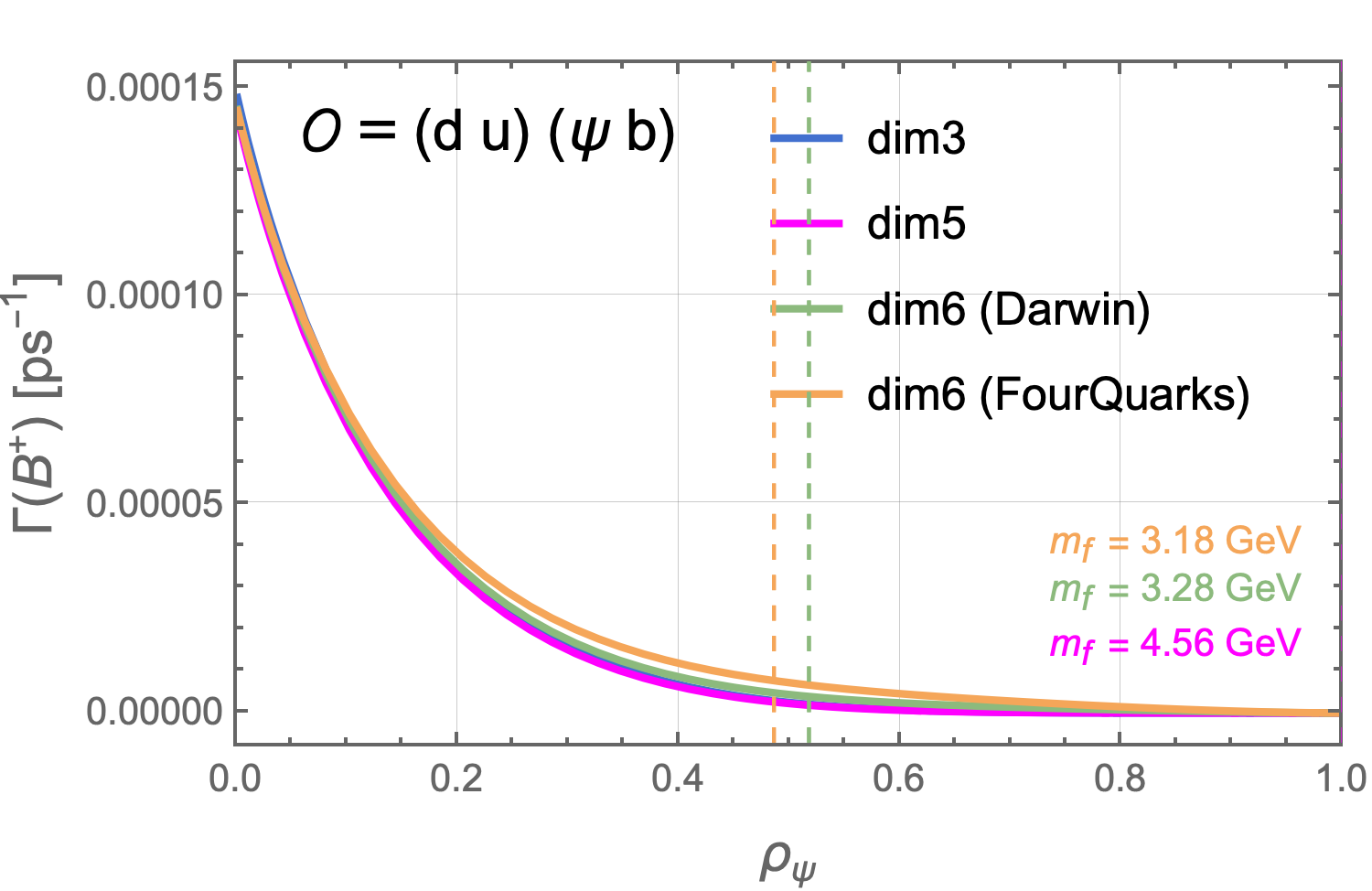}
    \end{subfigure}
        \caption{Total decay width of $\Gamma(B^+)$ arising purely from the Mesogenesis (BSM) operators as a function of $\rho_f= (m_f/m_b)^2$. The figure displays the cumulative sum of all terms up to the indicated dimension. The vertical lines indicate the value of $\rho_f$, where the contribution exceeds the leading dimension-three $\Gamma_3$ term (the corresponding $m_f$ values are also indicated).}\label{fig:DecayRatesTotal}
\end{figure}

For definiteness, Figs.~\ref{fig:DecayRates} and~\ref{fig:DecayRatesTotal}
present the results for two representative operators, which we discuss below; the corresponding plots for the remaining operators are provided in Appendix~\ref{app:D}. In the figures, the operators are labeled by their field content (e.g.\ $(b u)(\psi d)$), which directly corresponds to the operator structures defined in Eq.~\eqref{eq:Heff}. The values of the Wilson coefficients used in the plots correspond to their maximal values inferred from ATLAS and CMS searches for heavy colored particles, as discussed in Ref.~\cite{Alonso-Alvarez:2021qfd} (see Appendix~\ref{app:A}).

Key observations from the analysis are:

\begin{itemize}
    \item \textbf{Hierarchy of contributions:} Generally, we find
    $
     \tilde\Gamma_6 > \Gamma_6 > \Gamma_5,
    $
    i.e., the dimension-six four-quark contributions dominate the Darwin term, which in turn is larger than the dimension-five corrections. This mirrors the pattern observed in the SM, where the dimension-six four-quark effect arises from PI topologies~\cite{Lenz:2020oce,Egner:2024lay}. The dominance of $\tilde\Gamma_6$ is driven by both the phase-space/loop enhancement of four-quark spectator topologies (the familiar $16\pi^2$ factor, see Eq.~(\ref{eq:HQE})) and the specific colour/bag-parameter combinations (see Appendix~\ref{app:B}). Fig.~\ref{fig:Topologies} illustrates the relevant topologies of the dimension-six four-quark operators for the two representative cases.
    
    \item \textbf{Subleading contributions vs. leading term:} For sufficiently large $m_\psi$, subleading terms can exceed the leading dimension-three contribution. These regions are explicitly indicated by vertical lines in the plots, marking the $\rho_f$ values where $R_d > 1$. This behavior is expected once final-state masses approach $m_b$, and our analysis provides a quantitative determination of the parameter space where the HQE hierarchy breaks down. An analogous pattern emerges in the SM if the charm-quark mass is varied across the same range (see Appendix~\ref{app:B}).

    \item \textbf{Operator-dependent behavior:} For the operator $\mathcal{O} = (b\,u^c)(\psi \,d^c)$, the dimension-six four-quark contribution surpasses the leading dimension-three term over a substantial portion of the $\rho_\psi$ range. This is a generic feature of operators of the form $\mathcal{O} = (b\,q_1^c)(q_2\,q_3^c)$, where $q_1$ is the spectator quark and $q_2, q_3$ are arbitrary fermions. Such structures generate a $\overline{\text{PI}}$ topology (see Fig.~\ref{fig:Topologies}) that can yield a large negative contribution. A similar pattern is observed for $\Gamma(B_d)$ in operators from $\mathcal{L}_{2/3}$, e.g. $\mathcal{O} = (b\,d^c)(\psi\,u^c)$, where the spectator quark couples directly to the $b$-quark, again producing the $\overline{\text{PI}}$ topology.

    \item \textbf{Comparison to the SM:} In the SM, the operator $\mathcal{O} = (u\,b)(d\,u)$ has a similar form and generates a PI topology, but the spectator quark in the PI diagram is coupled to the $d$-quark rather than to the $b$-quark. For $\mathcal{O} = (u\,b)(dc)$, the $b$-quark couples to the spectator, but the resulting topology is not PI. Thus, the enhanced behavior in Mesogenesis arises from the unusual operator structures and the presence of charge-conjugated fields, which bypass certain topology-based suppressions present in the SM. A detailed operator-level comparison between the Mesogenesis $\overline{\rm PI}$ and the SM PI contributions is provided in Appendix~\ref{app:B}.

    \item \textbf{Comparison with exclusive LCSR results:} In the LCSR analysis of $B^+ \to p^+ \psi$~\cite{Boushmelev:2023huu}, the operator $\mathcal{O} = (b\,u^c)(\psi\,d^c)$ receives subleading twist contributions larger than the leading term for $m_\psi \gtrsim 3~\text{GeV}$, while the leading-twist result is a good approximation below this value. Our inclusive analysis shows a qualitatively similar behavior: the Darwin term dominates for $m_\psi \gtrsim 3~\text{GeV}$, while the dimension-six four-quark contribution surpasses the leading term already at $m_\psi \gtrsim 2~\text{GeV}$. For $\mathcal{O} = (d\,u^c)(\psi\,b^c)$, subleading effects remain smaller than the leading term, in contrast to our inclusive results. However, the high-twist vs. leading-twist differences in the LCSR analysis highlight the limitations of considering only leading contributions.
\end{itemize}
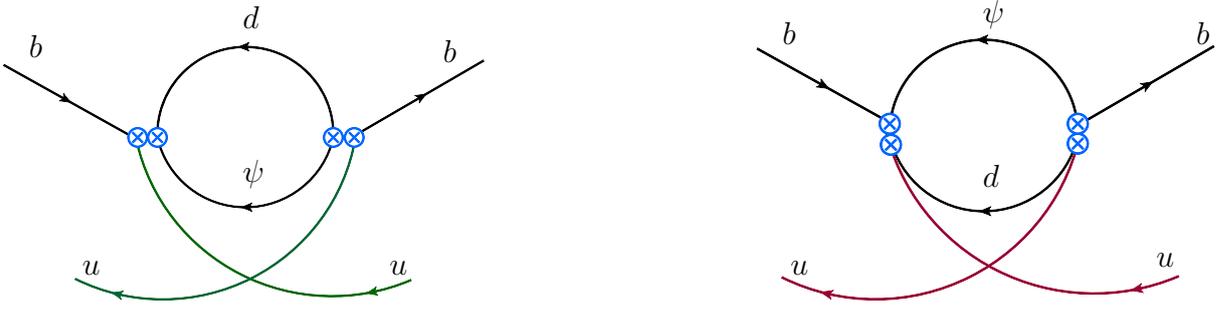
\begin{figure}[t]
  \centering
  \begin{minipage}{0.42\textwidth}
    \centering
    \setlength{\unitlength}{1pt}
    \begin{picture}(150,120)
      \put(0,0){\includegraphics[width=0.9\textwidth]{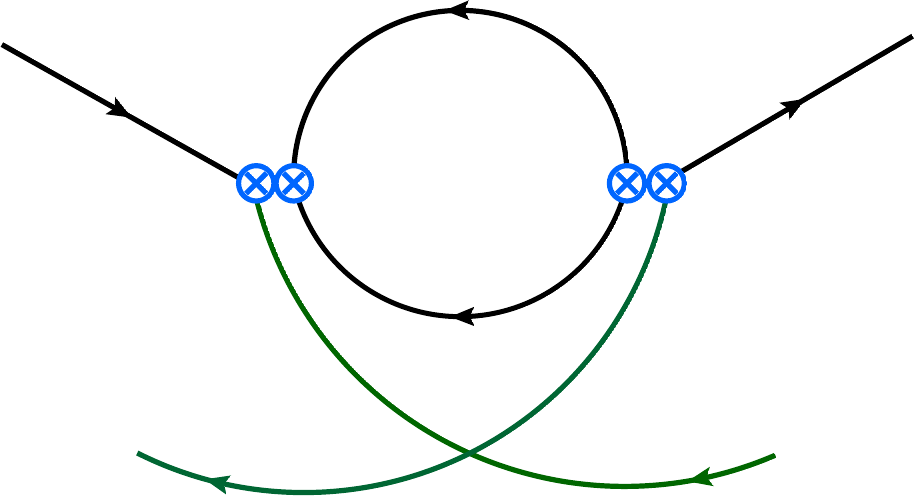}}
      \put(90,103){$d$}
      \put(90,45){$\psi$}
      \put(10,92){$b$}
      \put(30,10){$u$}
      \put(165,90){$b$}
      \put(145,10){$u$}
    \end{picture}
  \end{minipage}%
  \hfill
  \begin{minipage}{0.40\textwidth}
    \centering
    \setlength{\unitlength}{1pt}
    \begin{picture}(150,120)
      \put(0,0){\includegraphics[width=0.9\textwidth]{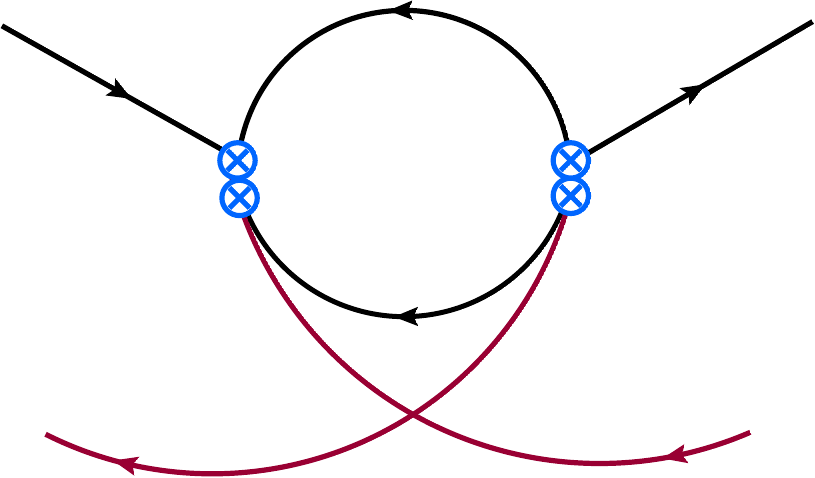}}
      \put(85,105){$\psi$}
      \put(85,43){$d$}
      \put(10,97){$b$}
      \put(13,10){$u$}
      \put(165,97){$b$}
      \put(150,13){$u$}
    \end{picture}
  \end{minipage}

  \caption{Topologies of the four-quark contributions to the total decay width $\Gamma(B^+)$ for the Mesogenesis operators: $\mathcal{O}=(b\,u)(\psi \,d)$ (left, $\overline{\text{PI}}$) and $\mathcal{O}=(d\,u)(\psi\,b)$ (right, $\text{PI}$).}

 \label{fig:Topologies}
\end{figure}
We therefore conclude that HQE hierarchy can be strongly operator-dependent, and that models with non-SM operators should be examined for such effects before deriving phenomenological conclusions. \\
We stress, however, that our findings are performed at LO. Experience from the SM shows that NLO QCD corrections and higher-dimensional terms can modify the relative sizes of individual contributions (see e.g. the discussions in Ref~\cite{Egner:2024lay}). The present ordering should therefore be regarded as an empirical LO result, pending confirmation from more complete studies.
\subsection{Exclusive Decays}
\label{lower limits}
\begin{figure}[ht]
    \centering
    \begin{subfigure}{0.49\textwidth}
        \centering
        \includegraphics[width=\textwidth]{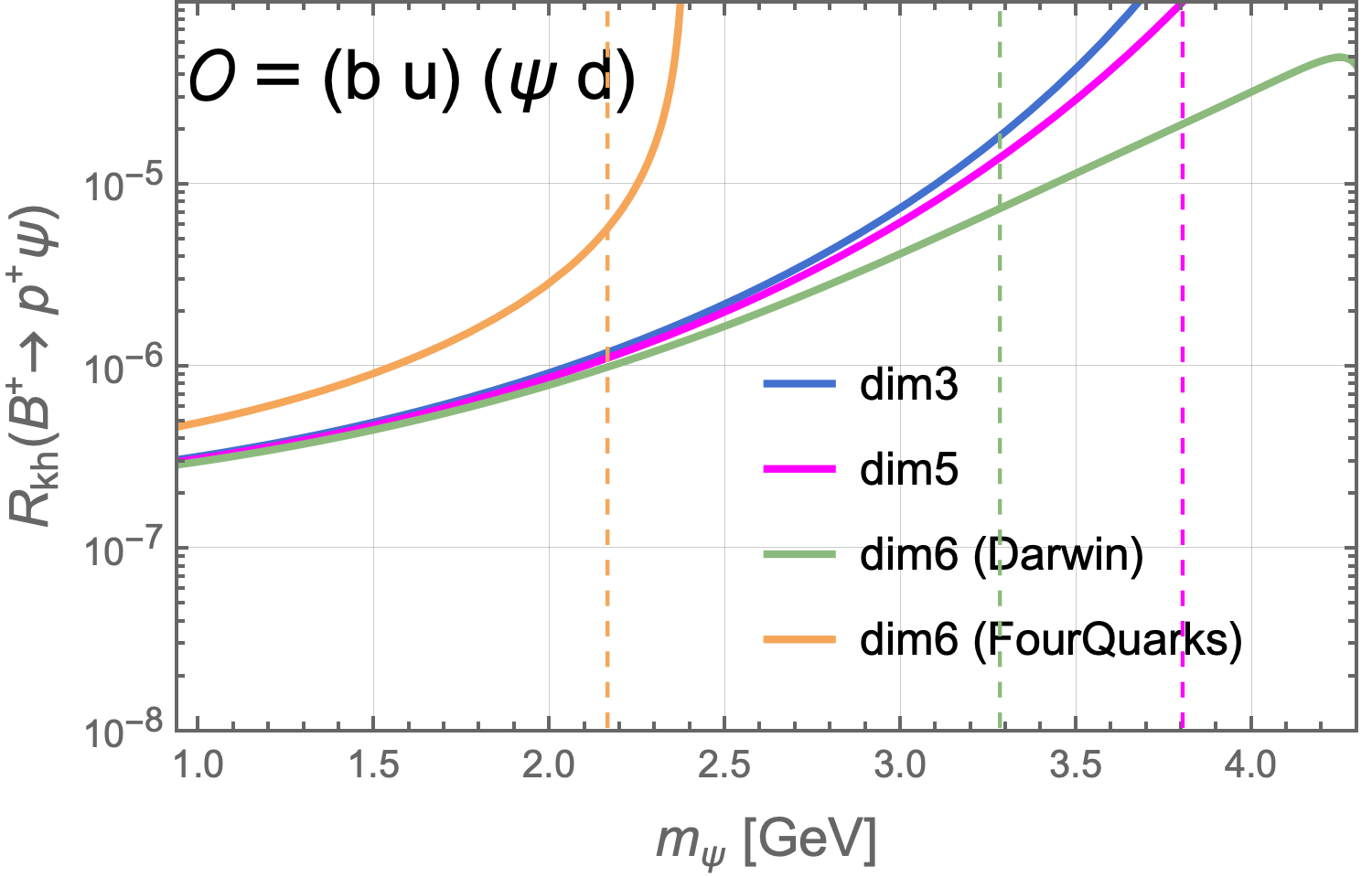}
    \end{subfigure}
    \hfill
    \begin{subfigure}{0.49\textwidth}
        \centering
        \includegraphics[width=\textwidth]{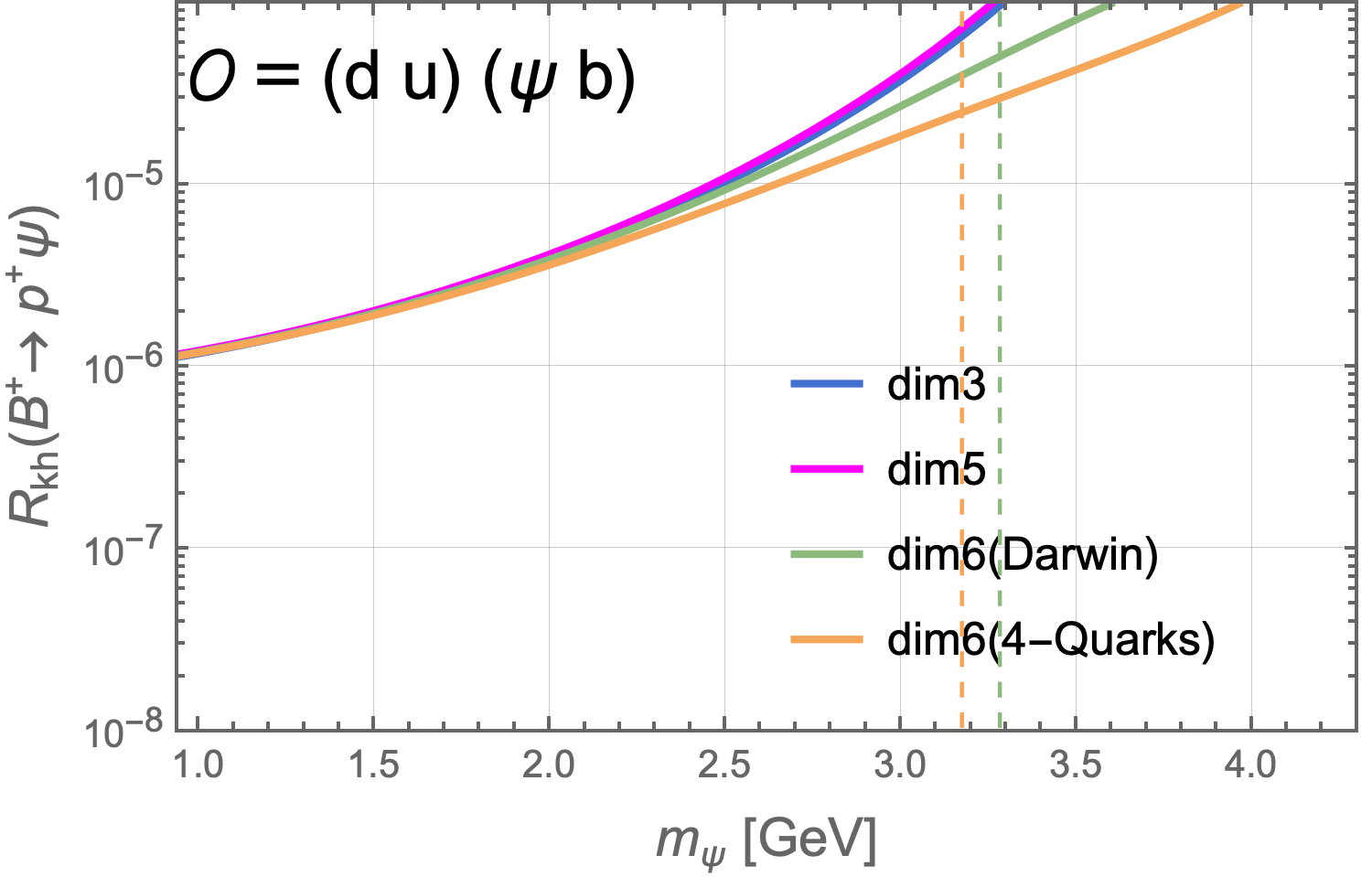}
    \end{subfigure}
       \caption{Lower limits for $Br (B^+ \to p^+ \psi)$ (given by $R_{Kh}$) from inclusive rates including subleading HQE terms up to the dimension indicated in the legend (e.g., ``dim5'' includes dim3 + dim5 terms). Vertical dashed lines indicate $m_\psi$ where subleading terms overtake the leading contribution.}
        \label{fig:RkhLowerLimits}
\end{figure}

\begin{figure}[ht]
    \centering
    \begin{subfigure}{0.49\textwidth}
        \centering
        \includegraphics[width=\textwidth]{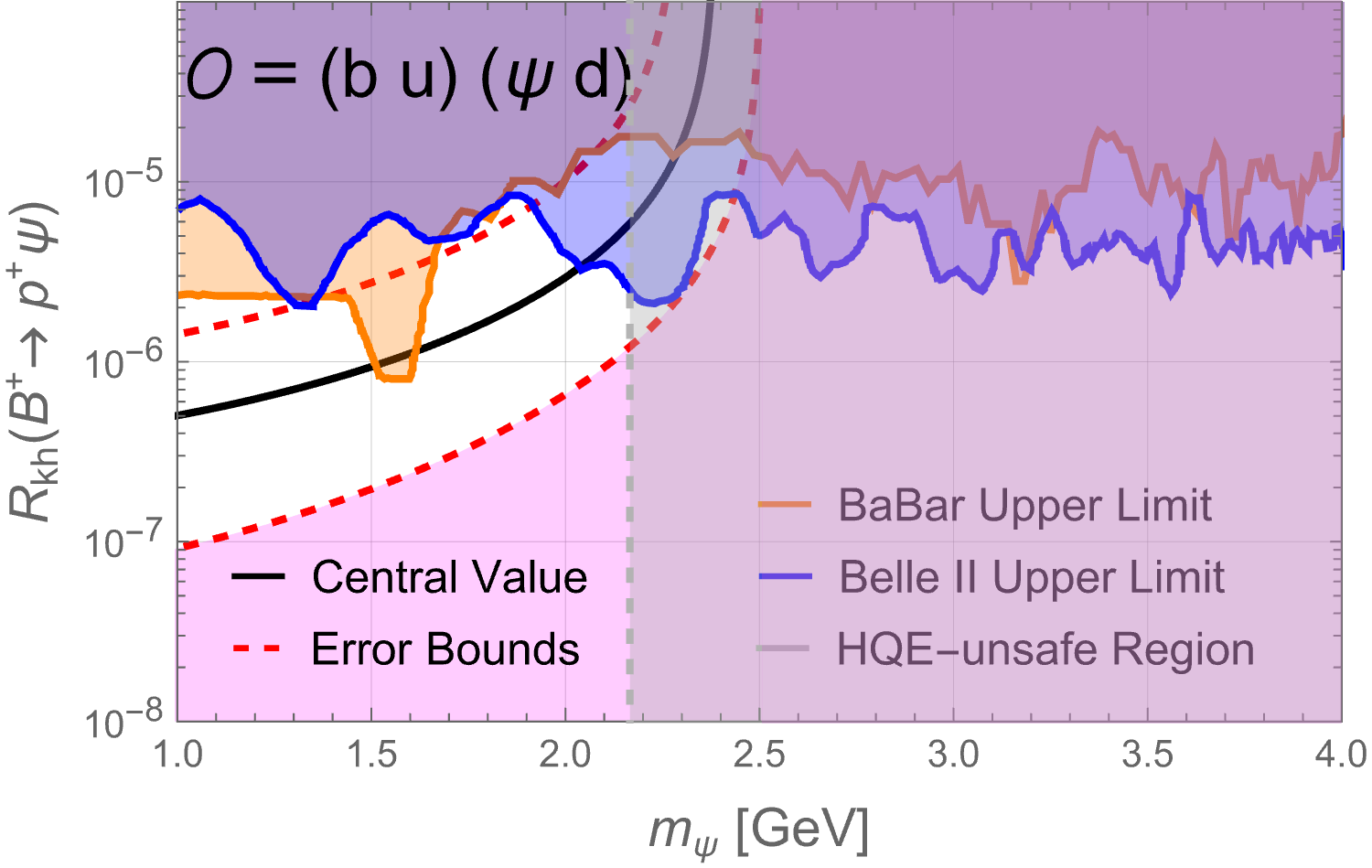}
    \end{subfigure}
    \hfill
    \begin{subfigure}{0.49\textwidth}
        \centering
        \includegraphics[width=\textwidth]{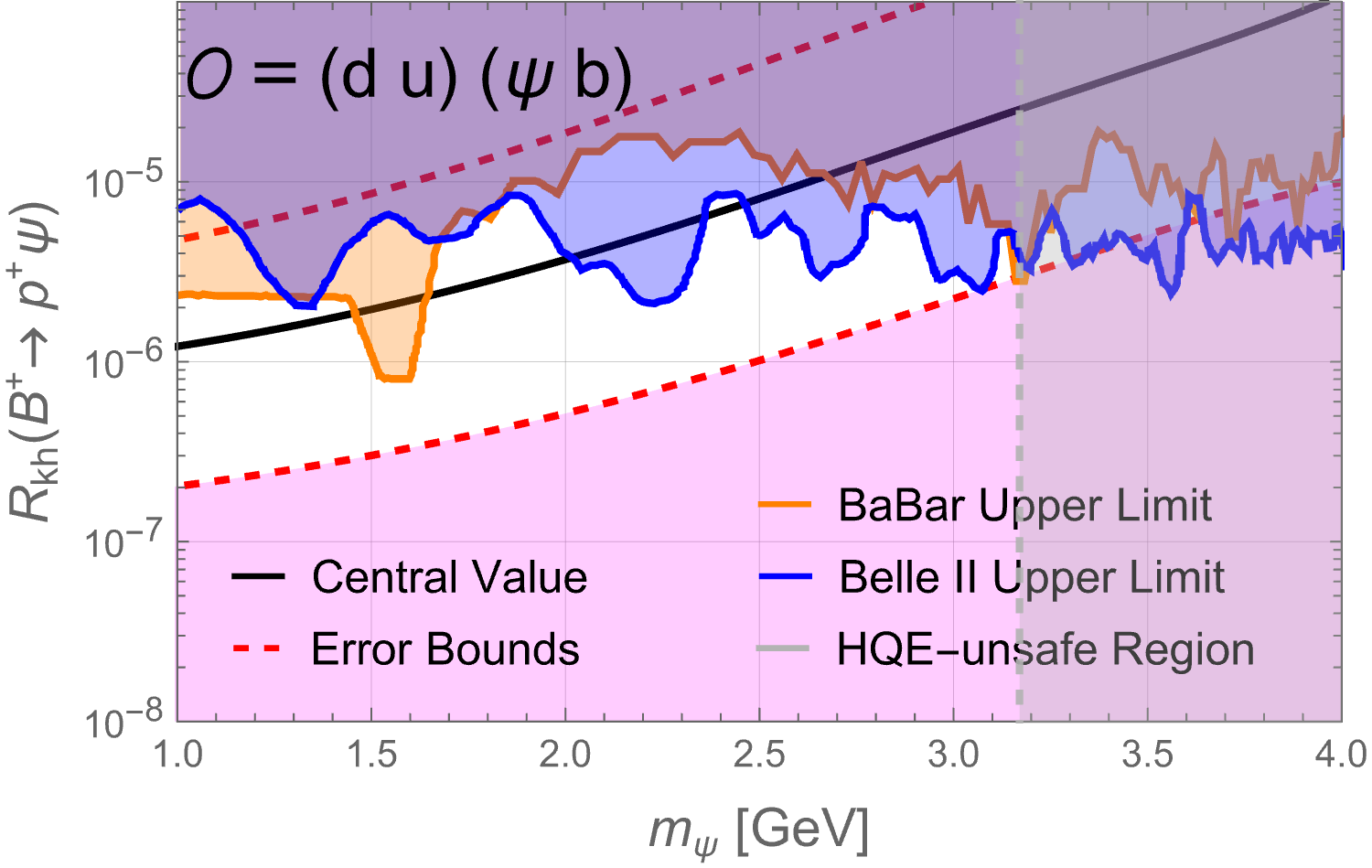}
    \end{subfigure}
    \hfill
    \begin{subfigure}{0.49\textwidth}
        \centering
        \includegraphics[width=\textwidth]{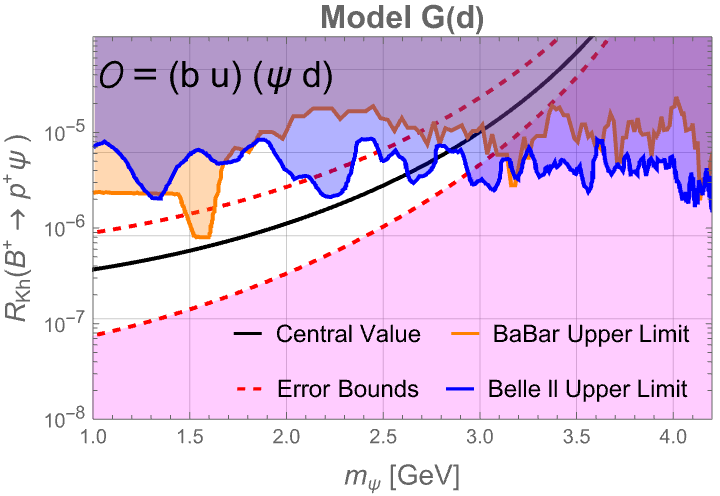}
    \end{subfigure}
     \hfill
    \begin{subfigure}{0.49\textwidth}
        \centering
        \includegraphics[width=\textwidth]{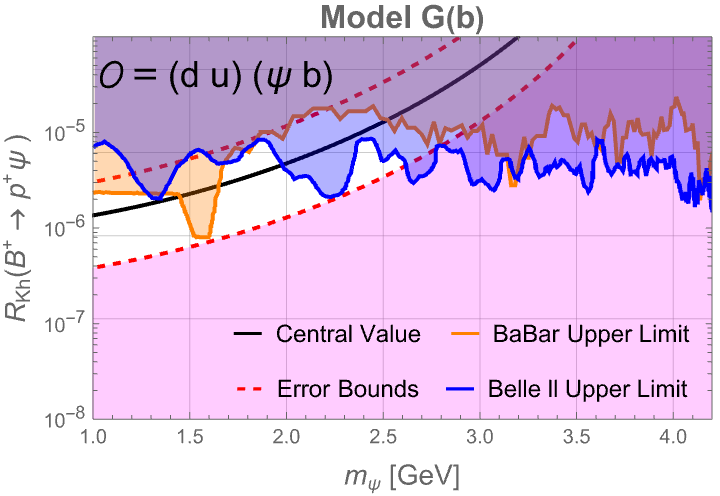}
    \end{subfigure}
       \caption{Predicted lower limits for $Br(B^+ \to p^+ \psi)$ (given by $R_{Kh}$) as a function of $m_\psi$. The upper plots show the resulting predictions from this work, which include subleading HQE contributions, while the lower plots show the lower limit obtained from including only the leading term ($\Gamma_3$) in the inclusive decay width, as derived in  Ref.~\cite{Lenz:2024rwi}.
Black: central values; red dashed: error bands; orange/blue: BaBar and Belle~II upper limits.  The white area denotes the parameter space allowed by current data. Vertical dashed lines indicate where subleading terms become comparable to the leading dimension–three contribution, beyond which the HQE truncation is unreliable (see text for details).}
        \label{fig:RkhLowerLimitsFinal}
\end{figure}

Given that in our framework each operator in the effective Hamiltonian is considered individually, and using the approximate condition on the inclusive branching ratio (Br) of $B$-meson decays required to generate the matter–antimatter asymmetry of the Universe~\cite{Alonso-Alvarez:2021qfd},  $ Br(B^+ \to \psi \, \mathcal{B} \, \mathcal{M}) > 10^{-4},$
where $\mathcal{B}$ denotes a SM baryon and $\mathcal{M}$ represents any number of light mesons\footnote{The condition ${Br}(B^+ \to \psi \, \mathcal{B} \, \mathcal{M}) > 10^{-4}$ should be understood as an order-of-magnitude estimate within the minimal Mesogenesis setup~\cite{Alonso-Alvarez:2021qfd}. In more general realizations of the model, this requirement can be relaxed, see e.g.~\cite{Elor:2024cea,Elor:2022jxy}. For instance, lowering it to $\sim 10^{-5}$ would reduce the corresponding lower bounds on exclusive decays by approximately one order of magnitude, thereby enlarging the allowed parameter space.}, a lower bound on the corresponding exclusive decay can be derived from the ratio of exclusive-to-inclusive branching fractions.
We define this ratio for a general exclusive final state as \( R_{\text{Kh}}(B^+ \to \psi \mathcal{B}) \)%
\footnote{Named after  A. Khodjamirian, who proposed its use in constraining exclusive decays.}. For the specific case of $B^+ \to p^+ \psi$ we have
\begin{equation}
    Br(B^+ \to p^+ \psi) 
    > \frac{Br(B^+ \to p^+ \psi)}{Br(B^+ \to \psi \, \mathcal{B} \, \mathcal{M})} \times 10^{-4}
    = \frac{\Gamma(B^+ \to p^+ \psi)}{\Gamma(B^+ \to \psi \, \mathcal{B} \, \mathcal{M})} \times 10^{-4}
    \equiv R_{{Kh}}(B^+ \to p^+ \psi).
    \label{eq:Rkh}
\end{equation}
In Fig.~\ref{fig:RkhLowerLimits}, we show the impact of the individual subleading HQE contributions on the inclusive decay width and, consequently, on the lower limits for the exclusive Br derived via Eq.~\eqref{eq:Rkh}. A full uncertainty analysis is presented in Fig.~\ref{fig:RkhLowerLimitsFinal}. For the operator $\mathcal{O} = (b\,u)(\psi \,d)$, once the dimension-six four-quark contribution is included, the inclusive decay width becomes negative for $m_\psi \gtrsim 2~\text{GeV}$, as shown in Figs.~\ref{fig:DecayRates} and \ref{fig:DecayRatesTotal}. This signals that the HQE prediction is no longer reliable in this region. 

In Fig.~\ref{fig:RkhLowerLimitsFinal} we present the predicted values of $R_{Kh}(B^+ \to p^+ \psi)$ for the two operators under study, shown as a function of $m_\psi,$ together with a conservative uncertainty analysis. For completeness and direct comparison, we also include the lower limit plots from our earlier work~\cite{Lenz:2024rwi}, which considered only the dimension-three contribution in inclusive decays.

To determine the uncertainty band,  we scan the inputs within their quoted ranges in Table~\ref{tab:LCSRinput} and Appendix~\ref{app:A}, taking for each parameter the three representative values $\{\text{central},\ \text{upper},\ \text{lower}\}$. For every $m_\psi$ we evaluate $R_{Kh}$ for all possible parameter combinations and take the global minimum and maximum as the lower and upper limits, respectively.
This procedure guarantees that the red dashed “error bound” curves in Fig.~\ref{fig:RkhLowerLimitsFinal} enclose any prediction obtainable within the stated input ranges, with the black curve representing the central values. 

We find that the dominant contributions to the uncertainties arise from the form factors $F^{(d)}_{B\to p_R}(0)$ and $F^{(b)}_{B\to p_R}(0)$. As detailed in the work of Boushmelev and Wald~\cite{Boushmelev:2023huu} (Section VI.A), we adopt their derivation and associated uncertainties for these quantities. To account for uncertainties in the inclusive parameters (notably $m_b$), our choice of mass scheme, and omitted higher-order corrections, we assign a 10\% theoretical uncertainty. Finally, we have not included additional parameter correlations in our uncertainty estimates beyond those explicitly considered in the Boushmelev and Wald analysis.

The resulting predictions are confronted with existing experimental constraints from BaBar~\cite{BaBar:2023dtq} (orange region) and Belle II~\cite{Belle-II:2026tyb} (blue region). 
 Since $R_{Kh}$ provides a lower bound on the exclusive branching ratio, parameter regions where $R_{Kh}$ exceeds the experimental upper limits on ${Br}(B^+ \to p^+ \psi)$ are excluded. The vertical dashed line indicates the value of $m_\psi$ at which subleading contributions become larger than the leading
dimension–three term, i.e. where $R_d\equiv|\Gamma_d/\Gamma_3|\gtrsim1$ (see Figs.~\ref{fig:DecayRates} and~\ref{fig:DecayRatesTotal}). This is  defined using the dimension-six four-quark term $\tilde\Gamma_6$, which provides the dominant power-suppressed contribution for both operators. Mass regions to the right of this line lie outside the domain where a fixed-order HQE truncated at the included dimensions can be considered quantitatively reliable; these are shaded accordingly in the plots. 

In conclusion, we find that the inclusion of power-suppressed terms, particularly the dimension-six four-quark contributions, can substantially affect the exclusive lower limits. At values of the dark-sector mass $m_\psi$ where subleading terms exceed the leading dimension-three contribution, reliable phenomenological conclusions can no longer be drawn. For $\mathcal{O} = (b\,u)(\psi \, d)$ this occurs for $m_\psi \gtrsim 2~\text{GeV}$, while for $\mathcal{O} = (d\,u)(\psi \, b)$ the breakdown sets in at $m_\psi \gtrsim 3~\text{GeV}$.
For $m_\psi$ values below these thresholds, the leading dimension-three term provides a good approximation to the inclusive rate. This extends and validates the approach of our earlier work~\cite{Lenz:2024rwi}, where only the leading contribution was used to set conservative lower bounds on exclusive decays.
\subsection{Lifetime Ratios}
In our previous work~\cite{Lenz:2024rwi}, we refrained from analyzing NP effects in
$\tau(B_s)/\tau(B_d)$ due to the dominance of $SU(3)_F$–breaking two–quark contributions and the
limited knowledge of the corresponding nonperturbative inputs, in particular the Darwin matrix
elements. Having now computed the Darwin contribution in our framework, we revisit this ratio as an
exploratory constraint on the Mesogenesis couplings.
\begin{figure}[ht]
    \centering
    \begin{subfigure}{0.43\textwidth}
        \centering
        \includegraphics[width=\textwidth]{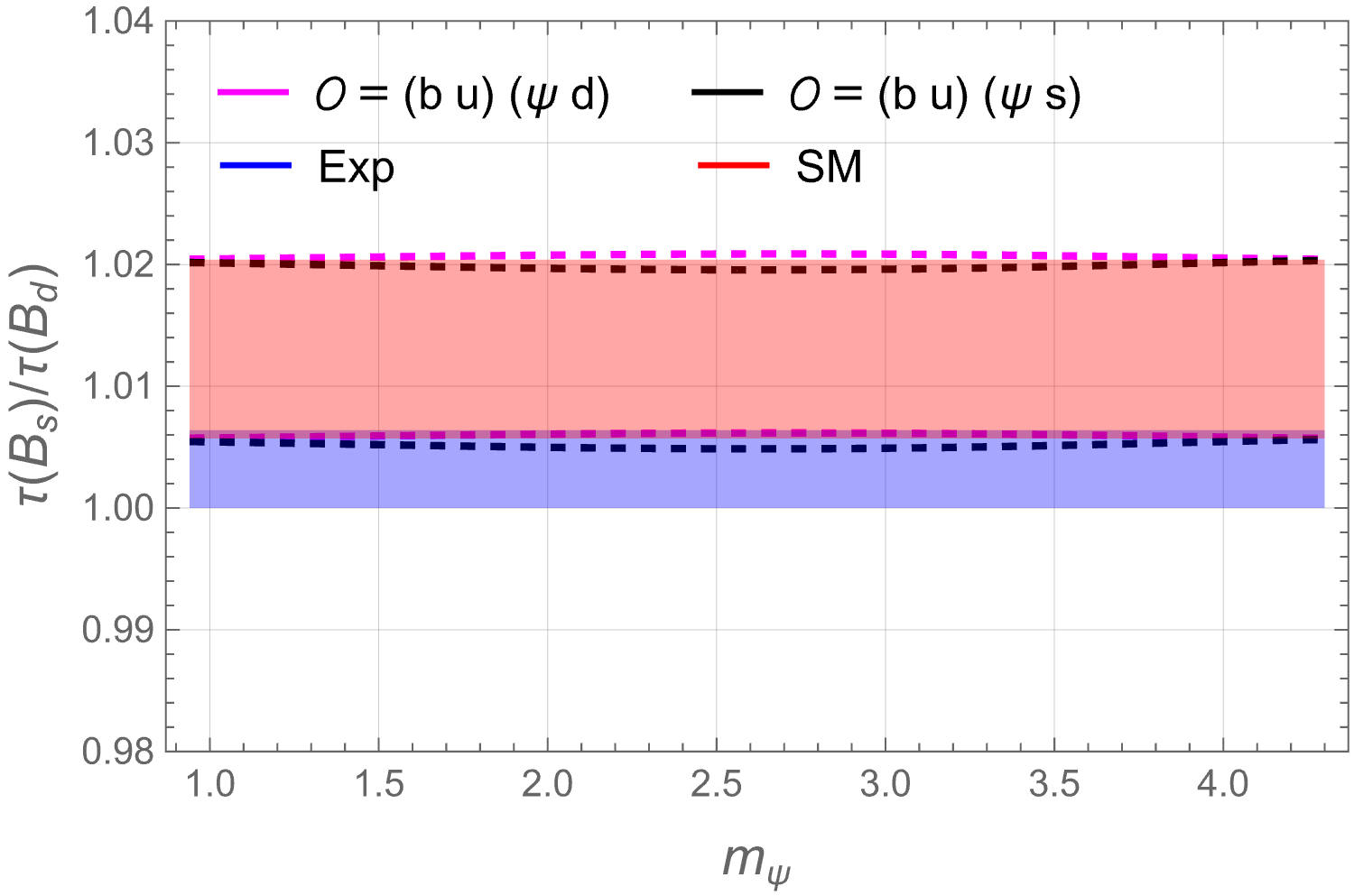}
    \end{subfigure}
    \hfill
    \begin{subfigure}{0.43\textwidth}
        \centering
        \includegraphics[width=\textwidth]{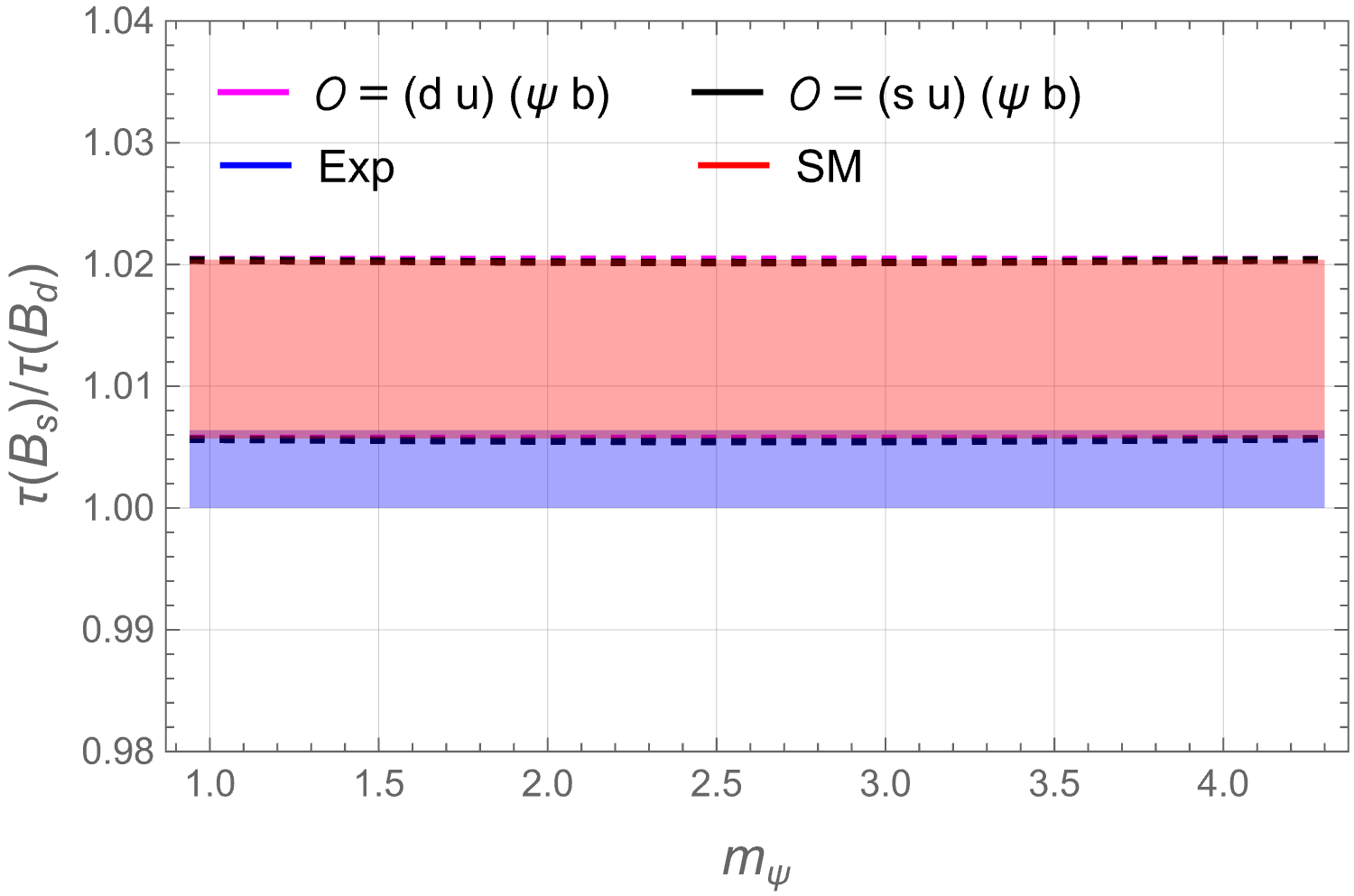}
    \end{subfigure}
      \caption{Lifetime ratio $\tau(B_s)/\tau(B_d)$ as a function of the dark fermion mass $m_\psi$, 
including both SM and SM+NP predictions for representative operators.  
The NP contribution is found to be negligible compared to present theoretical uncertainties.}
        \label{fig:BsBdG1G2}
\end{figure}
\begin{figure}[ht]
    \centering
    \begin{subfigure}{0.40\textwidth}
        \centering
        \includegraphics[width=\textwidth]{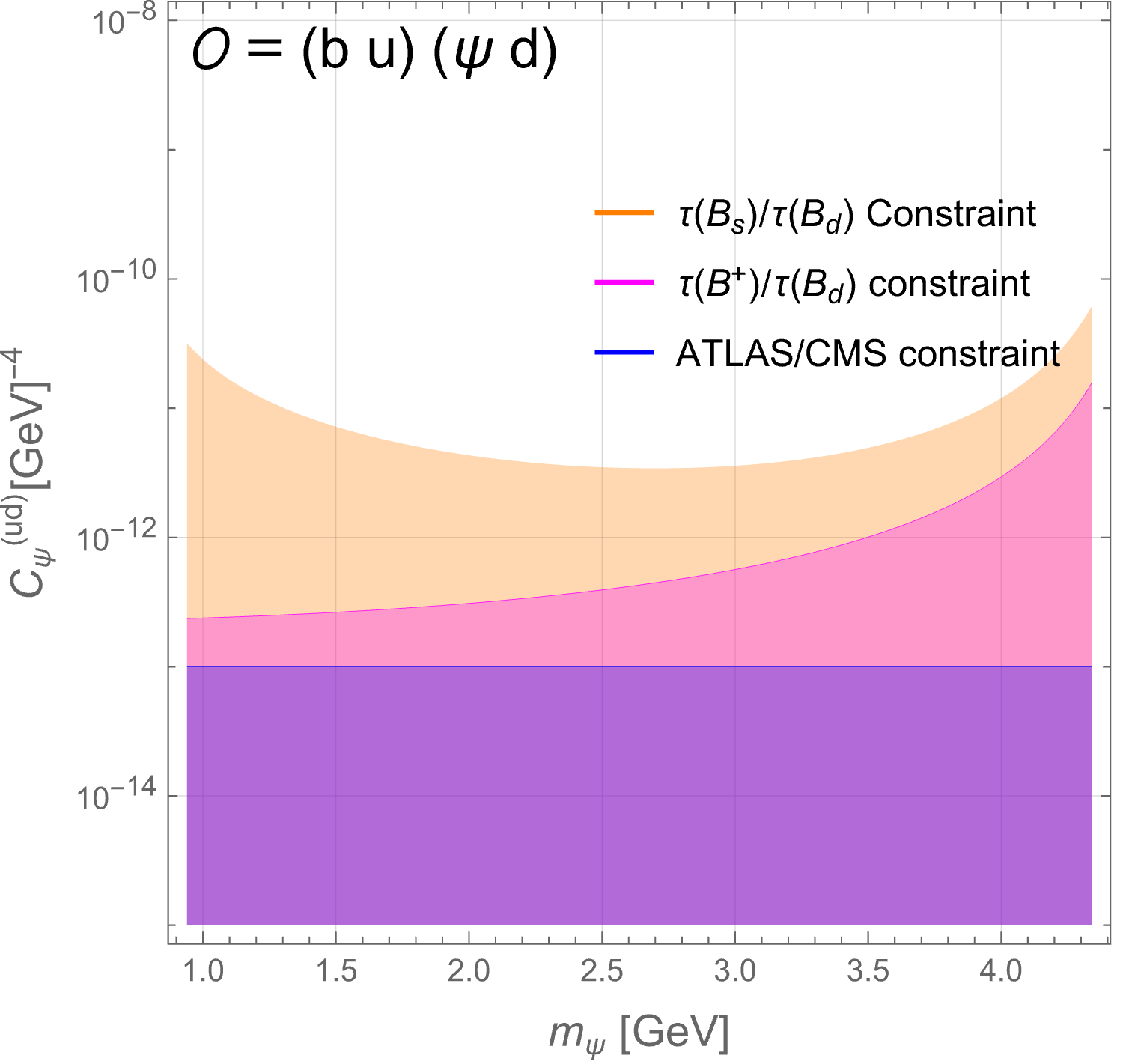}
    \end{subfigure}
    \hfill
    \begin{subfigure}{0.40\textwidth}
        \centering
        \includegraphics[width=\textwidth]{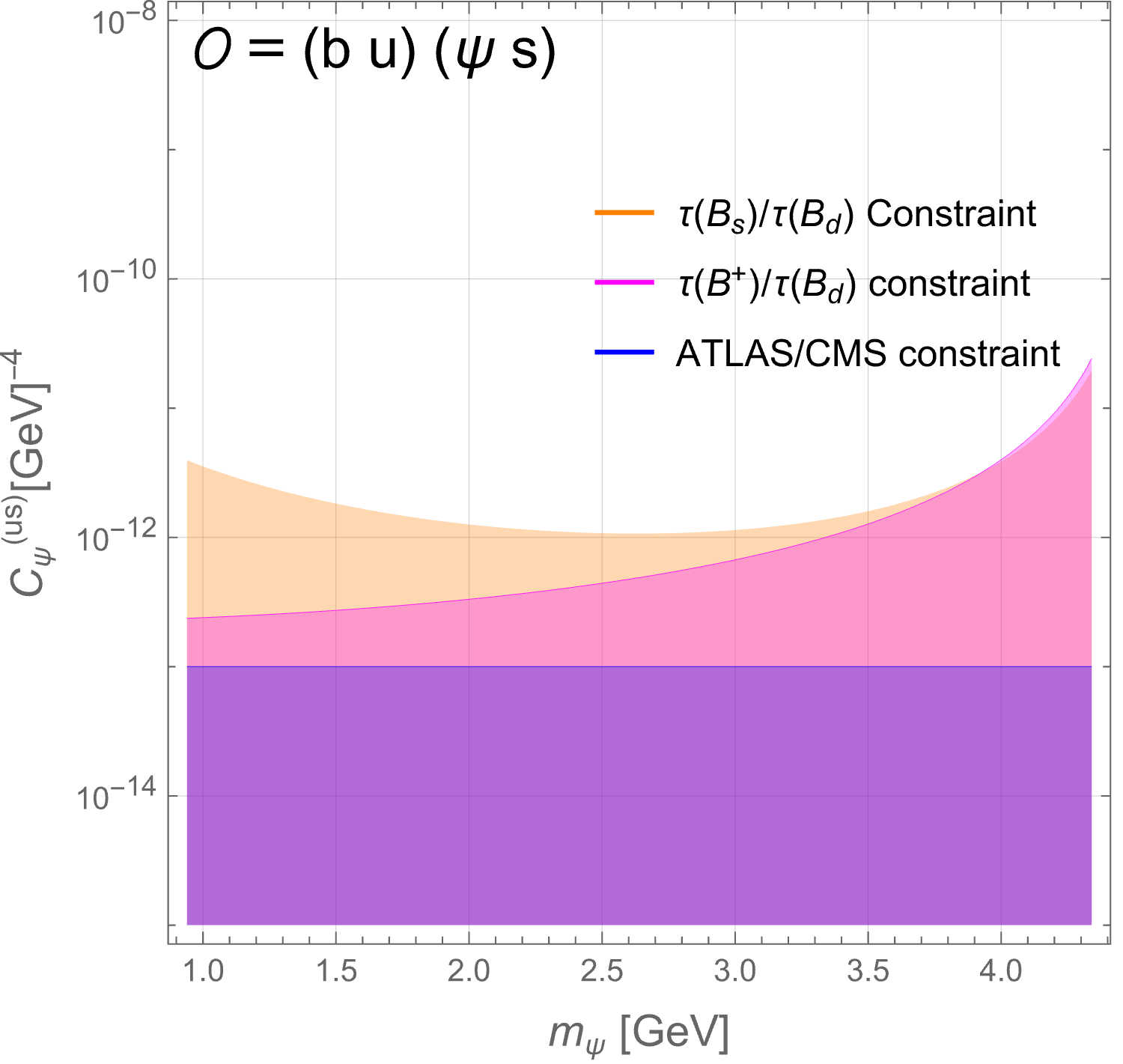}
    \end{subfigure}
    \begin{subfigure}{0.40\textwidth}
        \centering
        \includegraphics[width=\textwidth]{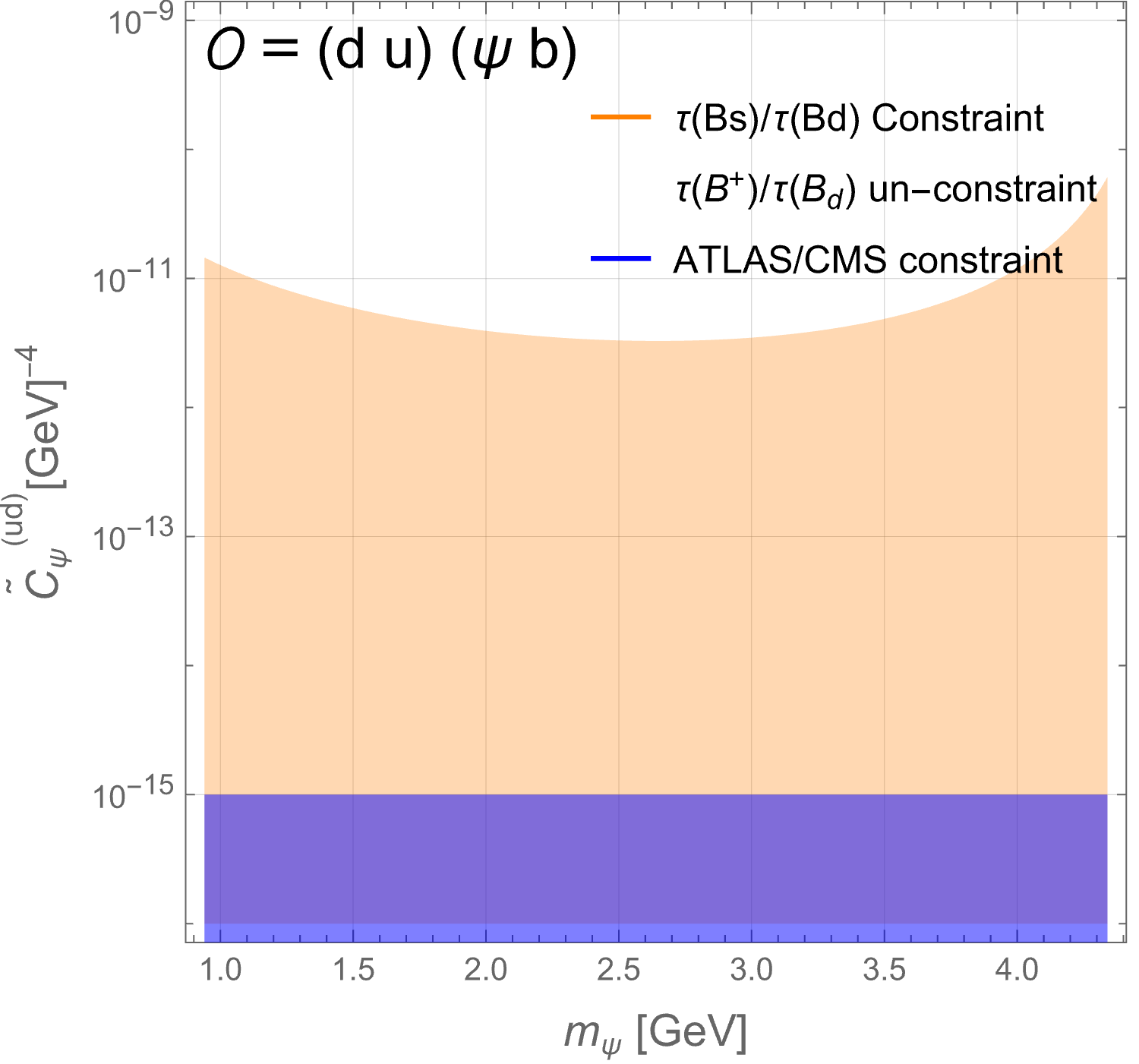}
    \end{subfigure}
      \hfill
    \begin{subfigure}{0.40\textwidth}
        \centering
        \includegraphics[width=\textwidth]{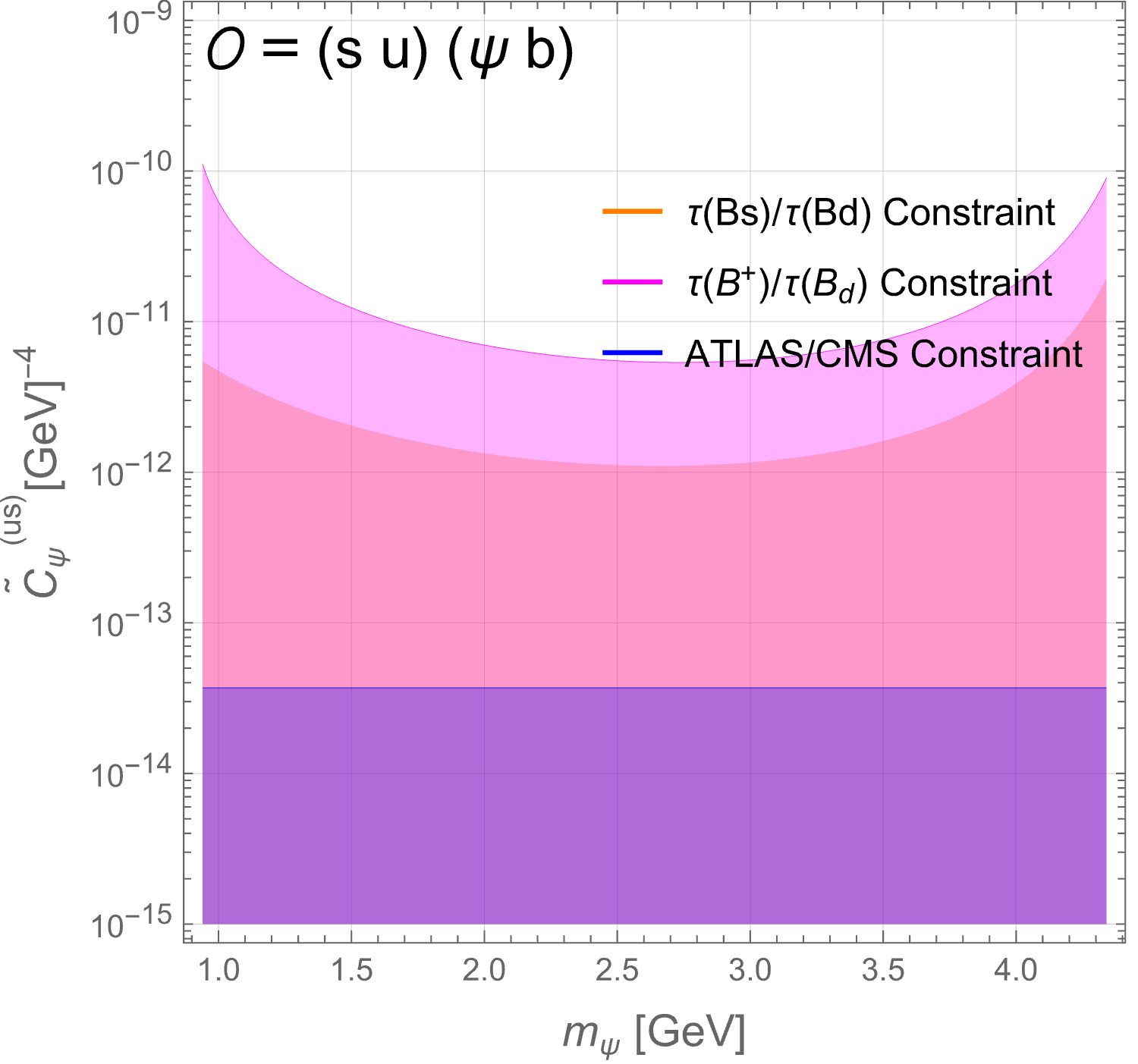}
    \end{subfigure}
       \caption{Constraints on the Wilson coefficients as a function of the dark fermion mass $m_\psi$, derived from the lifetime ratios $\tau(B_s)/\tau(B_d)$ (orange) and $\tau(B^+)/\tau(B_d)$ (magenta), compared to collider constraints (blue).}
        \label{fig:BsBdConstraintsG1}
\end{figure}
\\
We evaluate $\tau(B_s)/\tau(B_d)$ as a function of the dark-sector mass $m_\psi$ and compare with
experiment. In line with our study of $\tau(B^+)/\tau(B_d)$~\cite{Lenz:2024rwi}, we do not obtain
bounds stronger than those from collider searches~\cite{Alonso-Alvarez:2021qfd}. Using the collider
limits for the relevant Wilson coefficients, the NP shift of $\tau(B_s)/\tau(B_d)$ is tiny:
even at the ATLAS/CMS bound the total (SM+NP) prediction changes by at most $\sim 0.04\%$, far below current theoretical uncertainties dominated by $SU(3)_F$ breaking in two–quark matrix
elements (particularly the Darwin operator). We therefore treat $\tau(B_s)/\tau(B_d)$ chiefly as a consistency check at present precision. 
In Fig.~\ref{fig:BsBdG1G2}, we show the results for four representative operators, while the remaining cases are collected in Appendix~\ref{app:D}. The plots clearly illustrate that the NP shift relative to the SM prediction is negligible across the full $m_\psi$ range.
Fig.~\ref{fig:BsBdConstraintsG1} compares the resulting upper limits on the coupling (versus $m_\psi$)
from $\tau(B_s)/\tau(B_d)$, from $\tau(B^+)/\tau(B_d)$, and from collider searches. Across the full
$m_\psi$ range the collider bound is most stringent; $\tau(B^+)/\tau(B_d)$ is typically weaker, and
$\tau(B_s)/\tau(B_d)$ is the least constraining. However, this hierarchy is
operator–dependent: for operators that contribute to only one of $B^+$ or $B_d$ in the $\tau(B^+)/\tau(B_d)$ ratio, the cancellation inherent to the ratio suppresses sensitivity and the
$\tau(B^+)/\tau(B_d)$ constraint can be weaker than the one from $\tau(B_s)/\tau(B_d)$. Overall, current lifetime data are compatible with—but do not improve upon—collider constraints.
\section{Summary and Outlook}
\label{summary}
In this work, we investigated power-suppressed effects on $B$-meson decays within the Mesogenesis framework. We presented the calculation of inclusive decay rates in the HQE, including the dimension-six two-quark Darwin operator. Our analysis, incorporating the dimension-five chromomagnetic term, the Darwin contribution, and the dimension-six four-quark operators, demonstrates that these formally subleading effects can exceed the leading contribution in certain regions of the parameter space. For each operator, we pinpoint the critical values of the dark-sector mass $m_\psi$ where this breakdown of the HQE hierarchy occurs. This provides a quantitative guide for determining the region of validity for phenomenological constraints derived from these inclusive calculations.

We also analyze the lifetime ratio $\tau(B_s)/\tau(B_d)$  and find that it does not provide new constraints on the model's parameters beyond what is already established by collider data. We provide a comprehensive comparison of the sensitivity of this ratio with that of $\tau(B^+)/\tau(B_d)$ and collider signals. Furthermore, our study of the exclusive-to-inclusive ratio, which is affected by the  contribution of subleading terms, provides new lower bounds on exclusive decays $B^+ \to p^+ \psi$. 

The emergence of subleading effects as potentially dominant contributions underscores the need for dedicated studies beyond leading order. 
In particular, both (NLO) QCD corrections and higher-order power corrections would be valuable to reduce theoretical uncertainties and to test the stability of the HQE hierarchy more reliably.

Finally, we are pursuing an exclusive analysis based on LCSR with $B$-meson distribution amplitudes. 
Combining these exclusive results with the inclusive framework developed here will enable a more precise and systematic phenomenological study of Mesogenesis.
\section*{Acknowledgements}
I thank Alexander Lenz for his careful proofreading and commentary on the manuscript. I am also grateful to Maria Laura Piscopo and Alexey Pivovarov for their valuable insights and productive discussions. I further thank Daniel Marcantonio for drawing our attention to the updated Belle/Belle II analysis and the corresponding preprint. This project was supported by the Deutsche Forschungsgemeinschaft (DFG, German Research Foundation) under grant 396021762-TRR 257.

\appendix

\section{\boldmath Numerical Inputs}
\label{app:A}
For the input values of the bottom and charm quark masses we use~\cite{Finauri:2023kte}
\begin{equation}
m_b^{\rm kin}(1\GeV) = (4.573 \pm 0.012)\,\GeV, \quad
\overline{m}_c(2\GeV) = (1.090 \pm 0.010)\,\GeV.
\label{eq:mb-mc}
\end{equation}
The values for the non-perturbative matrix elements defined in Eq.~\eqref{eq:hadronic_parameters}, as well as the Bag parameters from the parameterization of the forward matrix element of the $\Delta B=0$ four-quark operators $\tilde{\cal{O}}_n$ in Eqs.~(\ref{eq:Q1-Q2}) and~(\ref{eq:Q3-Q4}) are shown in Tables \ref{tab:Non-perturbative-input} and \ref{tab:Bag-parameters}, respectively.\footnote{Note that the parameterizations of the SM and the Mesogenesis four-quark operators are the same since they differ only in chirality, see~\cite{Lenz:2024rwi}.} The maximum values of the Mesogenesis couplings in $\mathcal{H}_{-1/3}$, as constrained by ATLAS/CMS searches can be found in Ref.~\cite{Lenz:2024rwi} Appendix A. The decay constants are determined with high precision from lattice QCD~\cite{Aoki:2019cca},  
\begin{equation}
    f_B = (0.1900 \pm 0.0013)\,\GeV, 
    \qquad
    f_{B_s} = (0.2303 \pm 0.0013)\,\GeV \, .
    \label{eq:decay-constants}
\end{equation}
The $B_q$-meson masses, measured to sub-MeV accuracy, are taken from the PDG~\cite{PDG:2024}:  
\begin{equation}
m_{B^+} = 5.27934\,\GeV, \qquad
m_{B_d} = 5.27965\,\GeV, \qquad
m_{B_s} = 5.36688\,\GeV \, .
\end{equation}
For the CKM matrix elements, we rely on the results of the global fit provided by the CKMfitter collaboration~\cite{Charles:2004jd} (online update).
\begin{table}[h!]
\centering
\small
\renewcommand{\arraystretch}{1.4}
    \begin{tabular}{|c||c|c||c|c|}
    \hline 
    Parameter 
    & $B^{+}, B_{d}$ 
    & Source
    & $B_s$
    & Source \\
    \hline
    \hline
    $\mu_\pi^2 (B_q)$ [GeV$^2$]
    & $0.454 \pm  0.043 $
    & Exp. fit \cite{Finauri:2023kte}
    & $ 0.534 \pm  0.074 $
    & Exp. fit + Ref.~\cite{Egner:2024lay}
    \\
    \hline
    $\mu_G^2 (B_q)$ [GeV$^2$]
    & $0.274 \pm  0.053 $
    & Exp. fit \cite{Finauri:2023kte} 
    & $ 0.321 \pm  0.072$ 
    & Exp. fit + Ref.~\cite{Egner:2024lay}
    \\
    \hline
    $\rho_D^3 (B_q)$ [GeV$^3$]
    & $0.176 \pm  0.019 $
    & Exp. fit \cite{Finauri:2023kte}
    & $0.210 \pm  0.034$ 
    & Exp. fit +  Ref.~\cite{Egner:2024lay} \\
    \hline
    \end{tabular}
    \caption{Input values of the two-quark non-perturbative parameters.
    }
    \label{tab:Non-perturbative-input}
\end{table}

\begin{table}[th]
\centering
\small
\renewcommand{\arraystretch}{1.5} 
\begin{tabular}{|c||c|c|c|c|}
\hline
$\mu_0 = 1.5\GeV$    
& $ \tilde B_1^q$ 
& $ \tilde B_2^q$ 
& $ \tilde B_3^q$ 
& $ \tilde B_4^q$ 
\\
\hline
\hline
    $\langle B_{u,d} | \tilde O^{u,d}_i | B_{u,d} \rangle$ 
     & $1.0026^{+0.0246}_{-0.0221}$ 
     & $0.9982^{+0.0206}_{-0.0214}$ 
     & $-0.0057^{+0.0221}_{-0.0225}$ 
     & $-0.0014^{+0.0216}_{-0.0221}$
\\
\hline
     $\langle B_{s} | \tilde O^{s}_i | B_{s} \rangle$  
     & $1.0022^{+0.0246}_{-0.0221}$  
     & $0.9983^{+0.0246}_{-0.0221}$ 
     & $-0.0036^{+0.0265}_{-0.0270}$  
     & $-0.0009^{+0.0259}_{-0.0265}$
\\
\hline
\end{tabular}
\caption{Input values of the dimension-six Bag parameters at the renormalisation scale 
$\mu_0 = 1.5\GeV$, collected from Ref.~\cite{Egner:2024lay} and references therein.}
\label{tab:Bag-parameters}
\end{table}
\clearpage
\section{\boldmath Analytical Results}
\label{app:C}
\begin{figure}[ht]
    \centering
    \begin{subfigure}{0.43\textwidth}
        \centering
        \includegraphics[width=\textwidth]{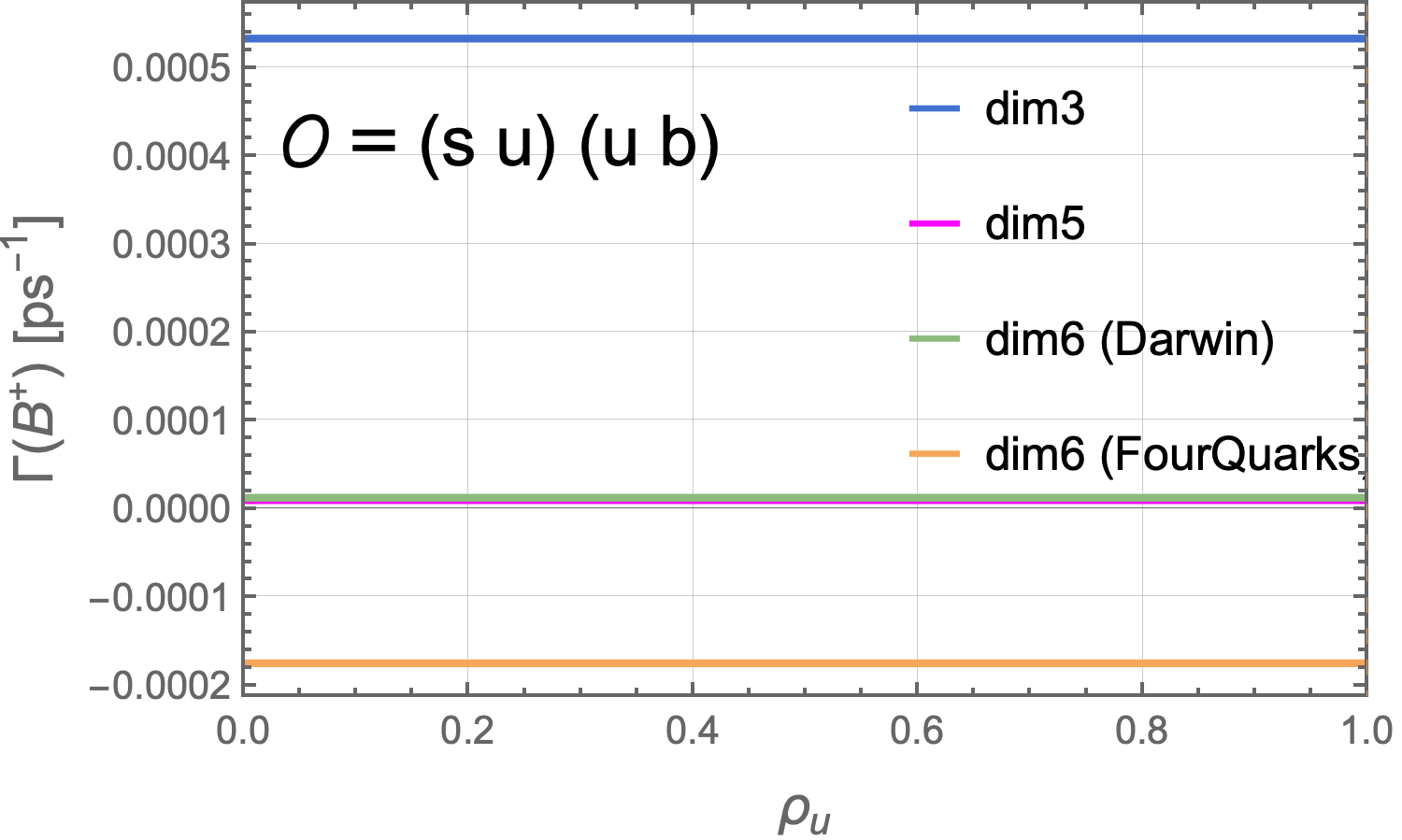}
    \end{subfigure}
    \hfill
    \begin{subfigure}{0.40\textwidth}
        \centering
        \includegraphics[width=\textwidth]{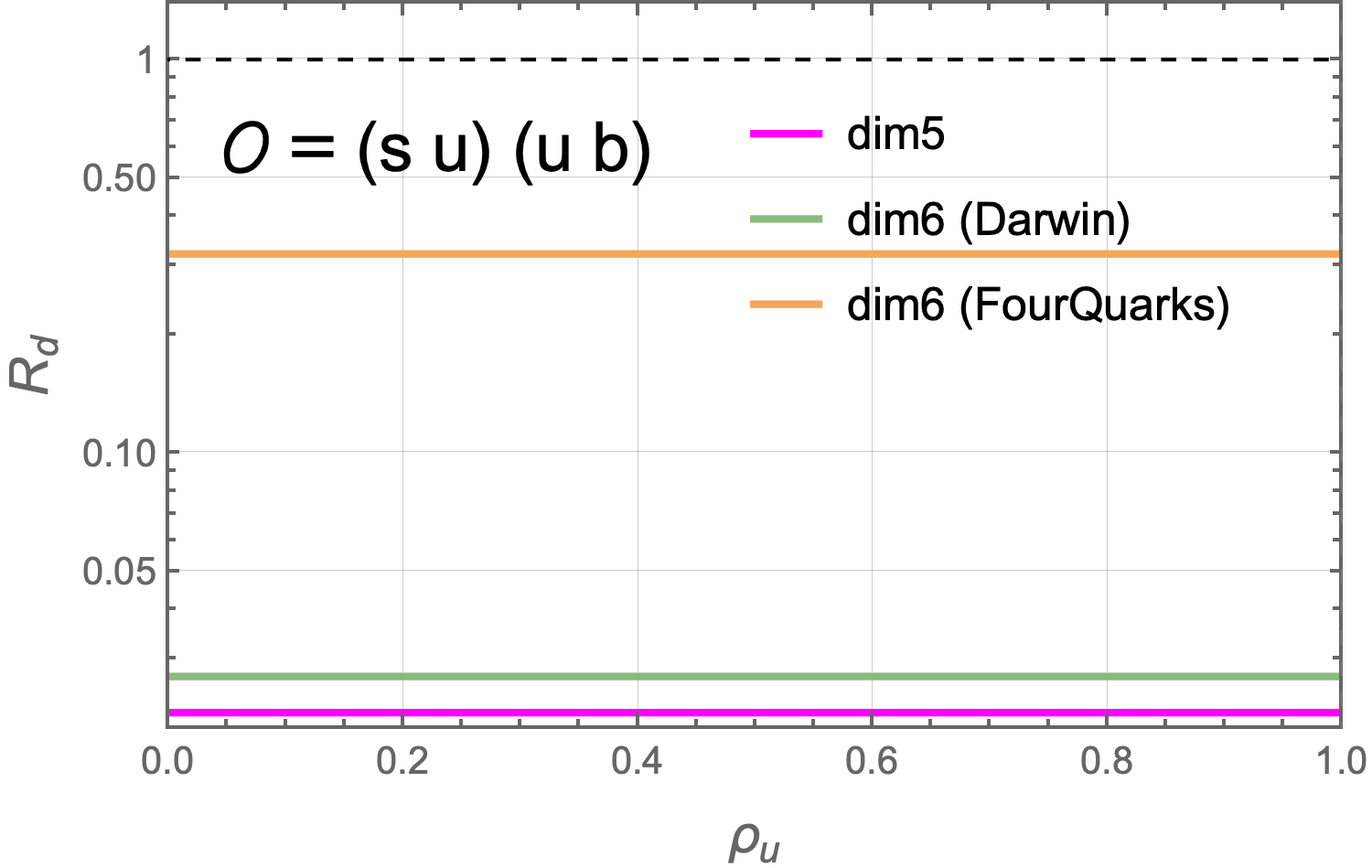}
    \end{subfigure}

     \medskip 
    
    \begin{subfigure}{0.43\textwidth}
        \centering
        \includegraphics[width=\textwidth]{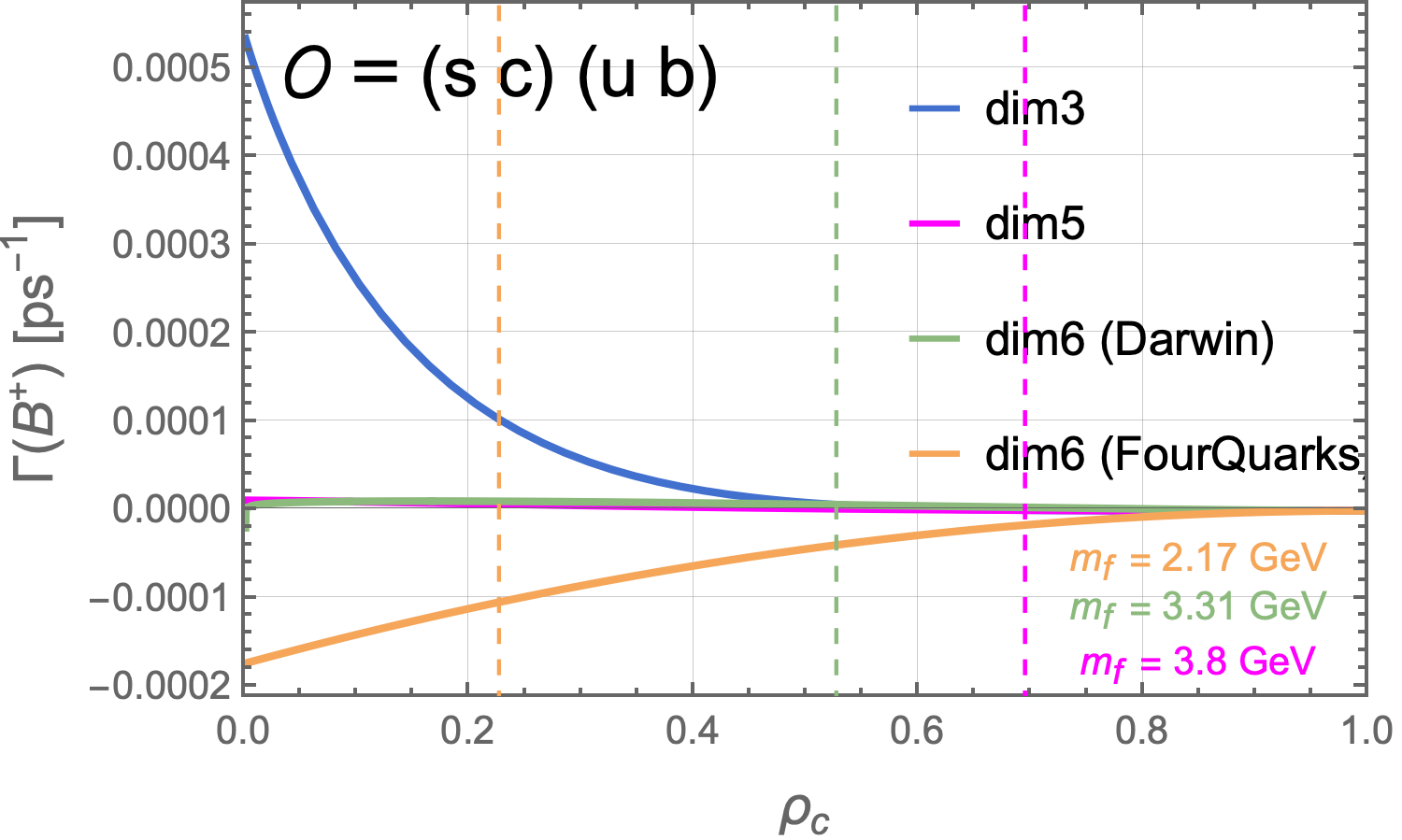}
    \end{subfigure}
    \hfill
    \begin{subfigure}{0.40\textwidth}
        \centering
        \includegraphics[width=\textwidth]{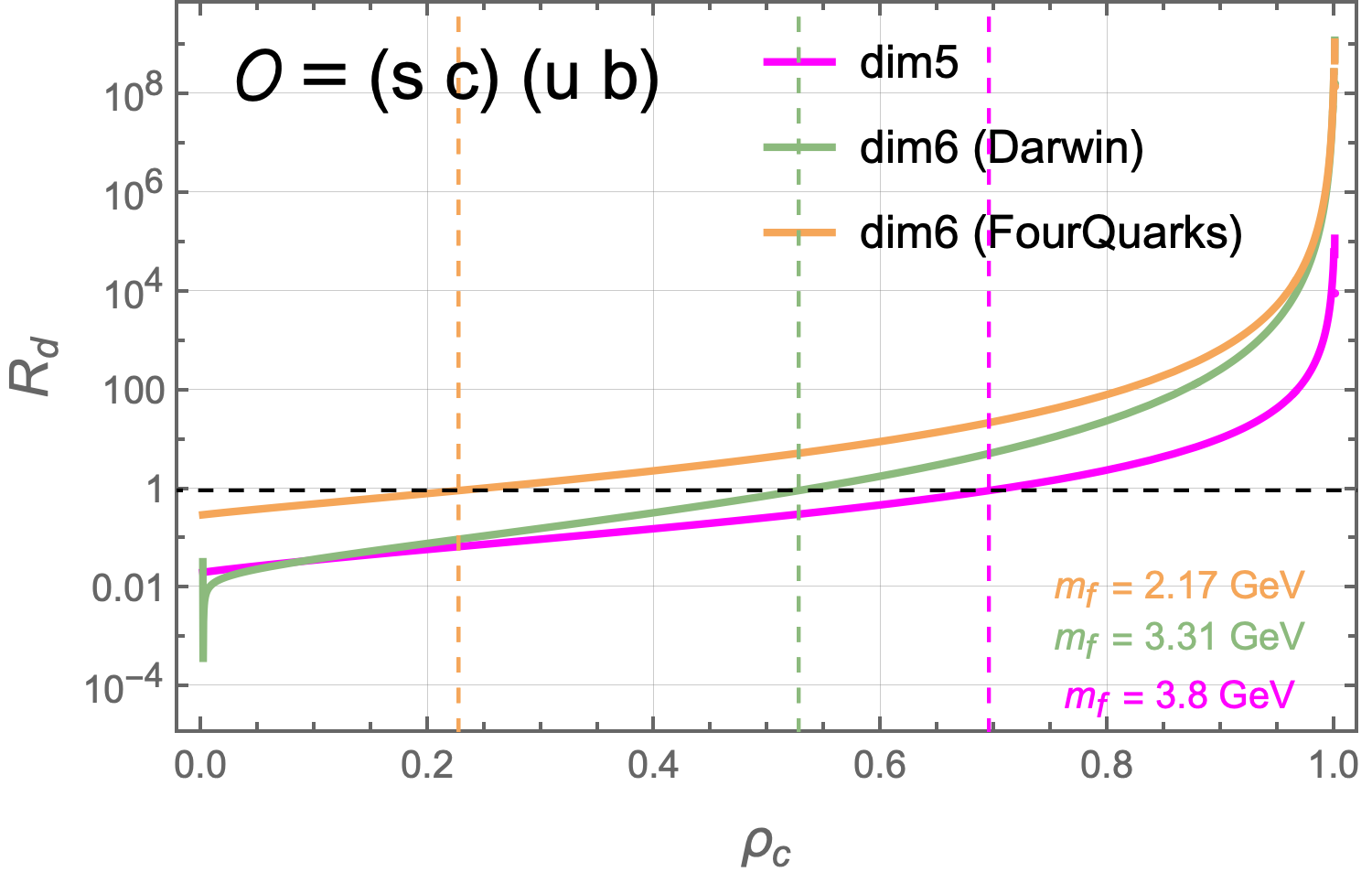}
    \end{subfigure}

    \medskip 
    
    \begin{subfigure}{0.43\textwidth}
        \centering
        \includegraphics[width=\textwidth]{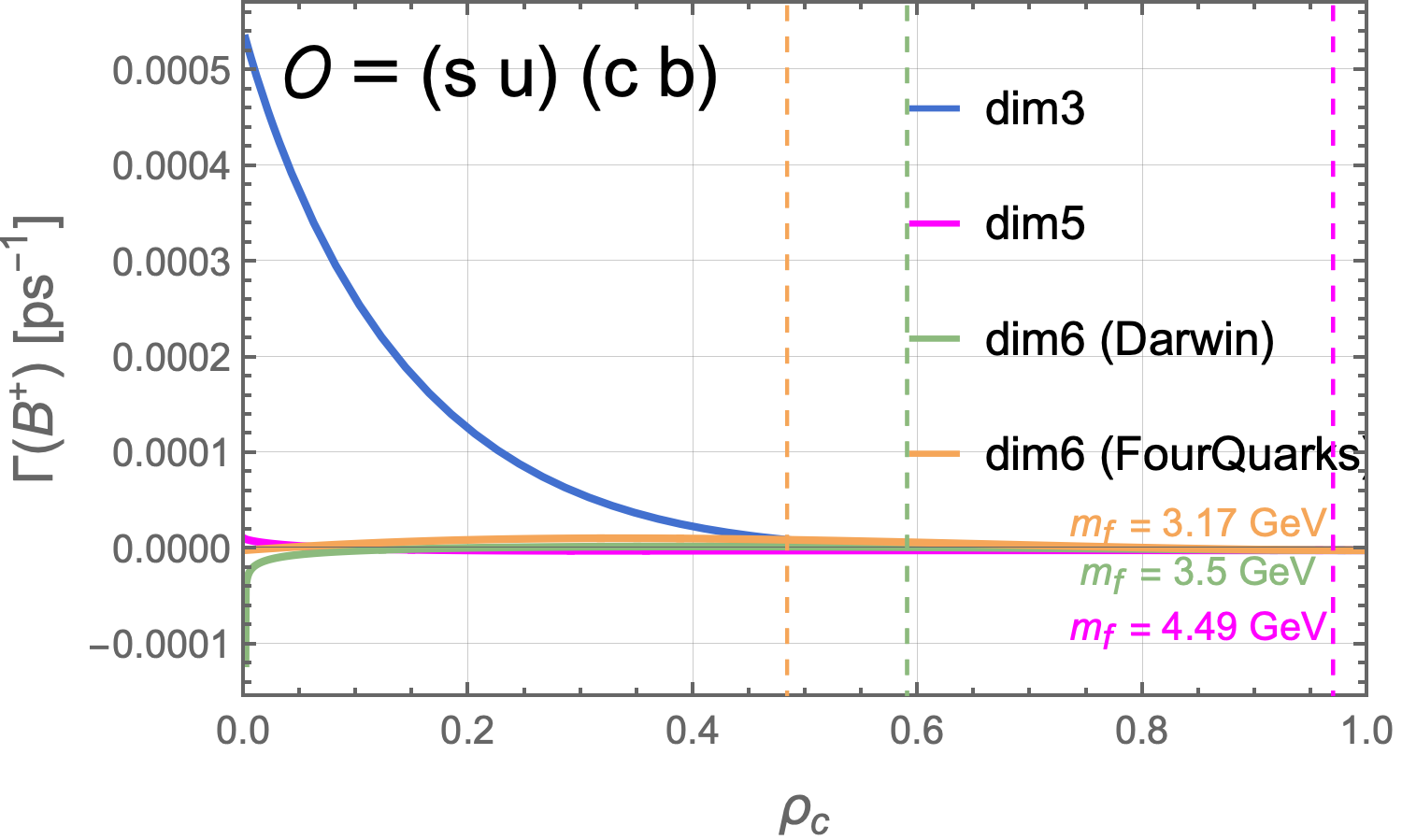}
    \end{subfigure}
    \hfill
    \begin{subfigure}{0.40\textwidth}
        \centering
        \includegraphics[width=\textwidth]{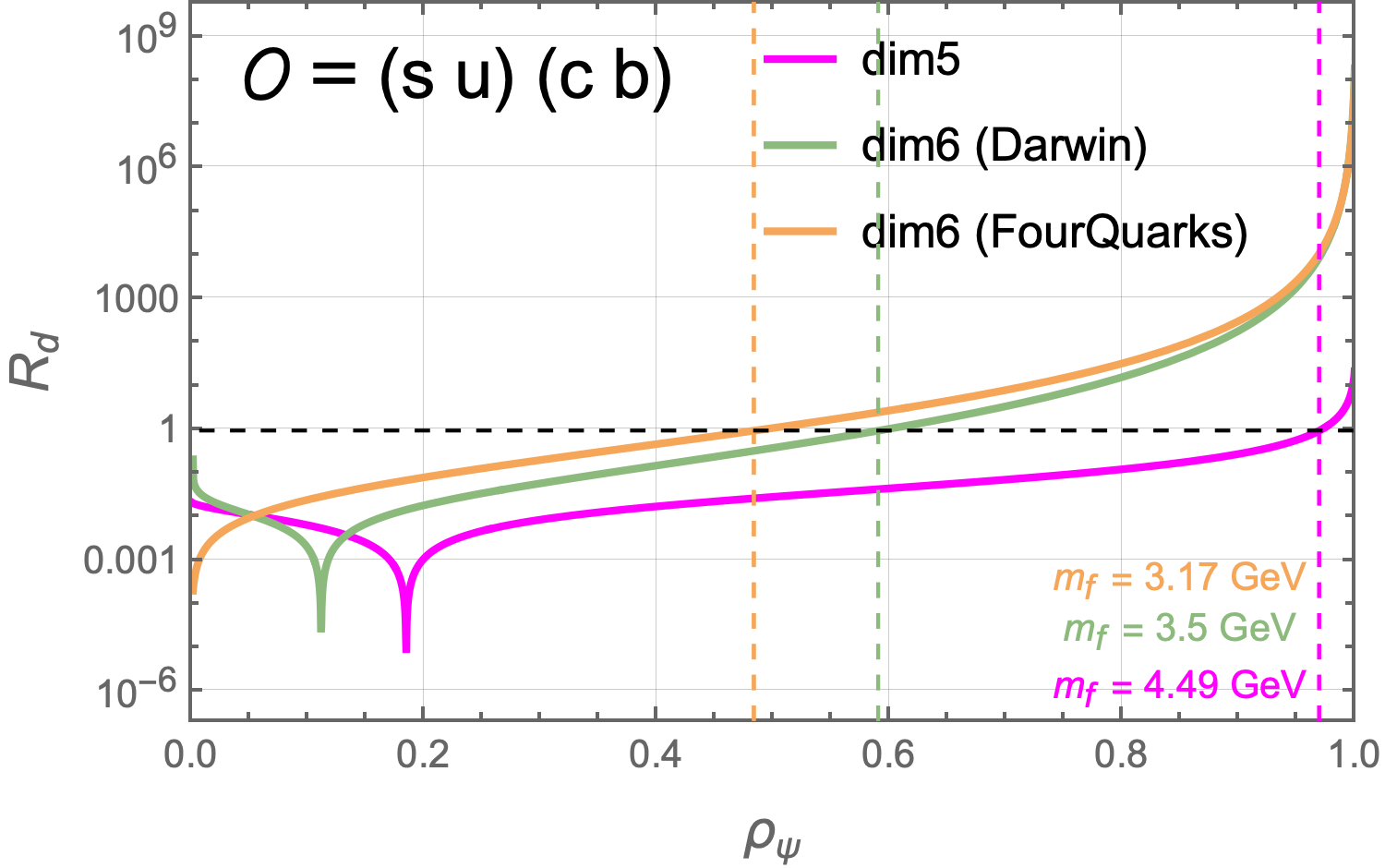}
    \end{subfigure}

      \medskip 
    
    \begin{subfigure}{0.43\textwidth}
        \centering
        \includegraphics[width=\textwidth]{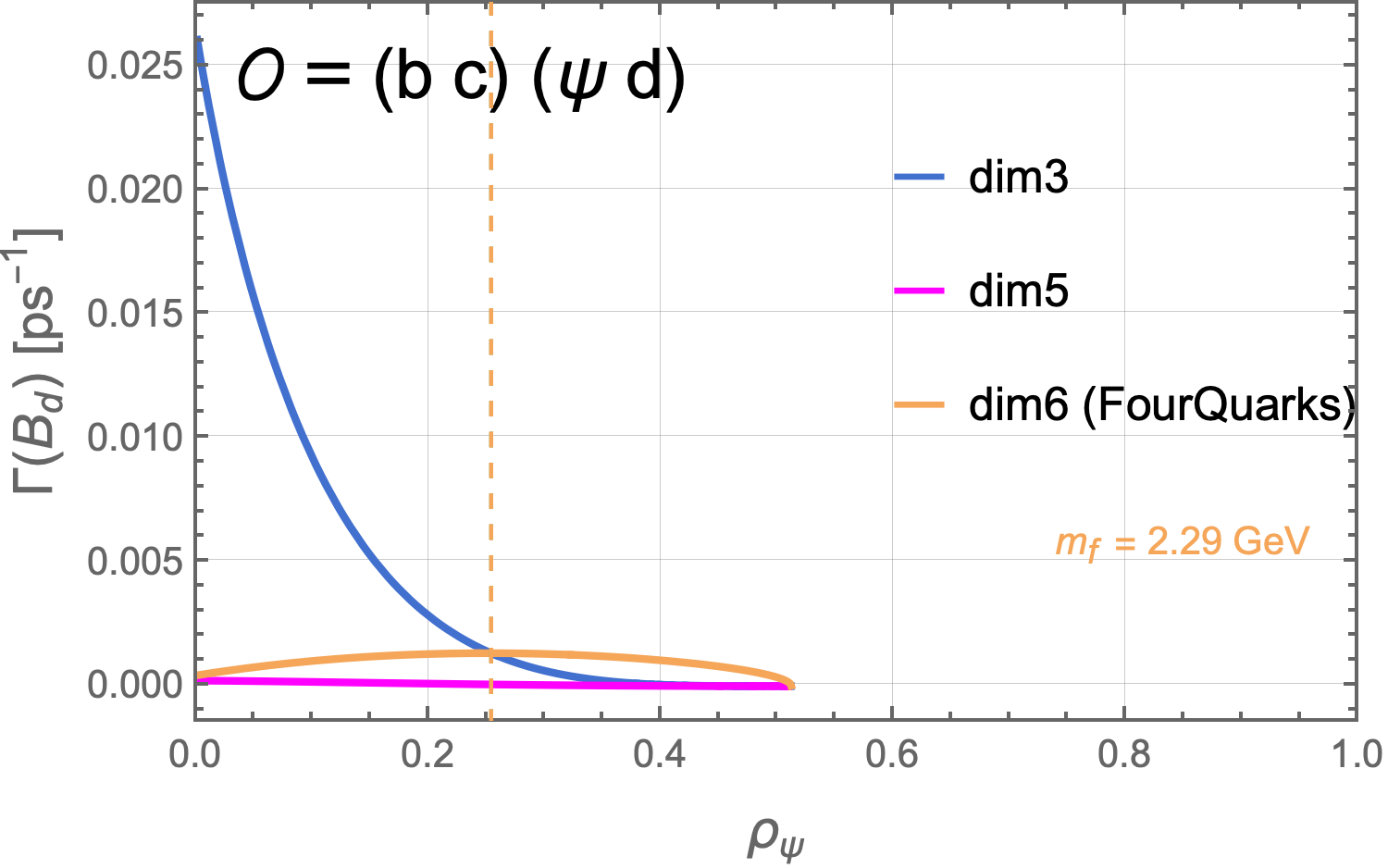}
    \end{subfigure}
    \hfill
    \begin{subfigure}{0.40\textwidth}
        \centering
        \includegraphics[width=\textwidth]{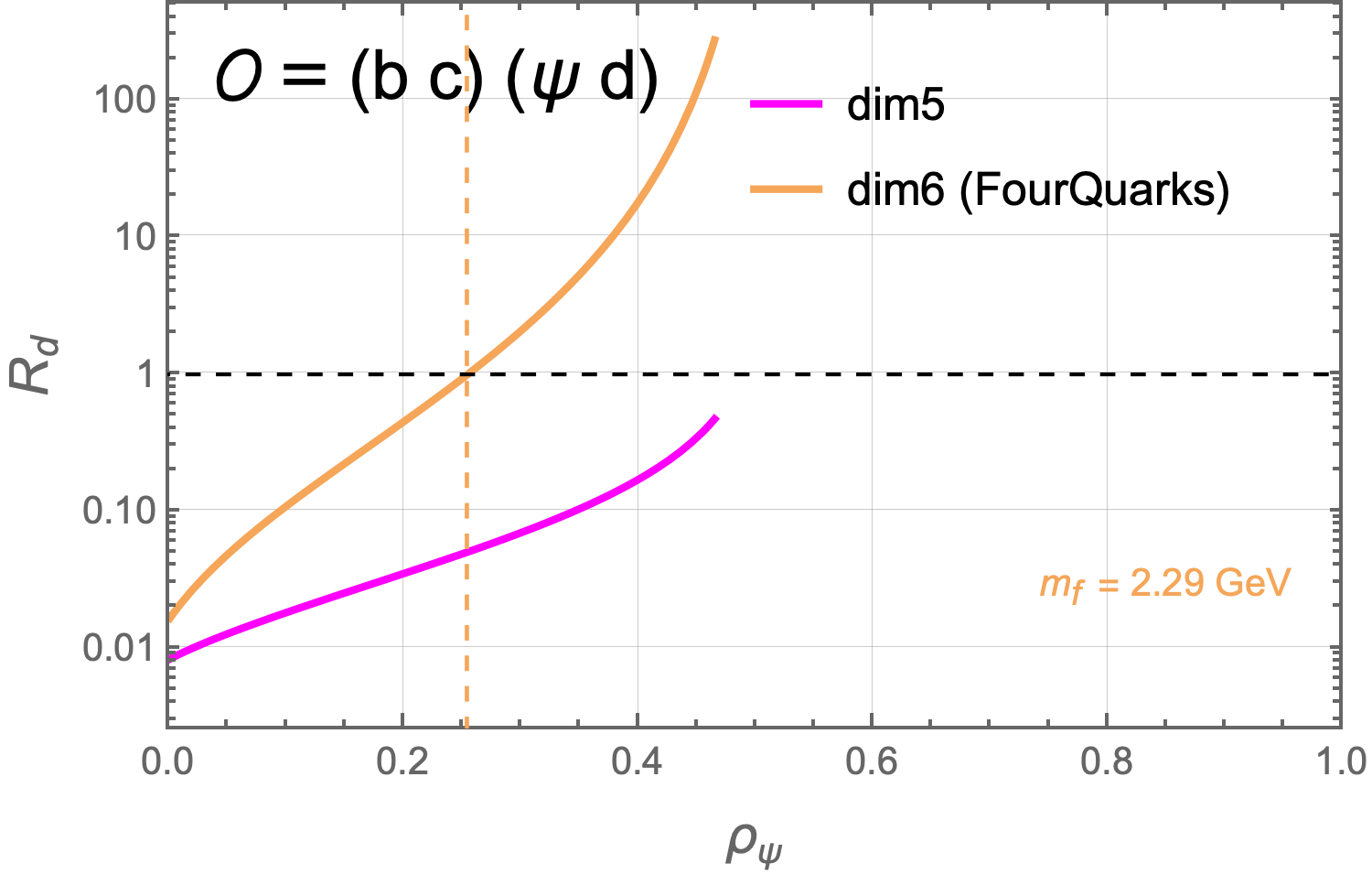}
    \end{subfigure}

    \caption{(Part 3): Additional plots for different operators with the same color coding as in the previous Figure.}
    \label{fig:DecayRates4}
\end{figure}

We present here the analytical expressions of the functions introduced in~\eqref{eq:MasterWidth} for the various operators in the effective Hamiltonian~\eqref{eq:Heff}. For operators involving two distinct massive final states (e.g. $c$ and $\psi$), the Darwin term requires two-scale master integrals, which are not available in closed form. Since such contributions are expected to be numerically subleading, and the dominant subleading effects arise from the dimension-6 four-quark operators, we restrict ourselves to the dimension-5 level in these cases.
The complete expressions for the dimension-3 and dimension-5 contributions are lengthy; they can be provided upon request from the corresponding author.  
The corresponding results for the dimension-3, dimension-5, and dimension-6 four-quark contributions of   $\mathcal{O}=(b\, c)(\psi \, d)$ are shown in Figs.~\ref{fig:DecayRates4} and \ref{fig:DecayRatesTotal2}. For $\mathcal{O}=(d\, c)(\psi \,b)$, we do not display the plots explicitly for brevity; its results coincide with those of $(b\, c)(\psi \,d)$, up to a negligible difference in the dimension-5 contribution.
 \\
For $C_{\psi\psi}^{(d)}$:
\begin{flalign}
f^{(\psi\psi d)}_0 &= \frac{1}{2} \left(\sqrt{1-4 \rho _{\psi }} \left(1-2 \rho _{\psi } \left(6
   \rho _{\psi }^2+\rho _{\psi }+7\right)\right)
   \right. \nonumber \\
& \quad \left.
   +12 \left(\rho _{\psi
   }^2-1\right) \rho _{\psi }^2 \log \left(\frac{2 \rho _{\psi }
   \left(\rho _{\psi }+\sqrt{1-4 \rho _{\psi }}-2\right)-\sqrt{1-4 \rho
   _{\psi }}+1}{2 \rho _{\psi }^2}\right)\right),& \\[4pt] 
   f^{(\psi\psi d)}_G &= \nonumber \frac{1}{4} \Bigg( 12 \rho _{\psi }^2 (1-5 \rho _{\psi }^2) \log \left(\frac{2 \rho _{\psi } (\rho _{\psi }-\sqrt{1-4 \rho _{\psi }}-2)+\sqrt{1-4 \rho _{\psi }}+1}{2 \rho _{\psi }^2}\right) \\
& -\sqrt{1-4 \rho _{\psi }} (10 \rho _{\psi } (6 \rho _{\psi }^2+\rho _{\psi }-1)+3) \Bigg),& \\[4pt]  \nonumber f^{(\psi\psi d)}_D &=
\frac{1}{3} \Bigg( -\sqrt{1-4 \rho _{\psi }} \left(2 \rho _{\psi } \left(\rho _{\psi } \left(30 \rho _{\psi }+17\right)-11\right)-9\right) \\ \nonumber
& -24 \sqrt{1-4 \rho _{\psi }} \log \left(\frac{1}{\rho _{\psi }}-4\right) -24 \log \left(2 \rho _{\psi }\right) \\  \nonumber
& -24 \left(\rho _{\psi }-1\right) \log \left(-2 \rho _{\psi }+\sqrt{1-4 \rho _{\psi }}+1\right) \\ \nonumber 
& -12 \rho _{\psi } \Bigg( \log \left(\frac{2 \rho _{\psi } \left(\rho _{\psi }-\sqrt{1-4 \rho _{\psi }}-2\right)+\sqrt{1-4 \rho _{\psi }}+1}{8 \rho _{\psi }^4}\right) \\
& +\rho _{\psi } \left(\rho _{\psi } \left(5 \rho _{\psi }+2\right)+1\right) \log \left(\frac{2 \rho _{\psi } \left(\rho _{\psi }-\sqrt{1-4 \rho _{\psi }}-2\right)+\sqrt{1-4 \rho _{\psi }}+1}{2 \rho _{\psi }^2}\right) \Bigg) \Bigg).
\end{flalign}\\
For $C_{cu}^{(d)}$:
\begin{flalign}
\nonumber f^{(cu d)}_0 &= -2 \left(\rho _c^4-8 \rho _c^3+8 \rho _c+12 \rho _c^2 \log \left(\rho
   _c\right)-1\right)
, \\ \nonumber f^{(cu d)}_G&=
-5 \rho _c^4+28 \rho _c^3-48 \rho _c^2+20 \rho _c-12 \left(\rho
   _c-2\right) \rho _c \log \left(\rho _c\right)+5, \\ \nonumber f^{(cu d)}_D&=
   \frac{4}{3} \left(-5 \rho _c^4+26 \rho _c^3-45 \rho _c^2-10 \rho _c-12
   \left(\rho _c-1\right){}^2 \left(\rho _c+1\right) \log \left(1-\rho
   _c\right)\right. \\ 
& \left.+6 \left(\rho _c^3-\rho _c^2+5 \rho _c+2\right) \log
   \left(\rho _c\right)+34\right)
\end{flalign}
For $C_{uc}^{(d)}$:
\begin{flalign}
\nonumber f^{(uc d)}_0 &= f^{(cu d)}_0, \\ \nonumber f^{(uc d)}_G&=
-5 \rho _c^4+16 \rho _c^3-16 \rho _c-12 \rho _c^2 \log \left(\rho
   _c\right)+5, \\ \nonumber f^{(uc d)}_D&=
  -\frac{2}{3} \left(10 \rho _c^4-15 \rho _c^3-18 \rho _c^2+71 \rho _c+12
   \left(\rho _c-1\right){}^2 \left(\rho _c+5\right) \log \left(1-\rho
   _c\right)\right. \\ 
& \left.-6 \left(\rho _c^3+3 \rho _c^2+1\right) \log \left(\rho
   _c\right)-48\right)
\end{flalign}
\\
For $C_{cc}^{(d)}$:
\begin{flalign}
f^{(cc d)}_0 &= \frac{1}{4} f^{(\psi\psi d)}_0 |_{\rho_\psi \to \rho_c}, \\ \nonumber f^{(cc d)}_G&=
\sqrt{1-4 \rho _c} \Bigg( 2 \rho _c \left(\left(7-30 \rho _c\right) \rho _c+7\right)+5 \Bigg) \\ \nonumber
& +12 \rho _c \left(\rho _c \left(1-5 \rho _c^2\right) \log \left(\frac{2 \rho _c \left(\rho _c-\sqrt{1-4 \rho _c}-2\right)+\sqrt{1-4 \rho _c}+1}{2 \rho _c^2}\right) \right. \\ \nonumber
& \left. +2 \left(\rho _c^2 \log \left(\frac{\rho _c}{\left(\sqrt{1-4 \rho _c}-3\right) \rho _c-\sqrt{1-4 \rho _c}+1}\right)+\log \left(4 \rho _c\right) \right. \right. \\ 
& \left. \left. +\left(\rho _c^2-2\right) \log \left(\sqrt{1-4 \rho _c}+1\right)\right)\right),& \\[4pt]  \nonumber f^{(cc d)}_D &= \frac{2}{3} \Bigg(6 \log \left(-32 \left(\sqrt{1-4 \rho _c}-1\right)\right)+12 \log \left(\rho _c\right) &\\ \nonumber
&+\sqrt{1-4 \rho _c} \left(\rho _c \left(2 \left(11-60 \rho _c\right) \rho _c-1\right)+12 \log \left(\frac{\rho _c}{1-4 \rho _c}\right)+72\right) &\\ \nonumber
& +6 \rho _c \left(\log \left(256 \rho _c^5\right)+\rho _c \left(\rho _c \left(20 \rho _c-7\right)+4\right) \log \left(-2 \left(\sqrt{1-4 \rho _c}-1\right) \rho _c\right)\right) &\\ \nonumber
& -6 \left(\rho _c \left(\rho _c \left(\rho _c \left(20 \rho _c-7\right)+4\right)+2\right)-1\right) \log \left(-2 \rho _c+\sqrt{1-4 \rho _c}+1\right) &\\
& -6 \left(\rho _c \left(\rho _c \left(\rho _c \left(20 \rho _c-7\right)+4\right)+6\right)+7\right) \log \left(\sqrt{1-4 \rho _c}+1\right)\Bigg). &
\end{flalign}
The above results also apply to the Wilson coefficients obtained by replacing $d \to s$. For operators with only light quarks $(u, d, s)$ in the final state, the corresponding expressions are obtained in the massless limit of the massive cases.
\pagebreak
\section{\boldmath Power-suppressed Effects in the SM: A Comparison}
\label{app:B}
In section~\ref{sec:InclusiveTotalLifetimes} we have shown that, within the Mesogenesis framework, subleading contributions can dominate over the leading dimension-three term in certain regions of the dark-sector mass $m_\psi$. 
For comparison, it is instructive to study the analogous behavior in the SM. Since in the Mesogenesis framework a single operator is expected to dominate at a time, we take into consideration the SM operator ${\cal O}=(u\,d)(c\, b)$, and hence consider the contribution to the inclusive decay $b \to c\bar{u}d$ which generates the familiar PI topology in the $B^+$ system, and investigate how the HQE contributions behave as a function of the scaled charm-quark mass $\rho_c = (m_c/m_b)^2$. The SM effective Hamiltonian for the operator ${\cal O}=(u\,d)(c \,b)$  reads (see e.g. the review~\cite{Buchalla:1995vs})
\begin{equation}
{\cal H}_{\rm eff}^{\rm SM}(x)
= 
\frac{4 G_F}{\sqrt 2} V_{cb} V_{ud}^*
\sum_{i = 1}^{2} 
C_i \, Q_i (x) 
+ {\rm h.c.},
\end{equation}
with
\begin{align}
Q_1 &= (\bar c^i \gamma_\mu P_L \, b^i) (\bar d^j \gamma^\mu P_L \, u^j)\,, &
Q_2 &= (\bar c^i \gamma_\mu P_L \, b^j) (\bar d^j \gamma^\mu P_L \, u^i)\,.
\end{align}
Using the LO values of the Wilson coefficients $C_1 (4.5) = 1.11$ and $C_2 (4.5) = -0.26$~\cite{Egner:2024lay} and the numerical inputs in Appendix~\Ref{app:A}, we show in  Fig.~\ref{fig:DecayRatesSM} the individual HQE contributions and their ratios to the leading term. 
We observe that by artificially varying $m_c$ across the kinematic range corresponding to $m_\psi$ in Mesogenesis, a qualitatively similar pattern emerges: certain subleading contributions exceed the leading term, thereby reproducing the nontrivial HQE hierarchy seen in the NP scenario. 
This comparison highlights that the phenomenon is not unique to Mesogenesis but rather reflects a more general feature of the HQE when quark-mass parameters are varied. 

At the same time, there are important quantitative differences between the SM and Mesogenesis. In the SM, the dimension-six four-quark contributions arising from PI topologies are generally smaller relative to the leading dimension-three term, whereas in Mesogenesis the $\overline{\rm PI}$ topology can generate a substantially larger subleading effect, as shown in  Figs.~\ref{fig:DecayRates} and~\ref{fig:DecayRatesTotal}.

To provide a clean benchmark for comparison with the Mesogenesis operator, we eliminate interference effects in the SM by setting $C_2 \to 0$. Switching off $C_2$ leaves the simplified structure
\begin{equation}
 {\cal H}_{b \to c u d}^{\rm SM}(x)
= 
\widetilde{G}_F \, (\bar c^i \gamma_\mu P_L \, b^i) (\bar d^j \gamma^\mu P_L \, u^j) (x) 
+ {\rm h.c.}\,,
\end{equation}
where $\widetilde{G}_F=\tfrac{4 G_F}{\sqrt 2} V_{cb} V_{ud}^* C_1$. 
The Mesogenesis operator of interest reads
\begin{equation}
\mathcal H^{-1/3}_{b\to \psi ud} (x) = 
   C_{\psi}^{(ud)} \, \epsilon_{ijk}\,(\bar u_R^{\,i} b_R^{c,j})(\bar\psi\, d_R^{c,k}) (x)+ {\rm h.c.}\,.
\end{equation}
Apart from the Wilson coefficients, the two Hamiltonians differ in their color and Dirac structures, which in turn affects the leading dimension-3 contribution as follows:
\begin{align}
\Gamma_3^{\rm SM}(B^+) 
= \Gamma_3^{\rm -1/3}(B^+) \,
   \cdot \mathop{4}\limits_{\substack{\downarrow \\ \text{Dirac}}} \,
   \cdot \mathop{\tfrac{3}{2}}\limits_{\substack{\downarrow \\ \text{Color}}}
   \quad ( |C_{\psi}^{(ud)}|^2 \to |\widetilde{G}_F|^2)\,.
\end{align}
These two effects carry over to the dimension-6 four-quark contribution. In the SM the PI contribution takes the form
\begin{equation}
\Gamma_{\rm PI}^{\rm SM} \;\propto\; 4 \cdot \left[\tfrac{1}{N_c}\,\tilde{Q}_1 + 2\,\tilde{Q}_3 \right],
\end{equation}
where $\tilde{Q}_{1,3}$ are defined in Eq.~(\ref{eq:Q1-Q2}),
while in Mesogenesis the contribution of the $\overline{\rm PI}$ topology in Fig.~\ref{fig:Topologies} becomes 
\begin{equation}
\Gamma_{\overline{\rm PI}}^{\rm -1/3} \;\propto\; \bigg[\tfrac{N_c-1}{N_c}\,\tilde{Q}_1 - 2\,\tilde{Q}_3 \bigg] \bigg|_{P_L \to P_R}.
\end{equation}
The SM again carries the factor $4$ from the Dirac structure, but the Mesogenesis operator generates a different color combination.  
Employing the parameterization of the forward matrix elements of the four-quark operators in terms of the bag parameters listed in Table~\ref{tab:Bag-parameters}, we find that the combination $((N_c-1)/N_c)\tilde{B}^{u}_1 - 2\tilde{B}^{u}_3$ evaluates numerically to roughly a factor of two larger (in magnitude) than the corresponding SM combination $(1/N_c)\tilde{B}^{u}_1 + 2\tilde{B}^{u}_3$.
It is precisely this modified bag-parameter combination that enhances the relative size of the dimension-6 four-quark term in Mesogenesis. 

\begin{figure}[h!]
    \centering
    \begin{subfigure}{0.47\textwidth}
        \centering
        \includegraphics[width=\textwidth]{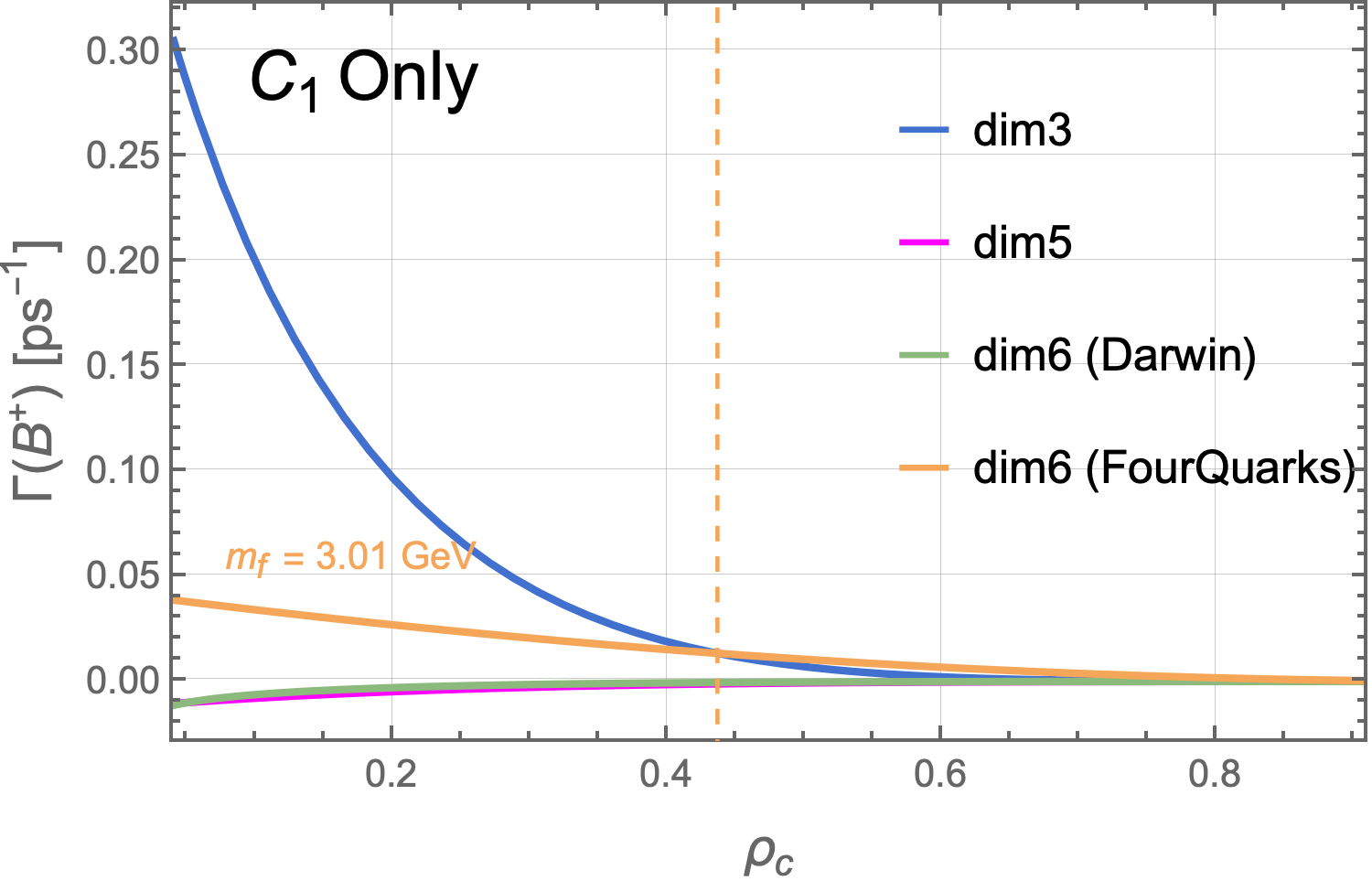}
    \end{subfigure}
    \hfill
    \begin{subfigure}{0.47\textwidth}
        \centering
        \includegraphics[width=\textwidth]{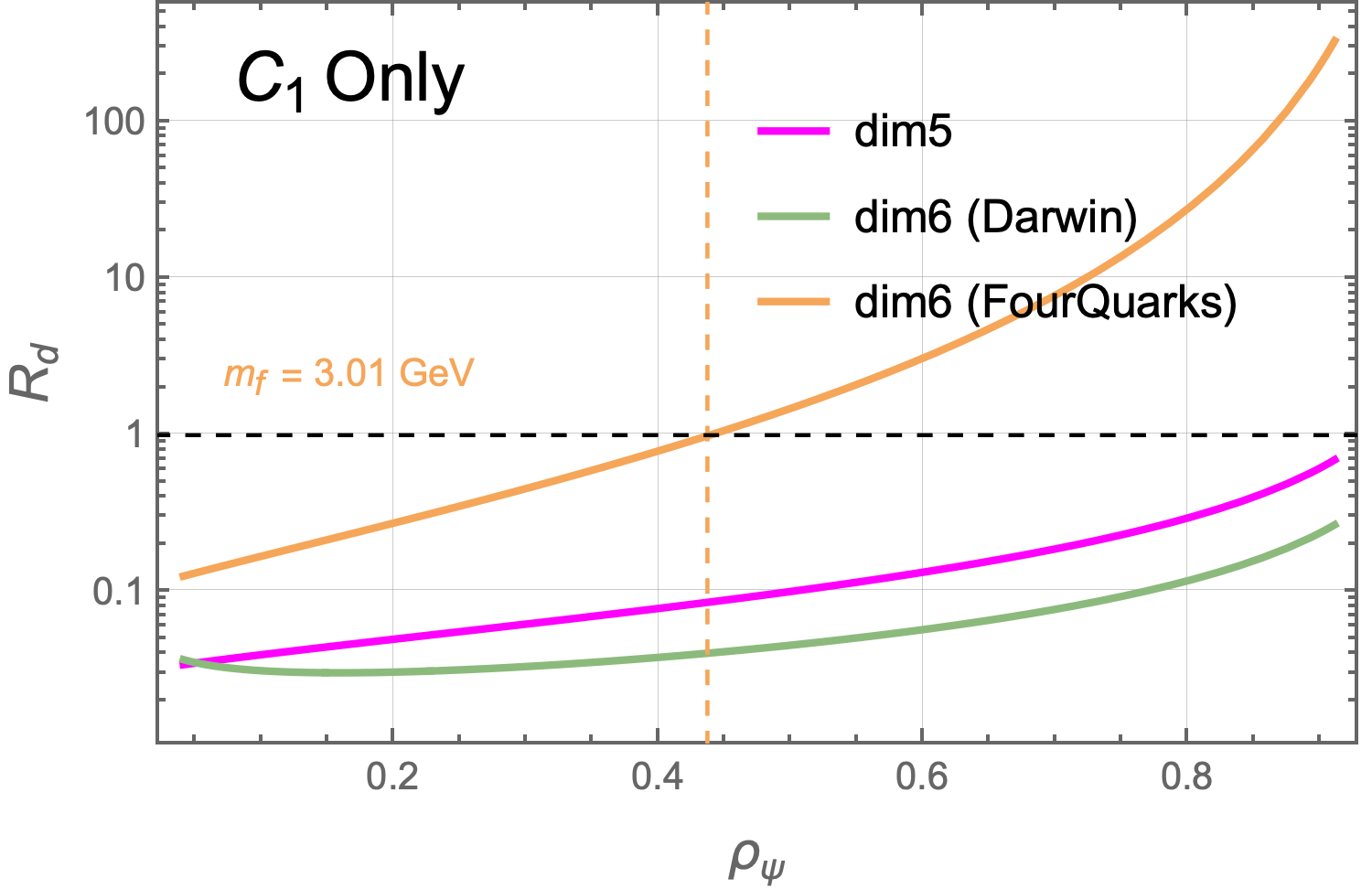}
    \end{subfigure}
    

   \begin{subfigure}{0.47\textwidth}
        \centering
        \includegraphics[width=\textwidth]{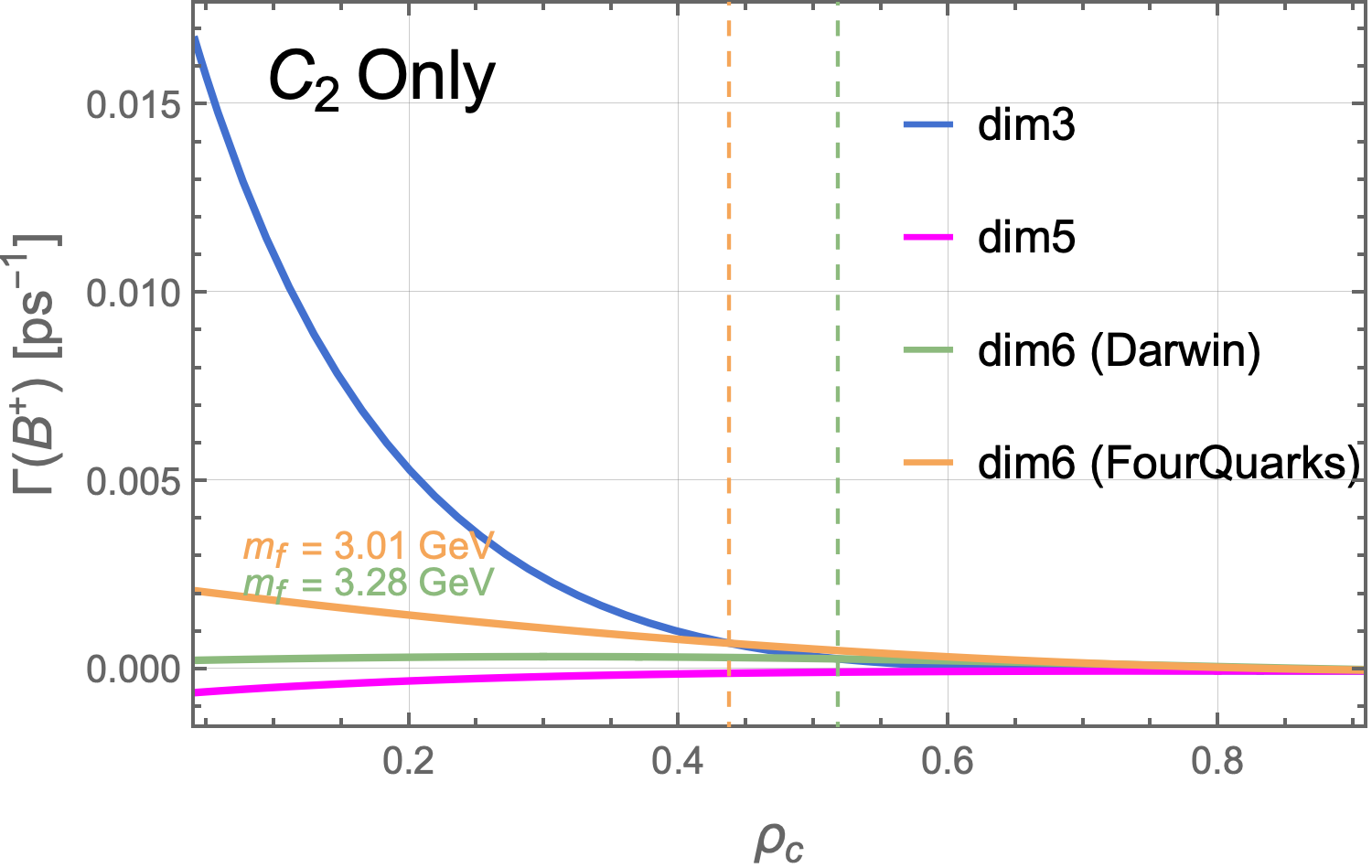}
    \end{subfigure}
    \hfill
    \begin{subfigure}{0.47\textwidth}
        \centering
        \includegraphics[width=\textwidth]{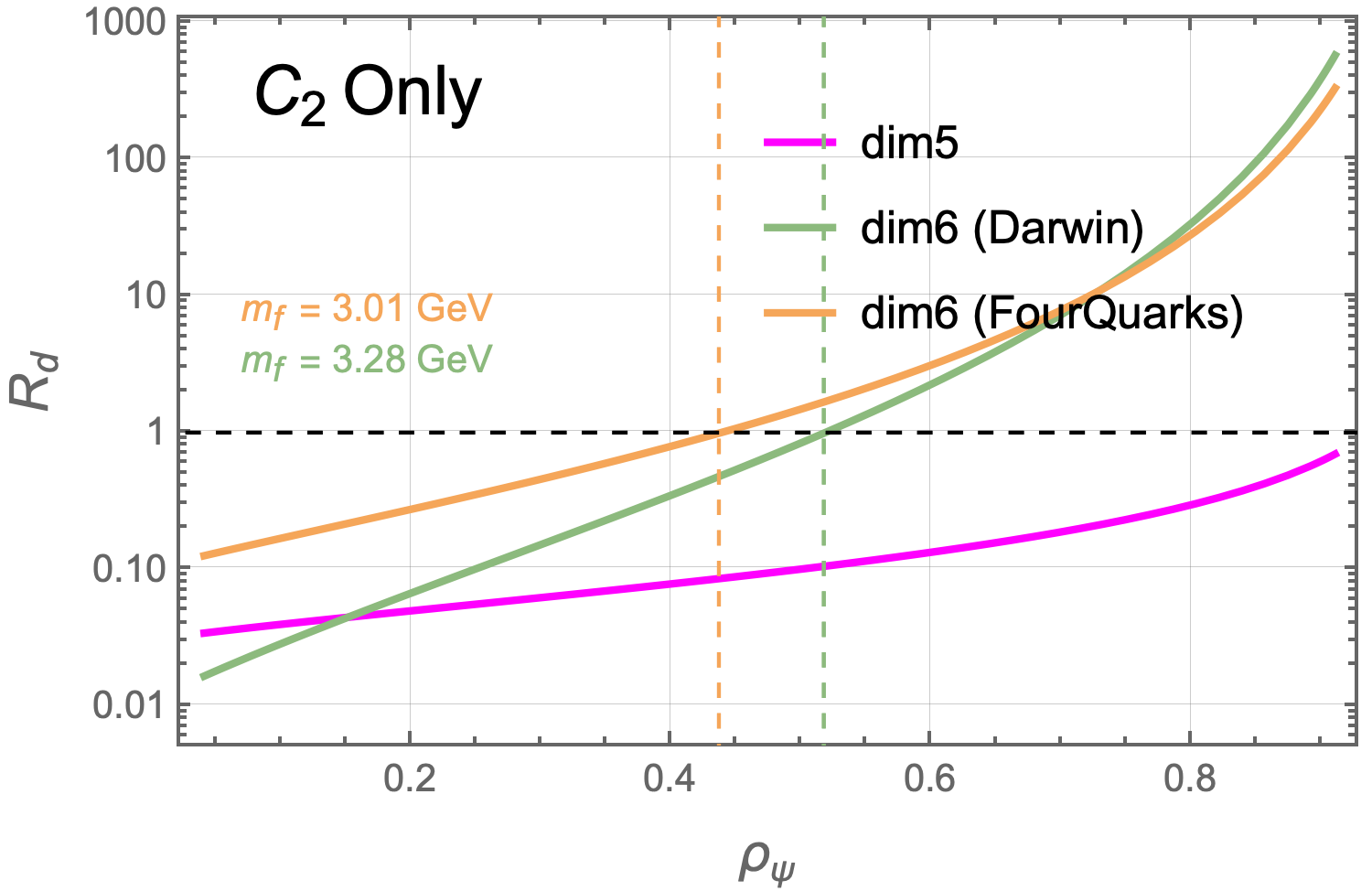}
    \end{subfigure}
\caption{Total decay width $\Gamma(B^+)$ arising from the SM operator $(c b)(d u)$, as a function of $\rho_f = (m_f/m_b)^2$. 
The plots show the individual HQE contributions (left) and the ratios $R_d = |\Gamma_d/\Gamma_3|$ (right). 
The upper panels display the case with only $C_1$ switched on, while the lower panels correspond to only $C_2$. 
Vertical lines mark the values of $\rho_f$ where subleading terms exceed the leading dimension-three width (the corresponding $m_f$ values are also indicated). }
    \label{fig:DecayRatesSMC1C2}
\end{figure}

Including operator interference in the SM via $C_2$ suppresses the four-quark effect even further.
Numerically, this reduces the dimension-6 contribution by about a factor of two and shifts the crossing point with the dimension-3 term from $\rho_c\simeq 0.4$ (with $C_2=0$) to $\rho_c\simeq 0.6$ (with both $C_1,C_2$ included), see Fig.~\ref{fig:DecayRatesSMC1C2}. 
By contrast, the Mesogenesis operators do not suffer from such cancellations, retaining the larger prefactor in the bag-parameter combination. 

Taken together, these observations show that the enhancement of the $\overline{\rm PI}$ contribution in Mesogenesis compared to the SM (relative to the dimension-three result) arises from two main effects:  
(i) different color structures, which lead to a larger bag-parameter combination in Mesogenesis, and  
(ii) the absence of destructive operator interference, which in the SM further suppresses the four-quark term.  
As a result, the ratio $\Gamma_{\rm dim6}/\Gamma_{\rm dim3}$ is systematically larger in Mesogenesis than in the SM, and the subleading terms can even dominate over the leading term in substantial regions of parameter space. For calculations of BSM operators with similar SM color but different Dirac structures, see Ref.~\cite{Lenz:2022pgw}.
\begin{figure}[t]
    \centering
    \begin{subfigure}{0.45\textwidth}
        \centering
        \includegraphics[width=\textwidth]{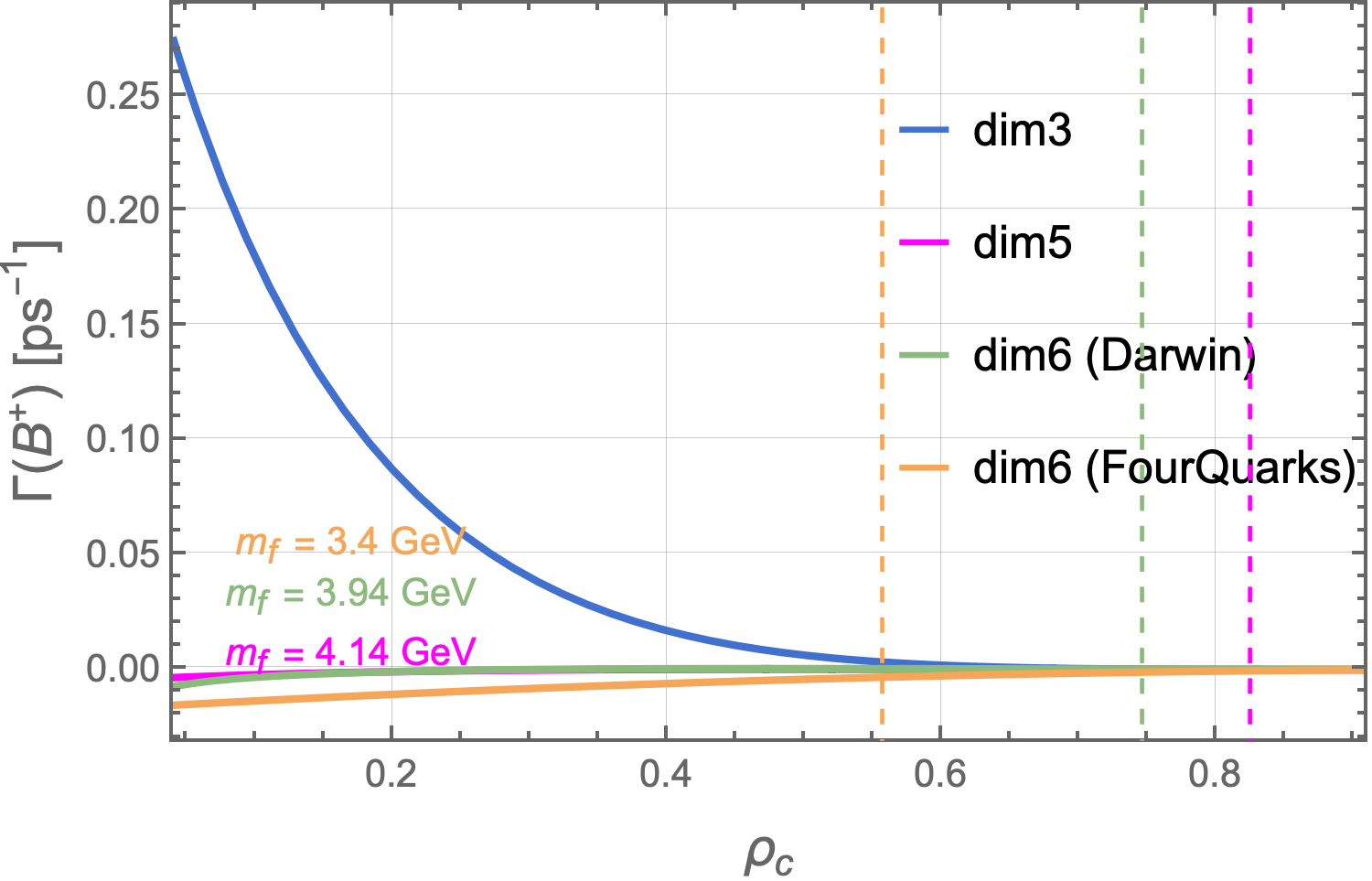}
    \end{subfigure}
    \hfill
    \begin{subfigure}{0.45\textwidth}
        \centering
        \includegraphics[width=\textwidth]{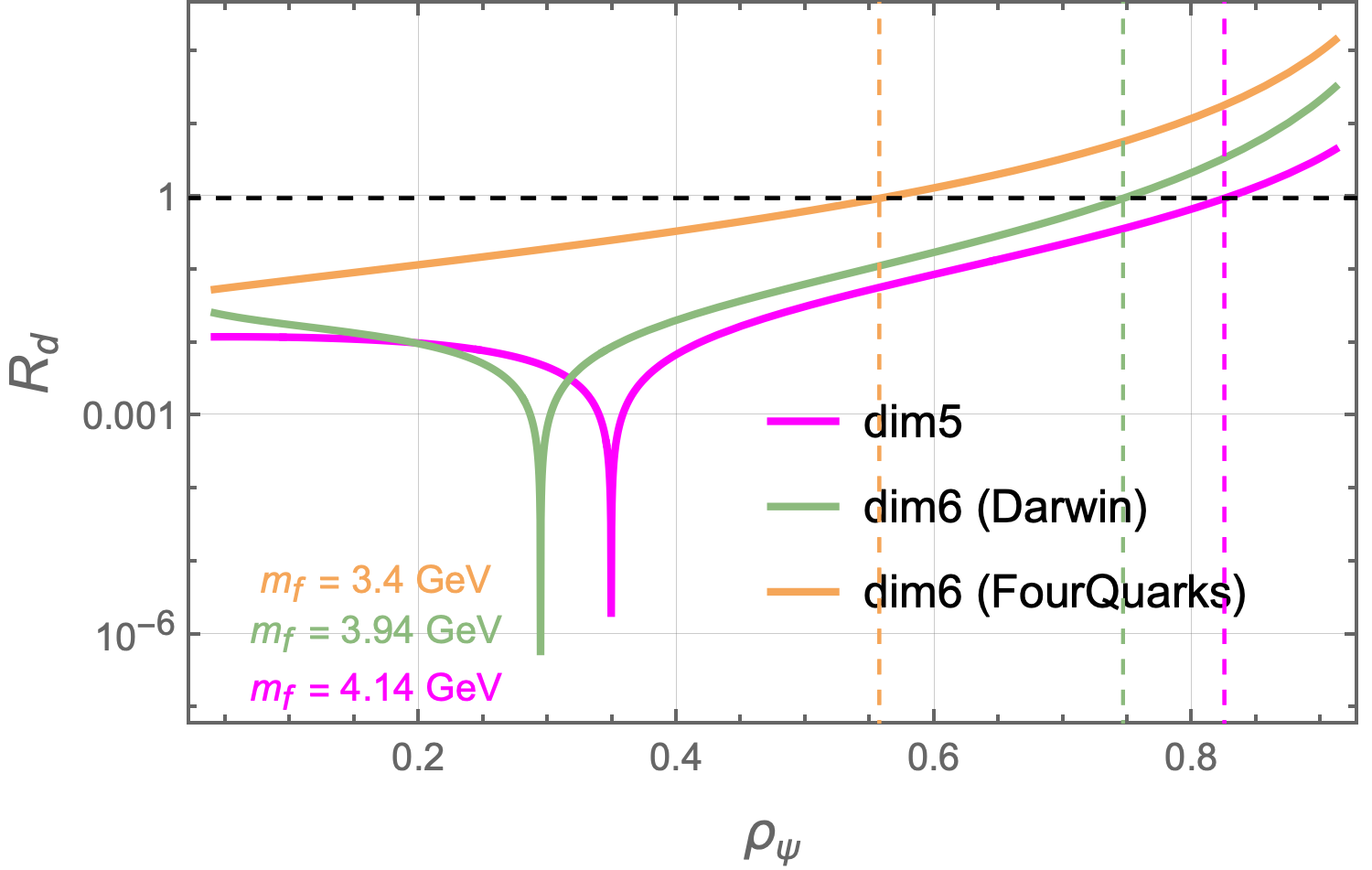}
    \end{subfigure}
    
\caption{Total decay width $\Gamma(B^+)$ arising from the SM operator $(c b)( d u)$ , shown as a function of $\rho_f = (m_f/m_b)^2$. 
The plots display the individual HQE contributions (left) and the corresponding ratios $R_d = |\Gamma_d/\Gamma_3|$ (right).
The vertical lines indicate the values of $\rho_f$ where subleading terms exceed the leading contribution (the corresponding $m_f$ values are also indicated). 
For the dimension-six four-quark contribution, which originates from the PI topology and is negative across the full range, the line marks the point where its \textit{absolute value} overtakes $\Gamma_3$. 
Both $C_1$ and $C_2$ are included. The sharp spikes in the $R_d$ plot occur where the numerator, $\Gamma_d$, changes sign. }
\label{fig:DecayRatesSM}
\end{figure}
\clearpage
\section{\boldmath Supplementary Plots}
This appendix collects the results for the remaining operators considered in our analysis. While the main text focuses on the representative cases ($\mathcal{O} = (b u)(\psi d)$ and $\mathcal{O} = (d u)(\psi b)$), the following plots provide a systematic overview and illustrate to what extent the observed features generalize across the operator basis.
\label{app:D}
\begin{figure}[ht]
    \centering
    \begin{subfigure}{0.43\textwidth}
        \centering
        \includegraphics[width=\textwidth]{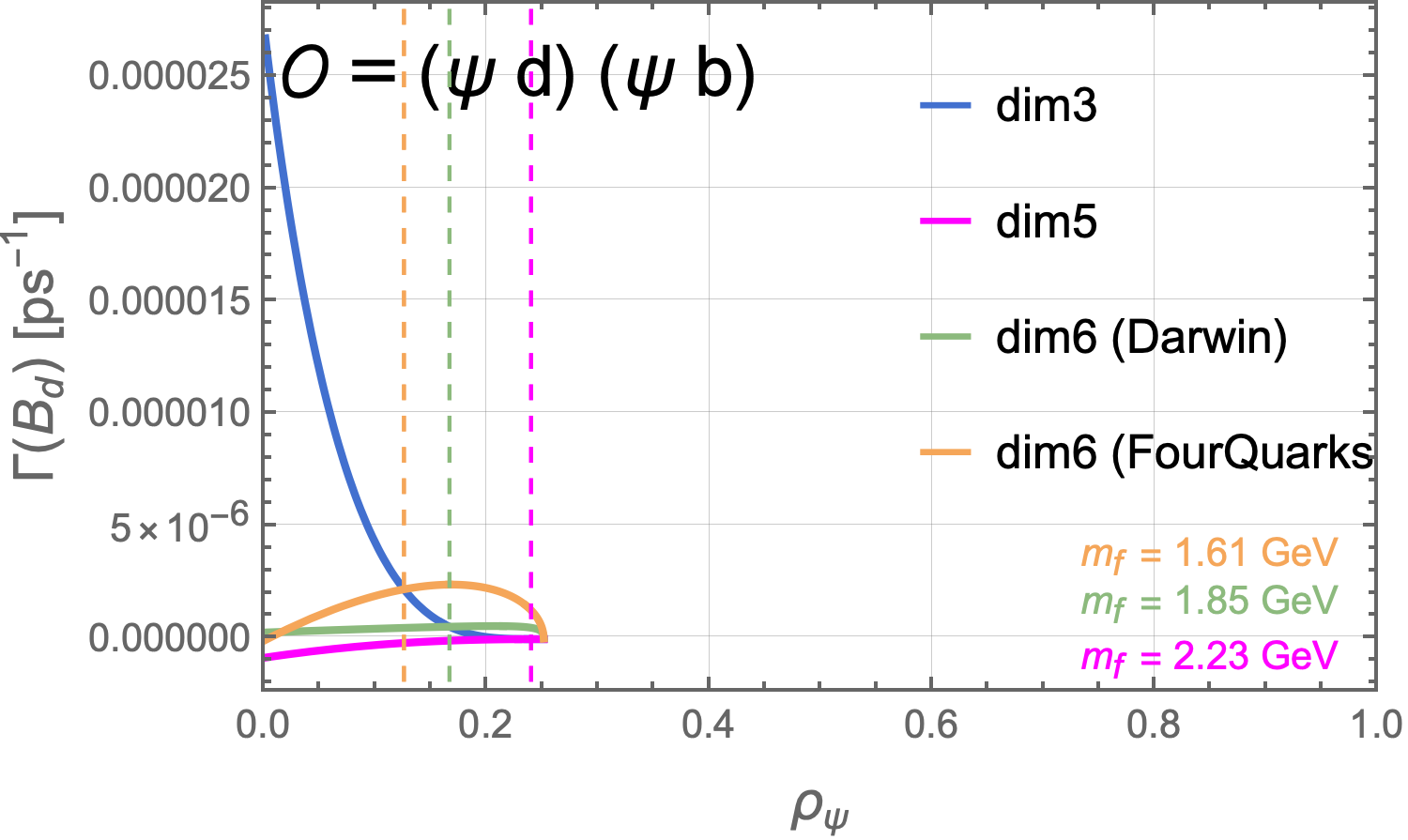}
    \end{subfigure}
    \hfill
    \begin{subfigure}{0.40\textwidth}
        \centering
        \includegraphics[width=\textwidth]{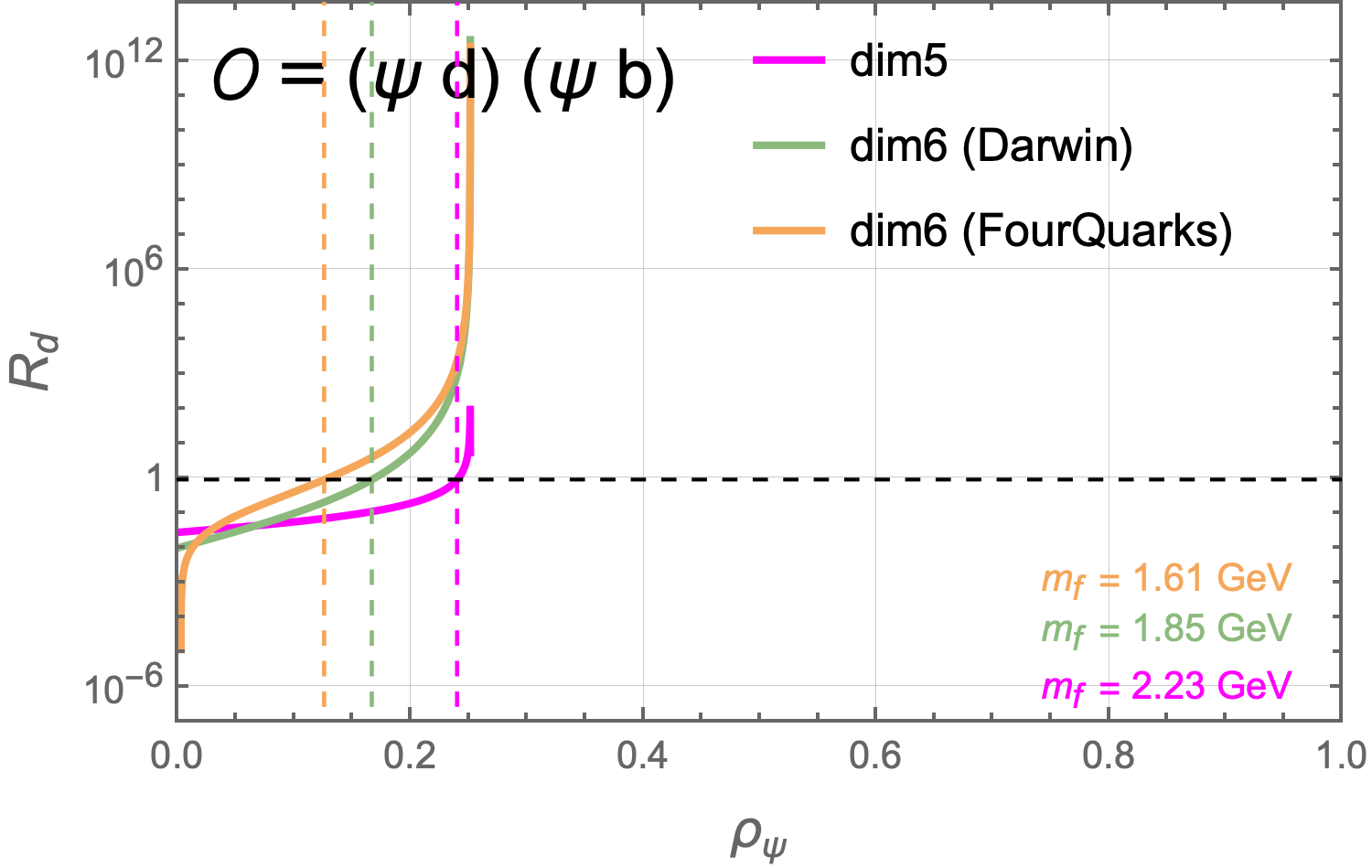}
    \end{subfigure}
    
    \medskip 
    
    \begin{subfigure}{0.43\textwidth}
        \centering
        \includegraphics[width=\textwidth]{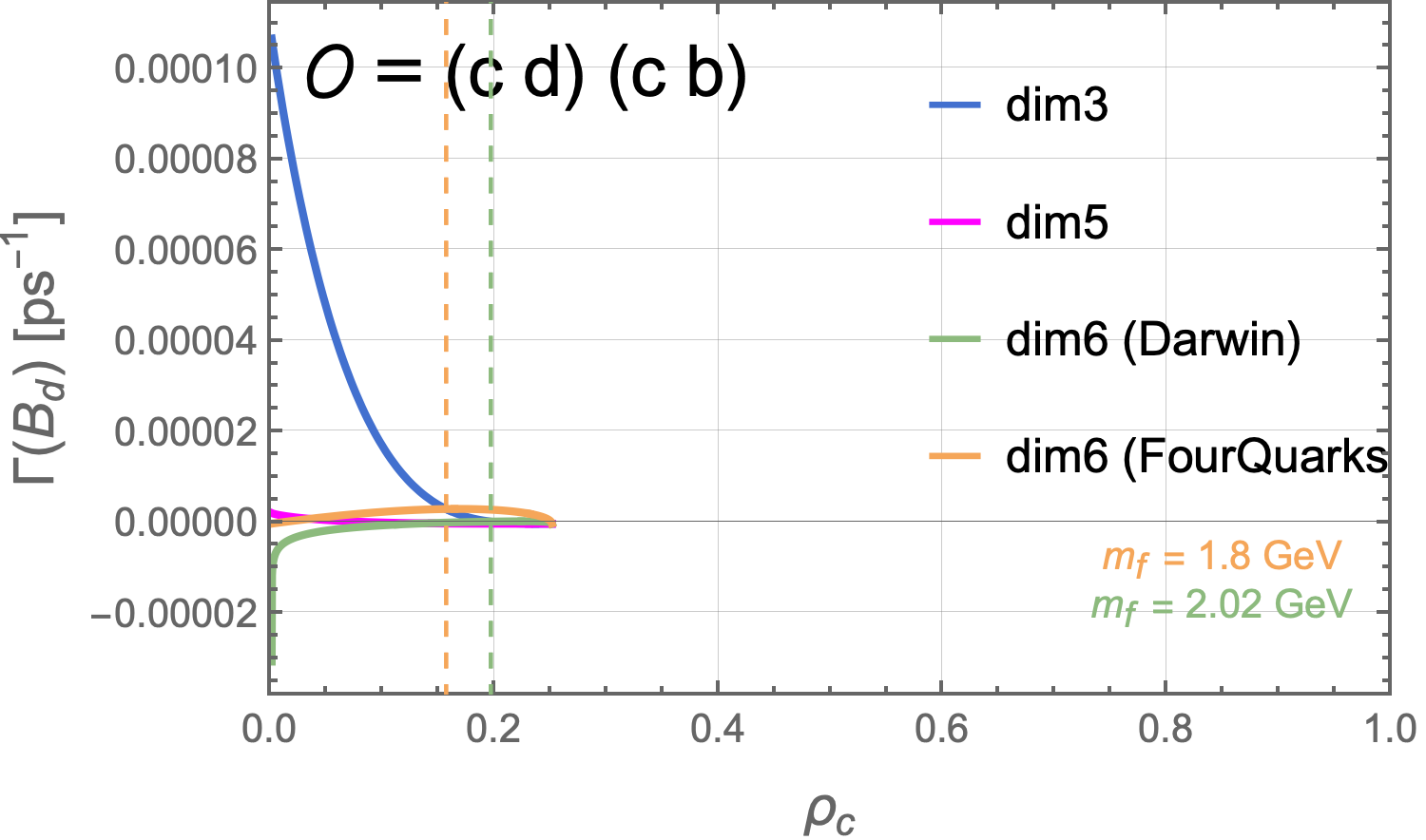}
    \end{subfigure}
    \hfill
    \begin{subfigure}{0.40\textwidth}
        \centering
        \includegraphics[width=\textwidth]{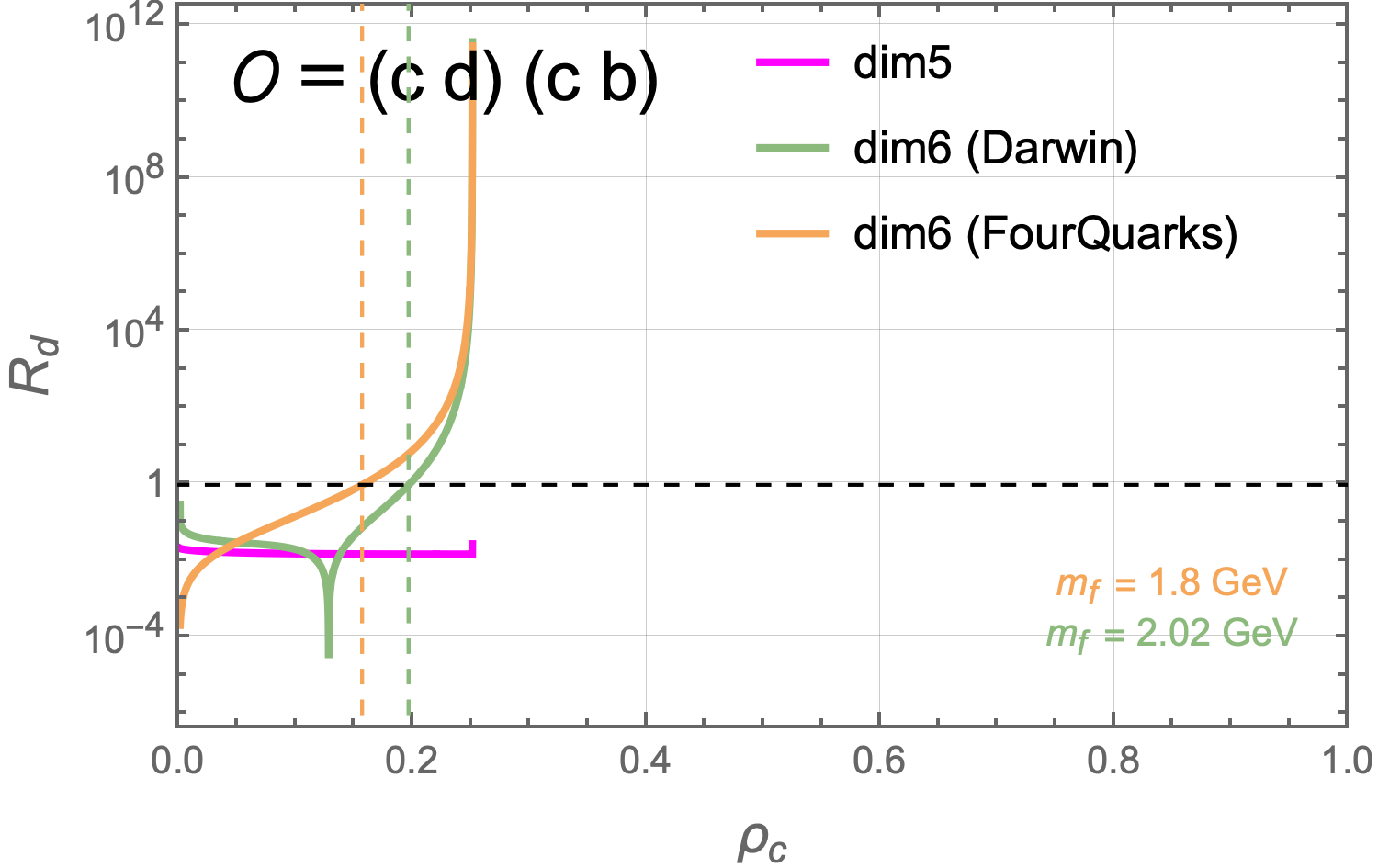}
    \end{subfigure}

     \medskip 
    
    \begin{subfigure}{0.43\textwidth}
        \centering
        \includegraphics[width=\textwidth]{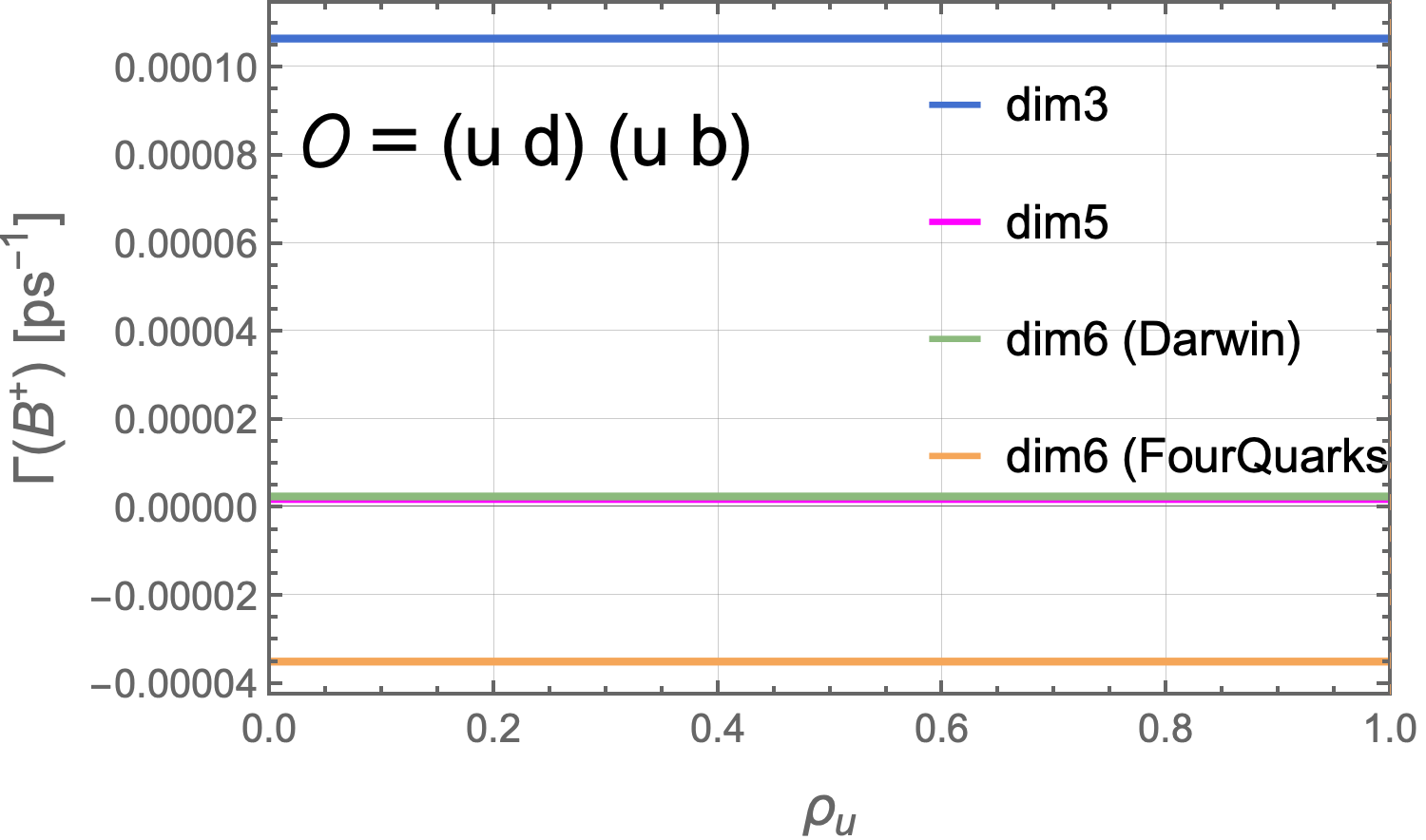}
    \end{subfigure}
    \hfill
    \begin{subfigure}{0.40\textwidth}
        \centering
        \includegraphics[width=\textwidth]{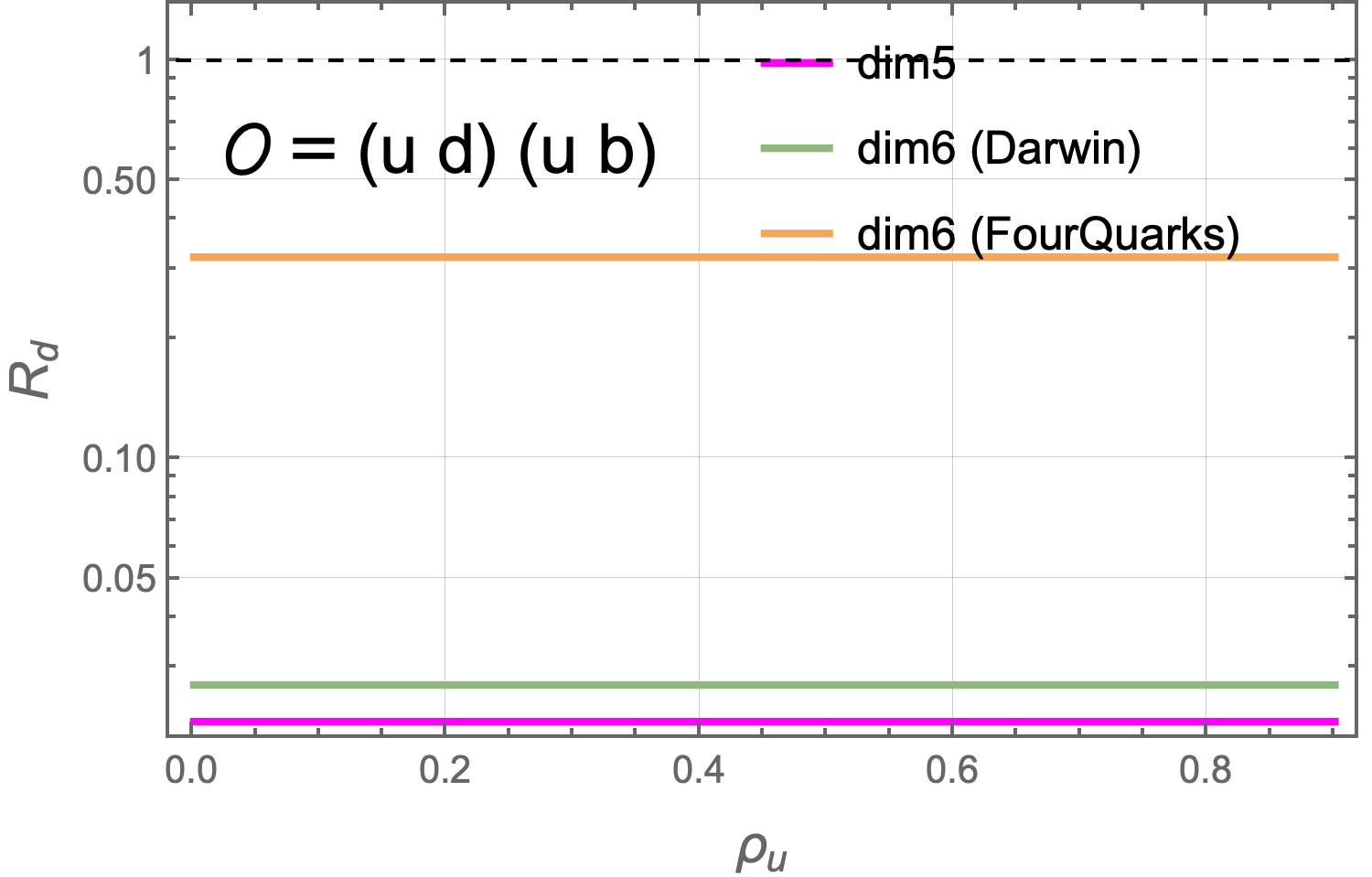}
    \end{subfigure}
    \caption{(Part 1) Total decay width of $\Gamma(B^+)$ arising purely from the Mesogenesis (BSM) operators as a function of $\rho_f= (m_f/m_b)^2$. The left and right columns display the individual contributions from the HQE expansion and the ratio $R_d=|\Gamma_d/\Gamma_3|$,where the numerator corresponds to one of the subleading contributions as indicated on the plot, respectively. The vertical lines indicate the value of $\rho_f$, where the contribution exceeds the leading dimension-three $\Gamma_3$ term (the corresponding $m_f$ values are also indicated). The plots terminate at specific $\rho_f$ values because the total width is proportional to the imaginary part of the forward amplitude, see Eq.~(\ref{eq:GammaB}), requiring the loop particles to be \textit{on-shell}. The sharp spikes in the $R_d$ plot occur where the numerator, $\Gamma_d$, changes sign.}
    \label{fig:DecayRates2}
\end{figure}

\begin{figure}[ht]
    \centering
    \begin{subfigure}{0.43\textwidth}
        \centering
        \includegraphics[width=\textwidth]{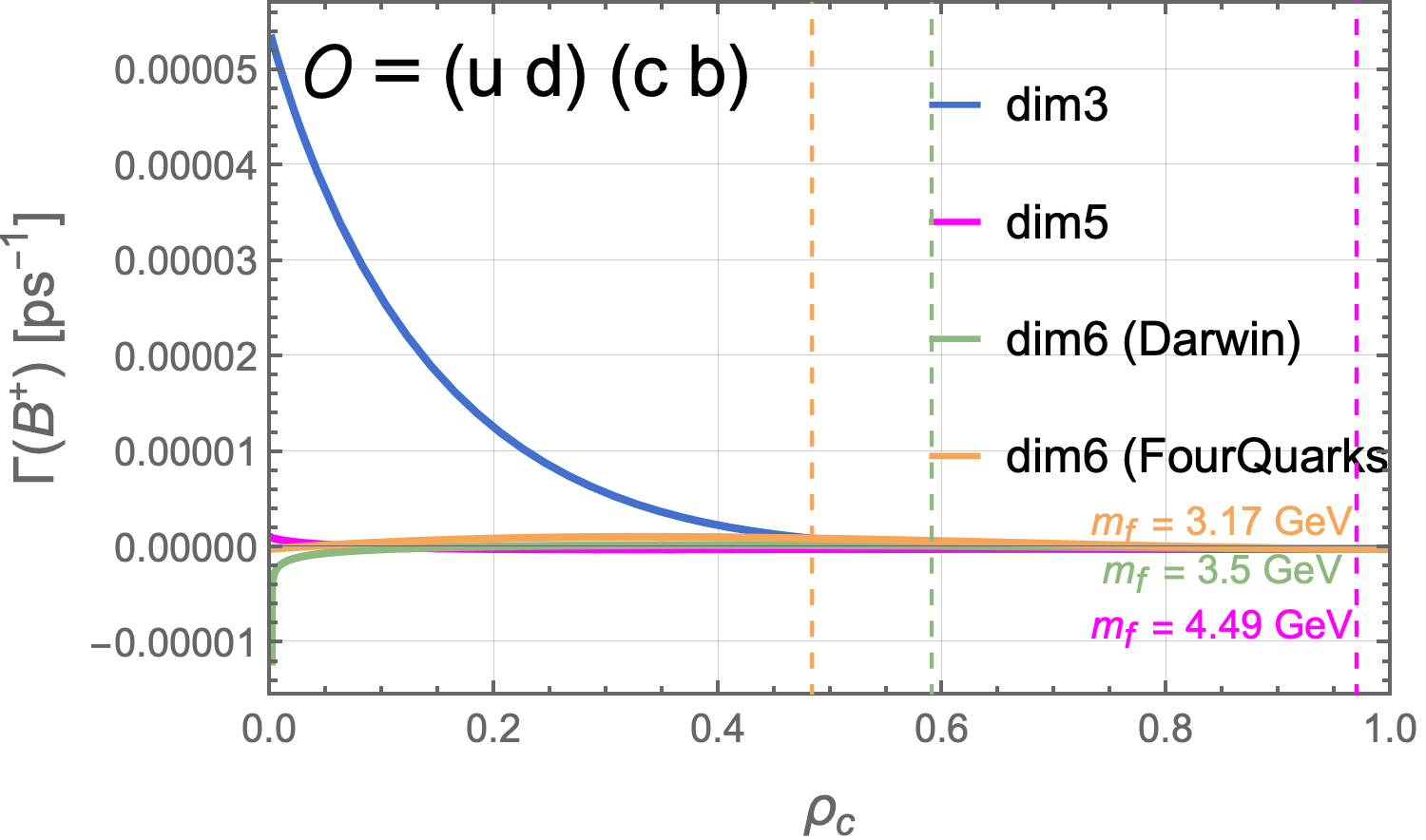}
    \end{subfigure}
    \hfill
    \begin{subfigure}{0.40\textwidth}
        \centering
        \includegraphics[width=\textwidth]{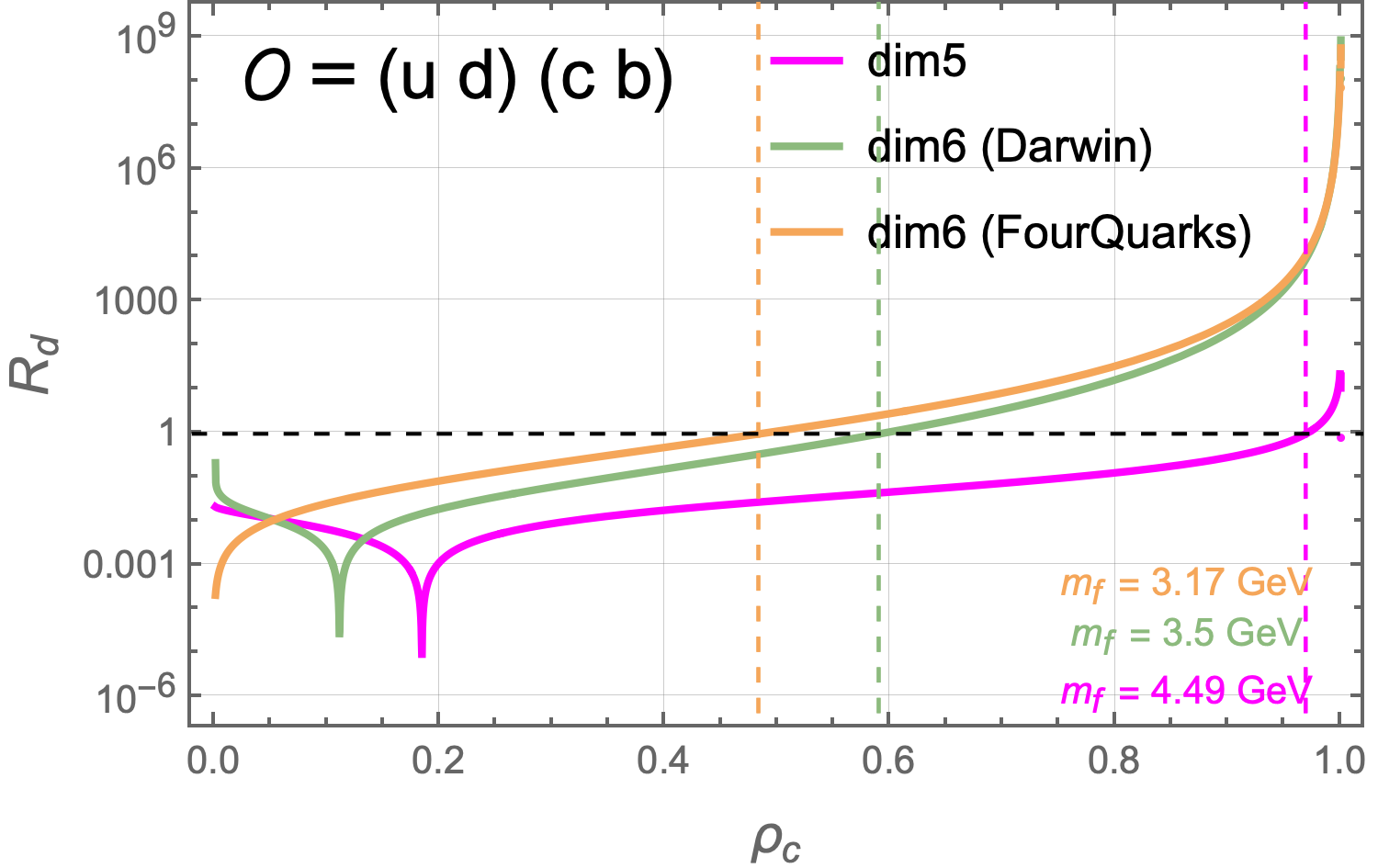}
    \end{subfigure}

     \medskip 
    
    \begin{subfigure}{0.43\textwidth}
        \centering
        \includegraphics[width=\textwidth]{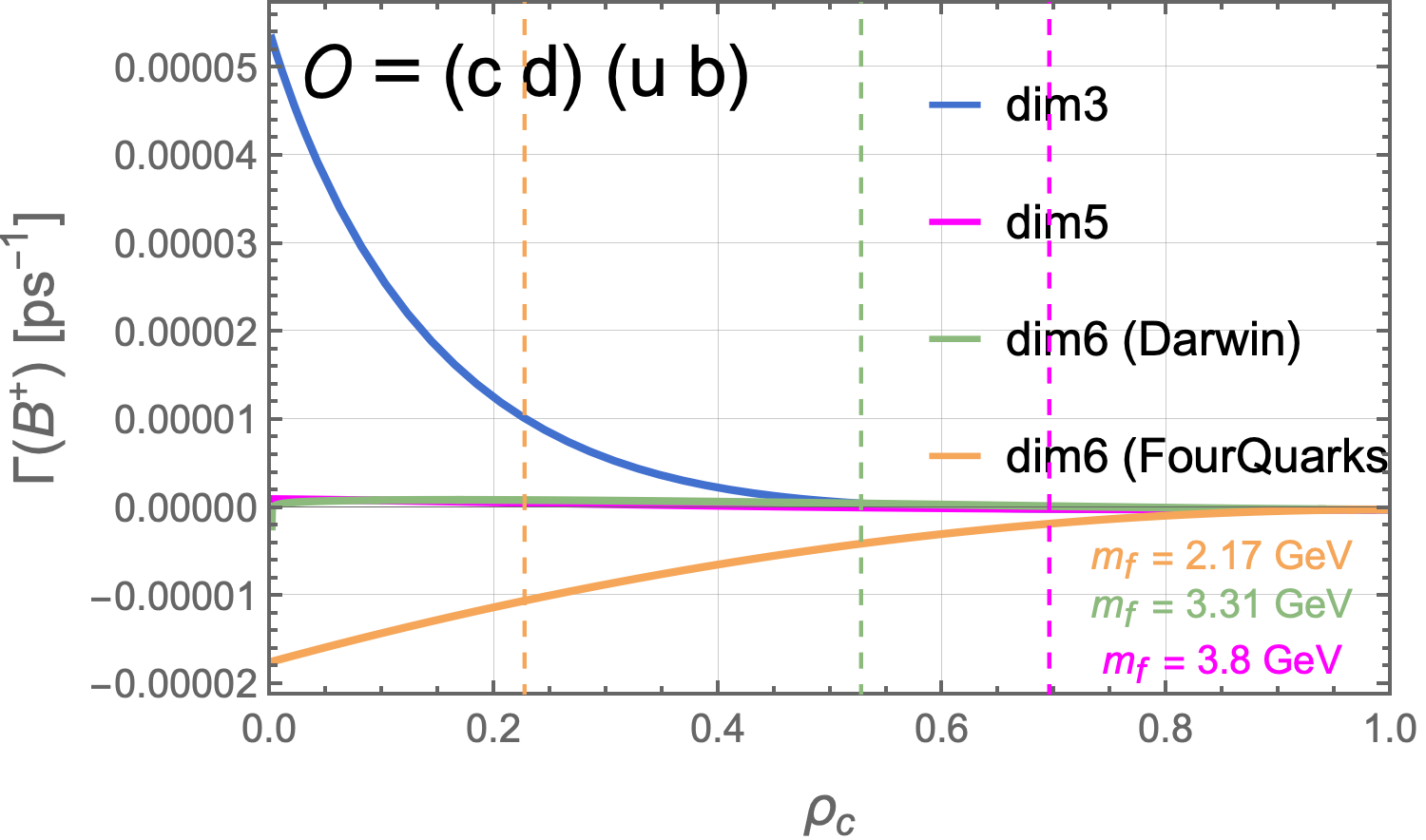}
    \end{subfigure}
    \hfill
    \begin{subfigure}{0.40\textwidth}
        \centering
        \includegraphics[width=\textwidth]{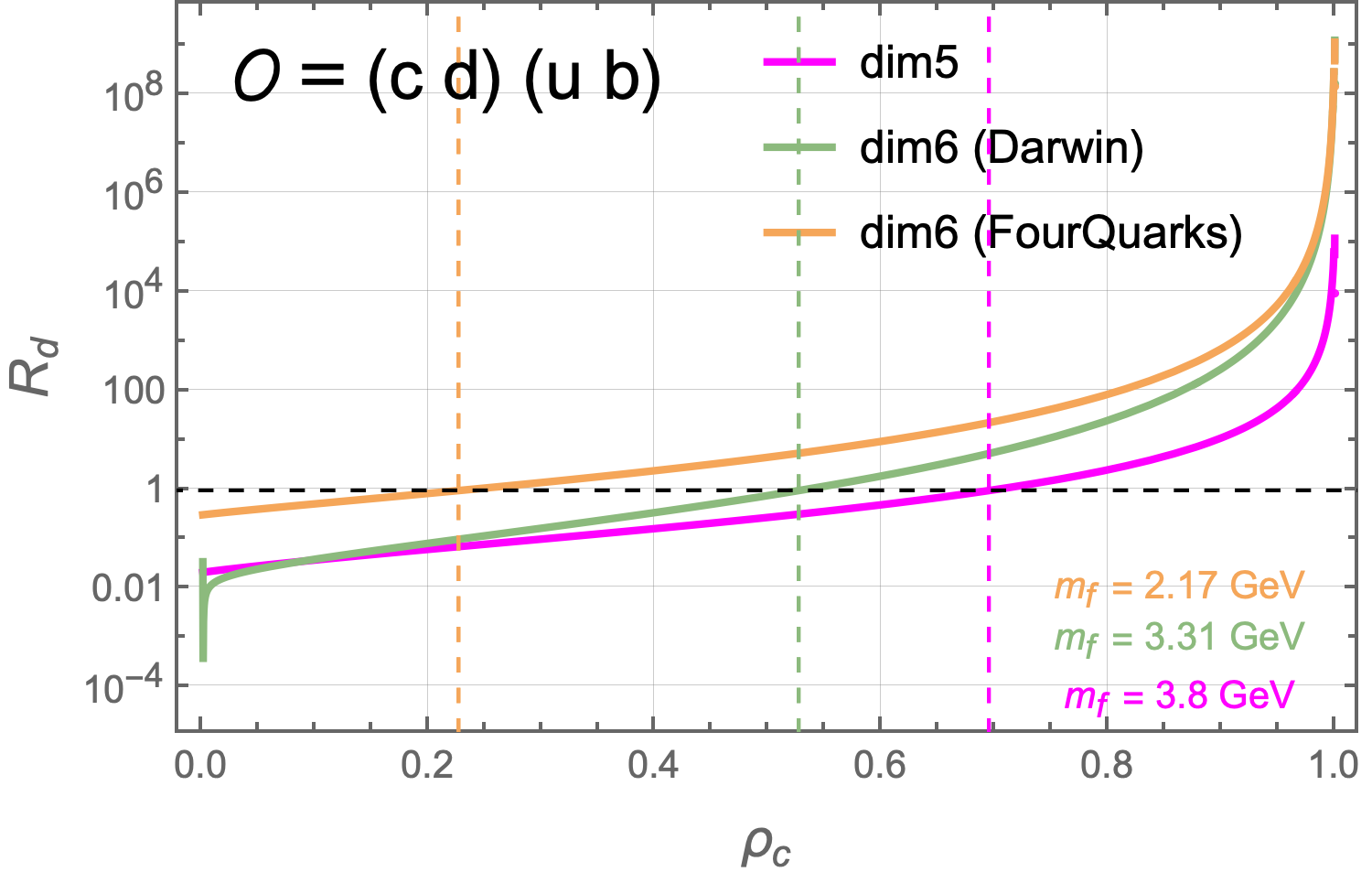}
    \end{subfigure}

    \medskip 
    
    \begin{subfigure}{0.43\textwidth}
        \centering
        \includegraphics[width=\textwidth]{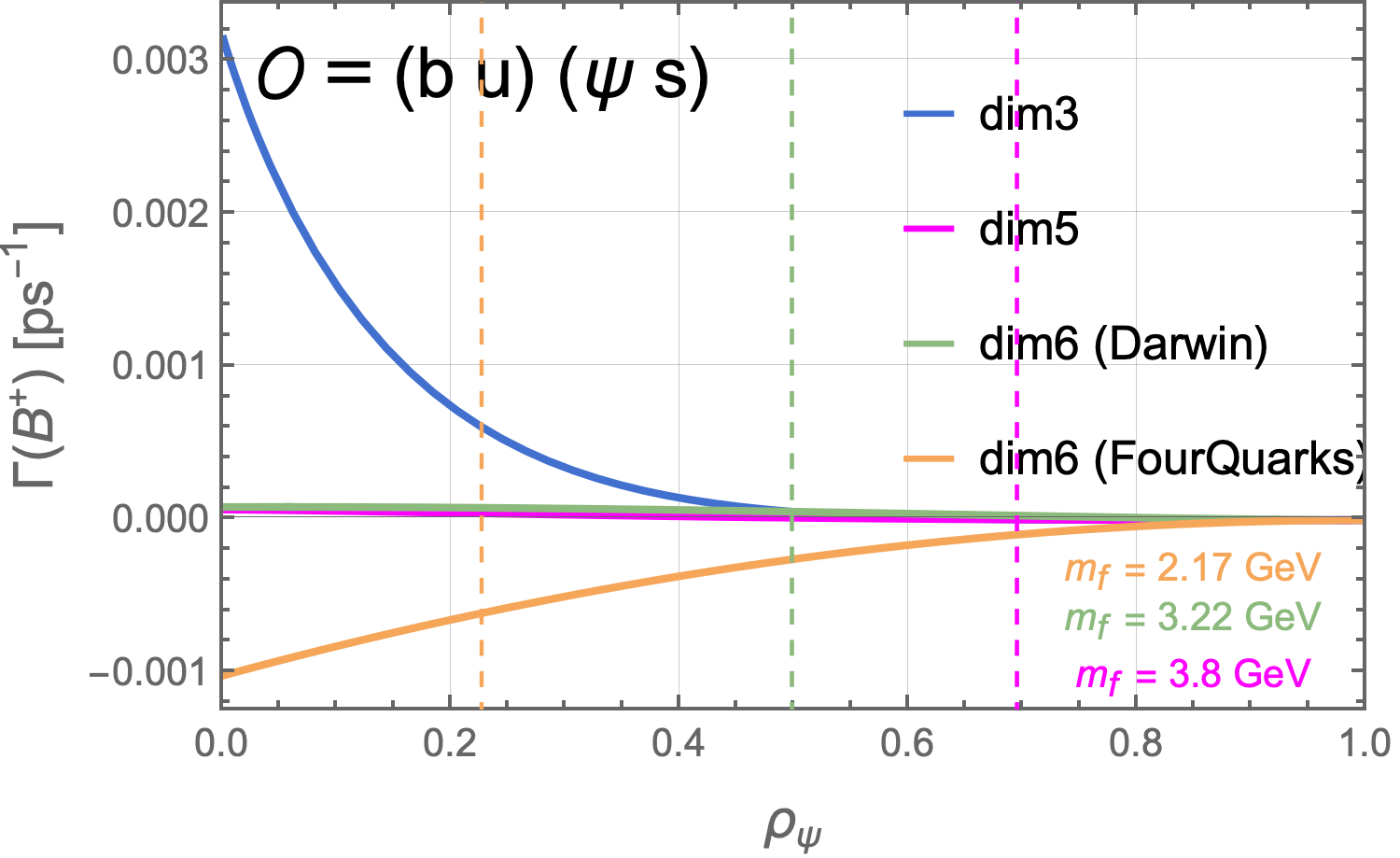}
    \end{subfigure}
    \hfill
    \begin{subfigure}{0.40\textwidth}
        \centering
        \includegraphics[width=\textwidth]{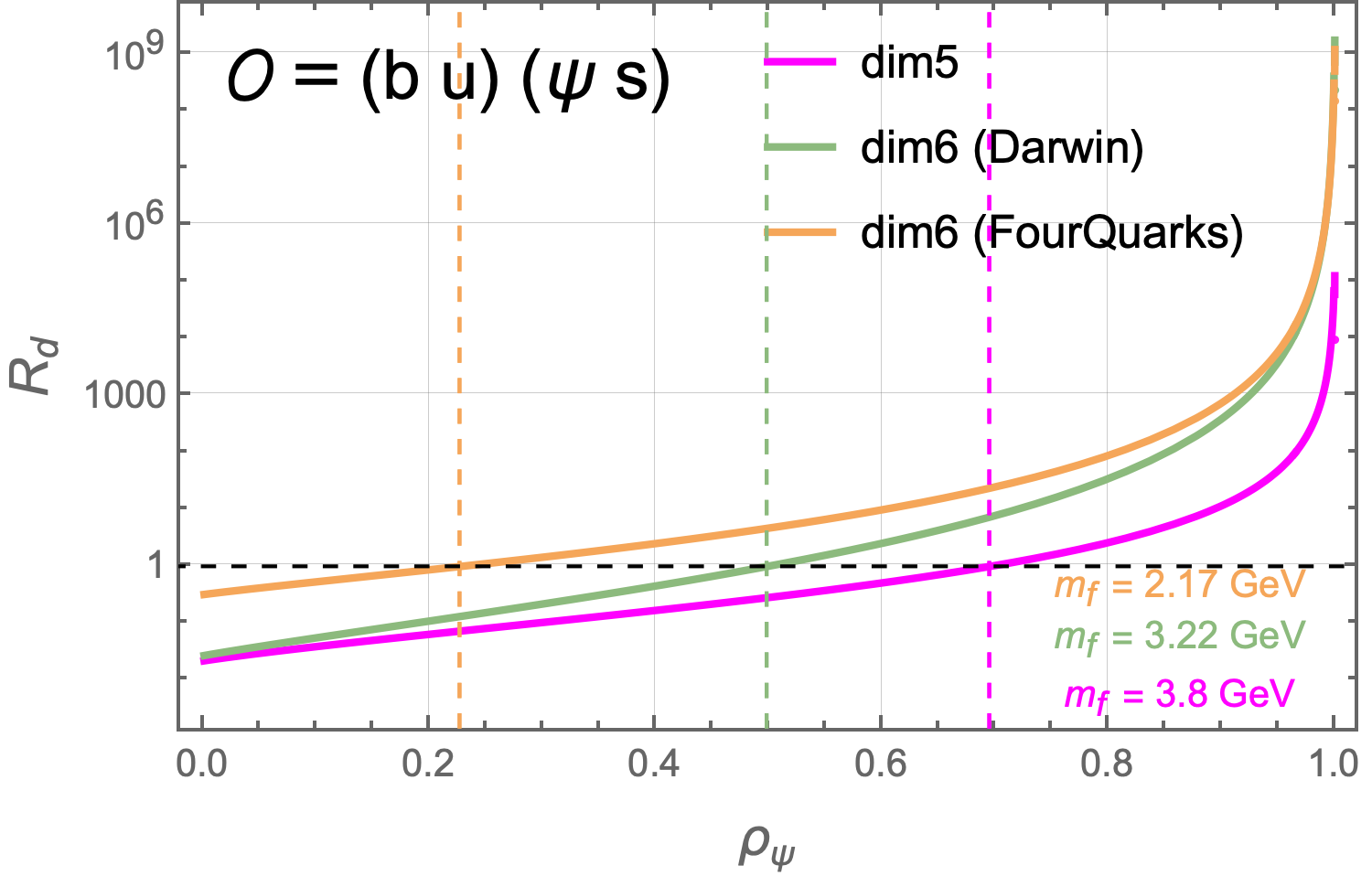}
    \end{subfigure}

    \medskip 
    
    \begin{subfigure}{0.43\textwidth}
        \centering
        \includegraphics[width=\textwidth]{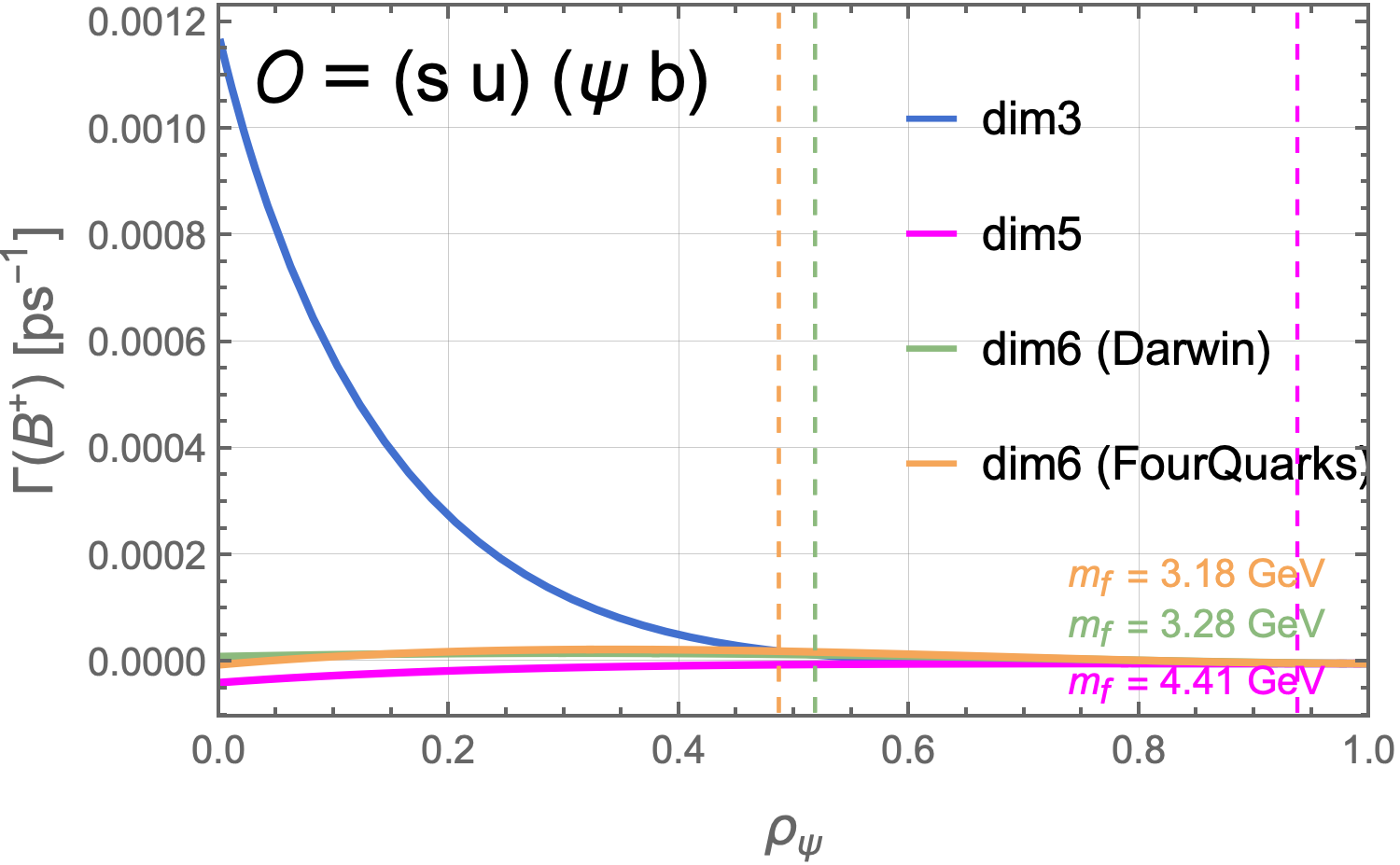}
    \end{subfigure}
    \hfill
    \begin{subfigure}{0.40\textwidth}
        \centering
        \includegraphics[width=\textwidth]{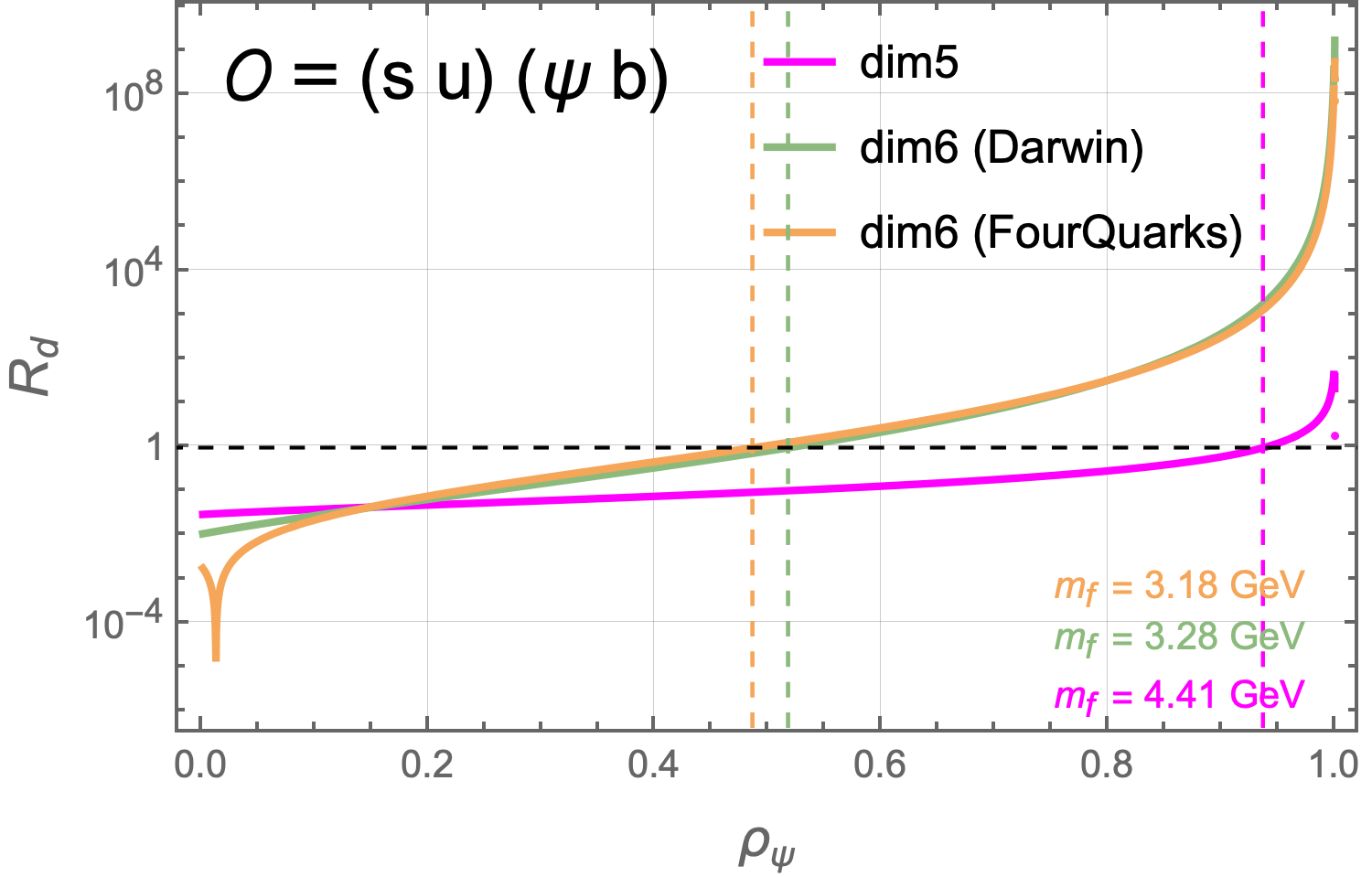}
    \end{subfigure}

    \caption{(Part 2): Additional plots for different operators with the same color coding as in the previous Figure.}
    \label{fig:DecayRates3}
\end{figure}

\begin{figure}[ht]
    \centering
    \begin{subfigure}{0.43\textwidth}
        \centering
        \includegraphics[width=\textwidth]{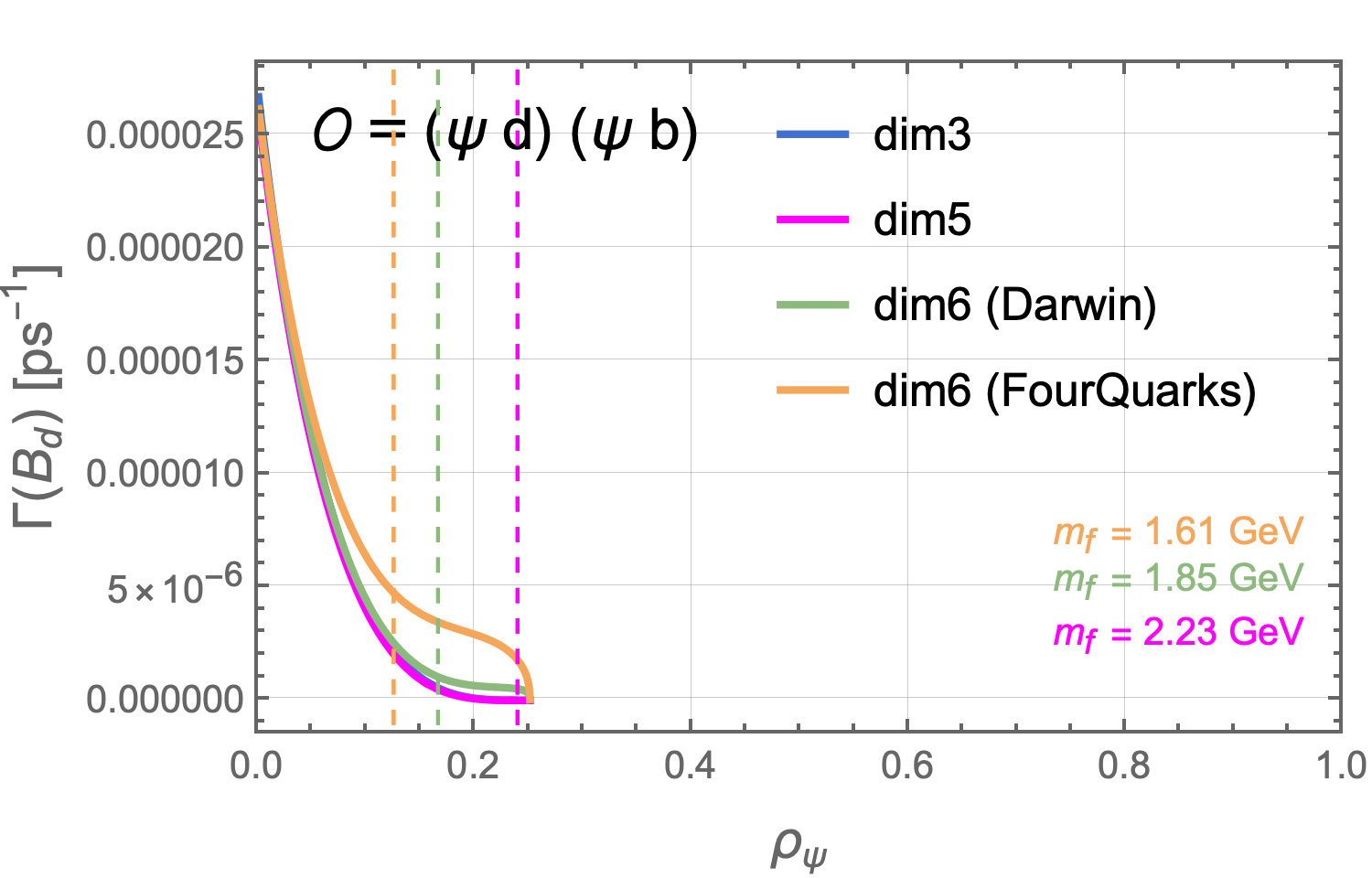}
    \end{subfigure}
    \hfill
    \begin{subfigure}{0.40\textwidth}
        \centering
        \includegraphics[width=\textwidth]{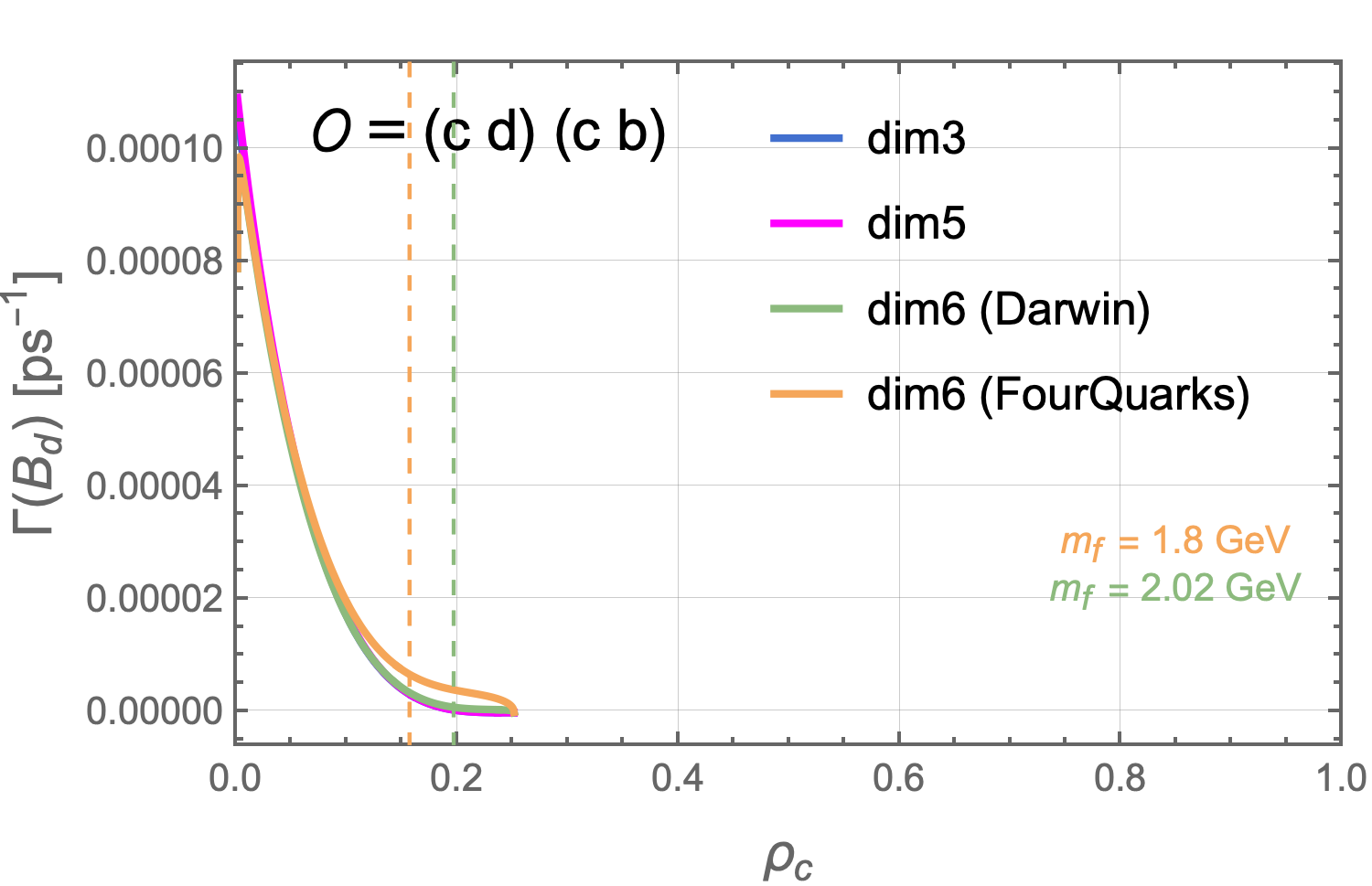}
    \end{subfigure}
     \hfill
    \begin{subfigure}{0.43\textwidth}
        \centering
        \includegraphics[width=\textwidth]{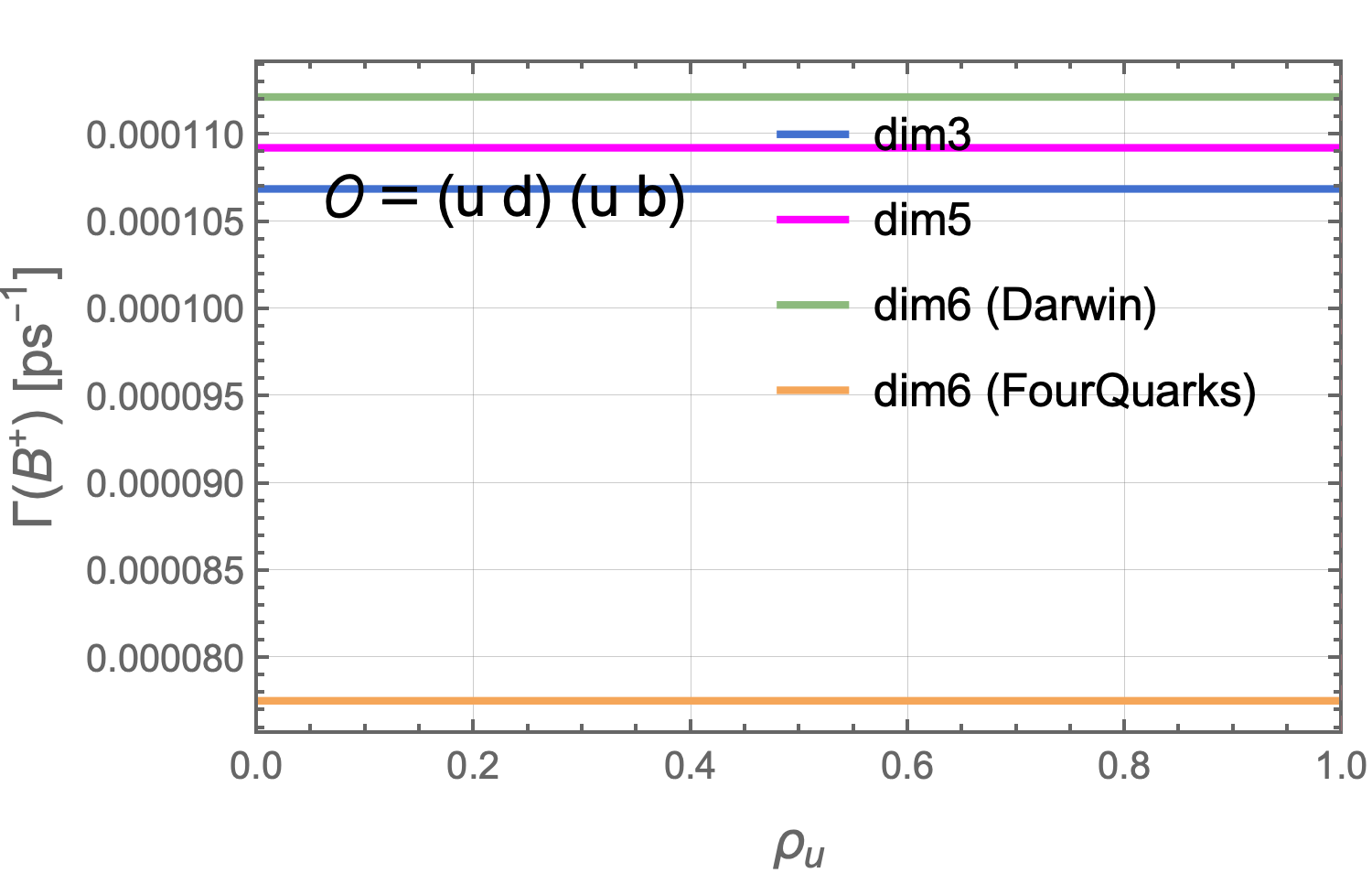}
    \end{subfigure}
     \hfill
    \begin{subfigure}{0.40\textwidth}
        \centering
        \includegraphics[width=\textwidth]{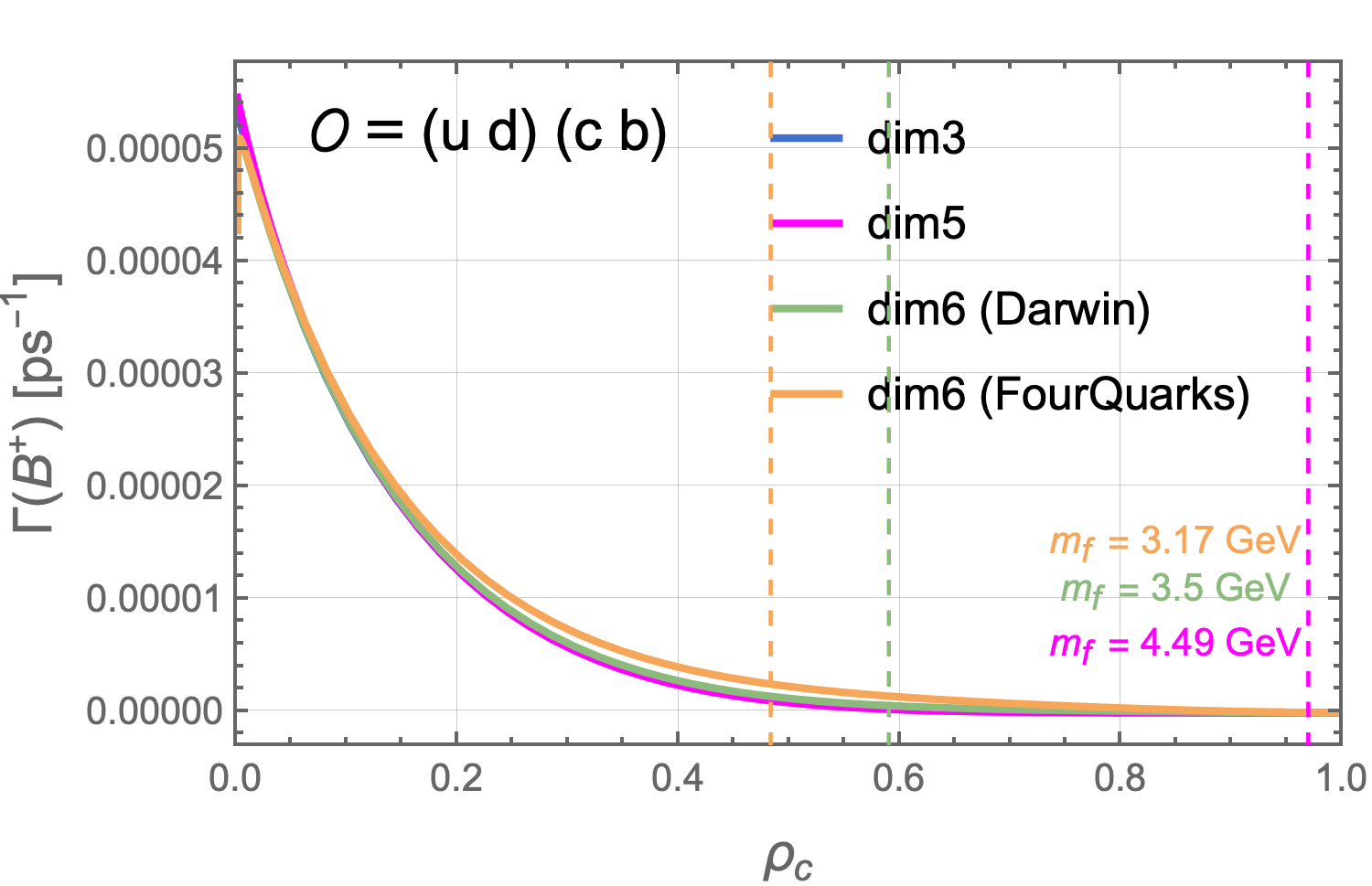}
    \end{subfigure}
     \hfill
    \begin{subfigure}{0.43\textwidth}
        \centering
        \includegraphics[width=\textwidth]{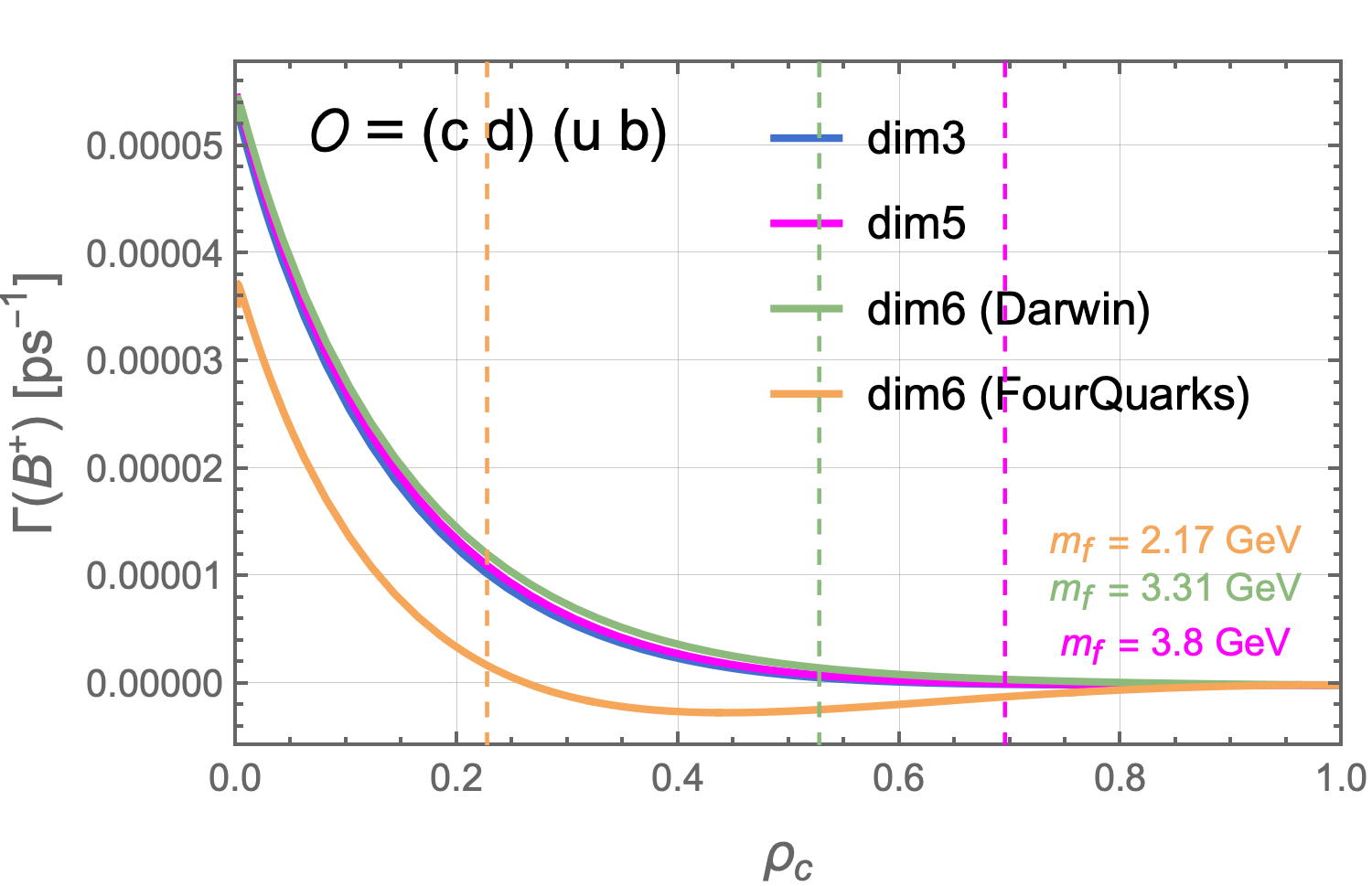}
    \end{subfigure}
     \hfill
    \begin{subfigure}{0.40\textwidth}
        \centering
        \includegraphics[width=\textwidth]{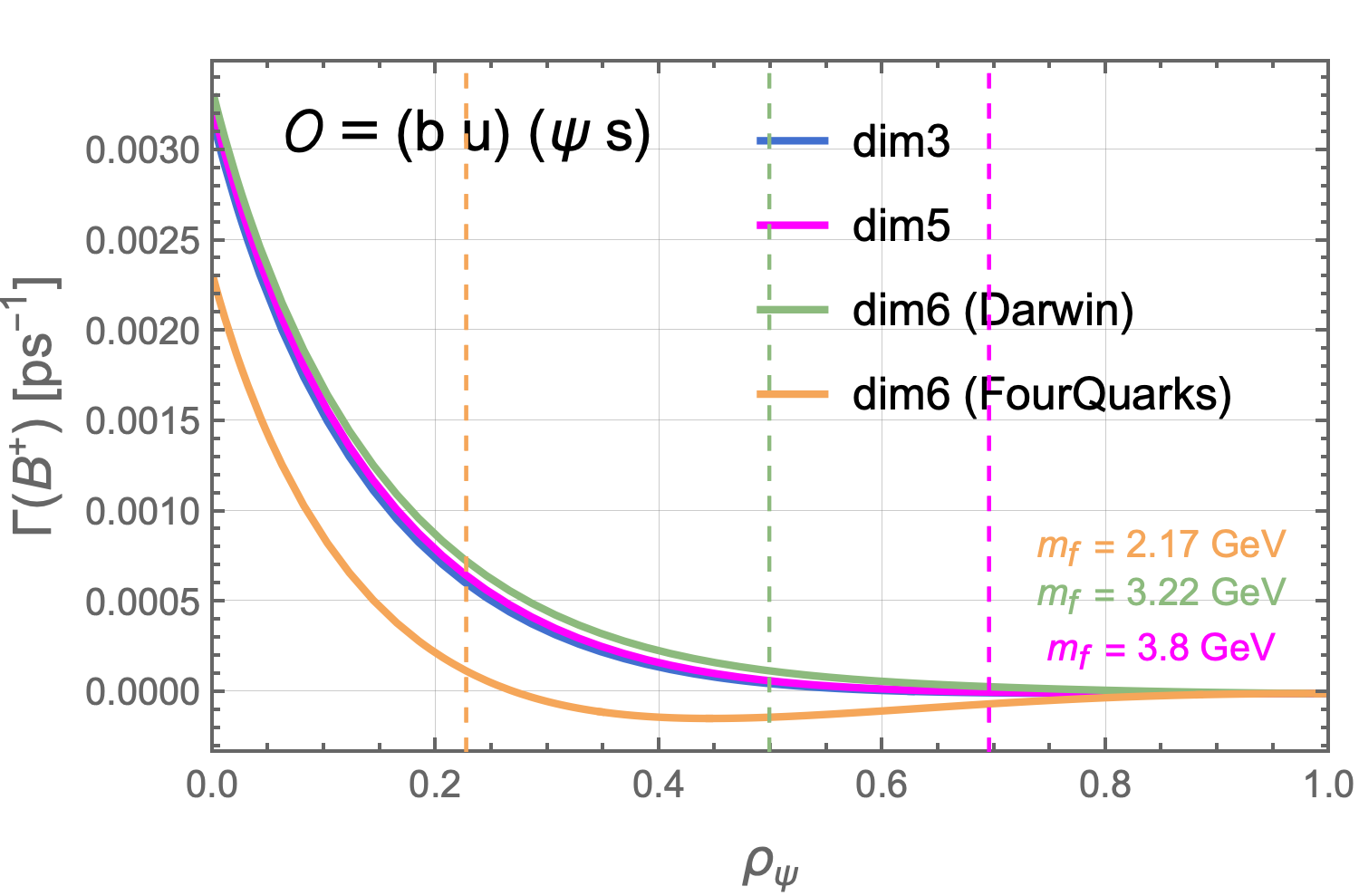}
    \end{subfigure}
     \hfill
    \begin{subfigure}{0.43\textwidth}
        \centering
        \includegraphics[width=\textwidth]{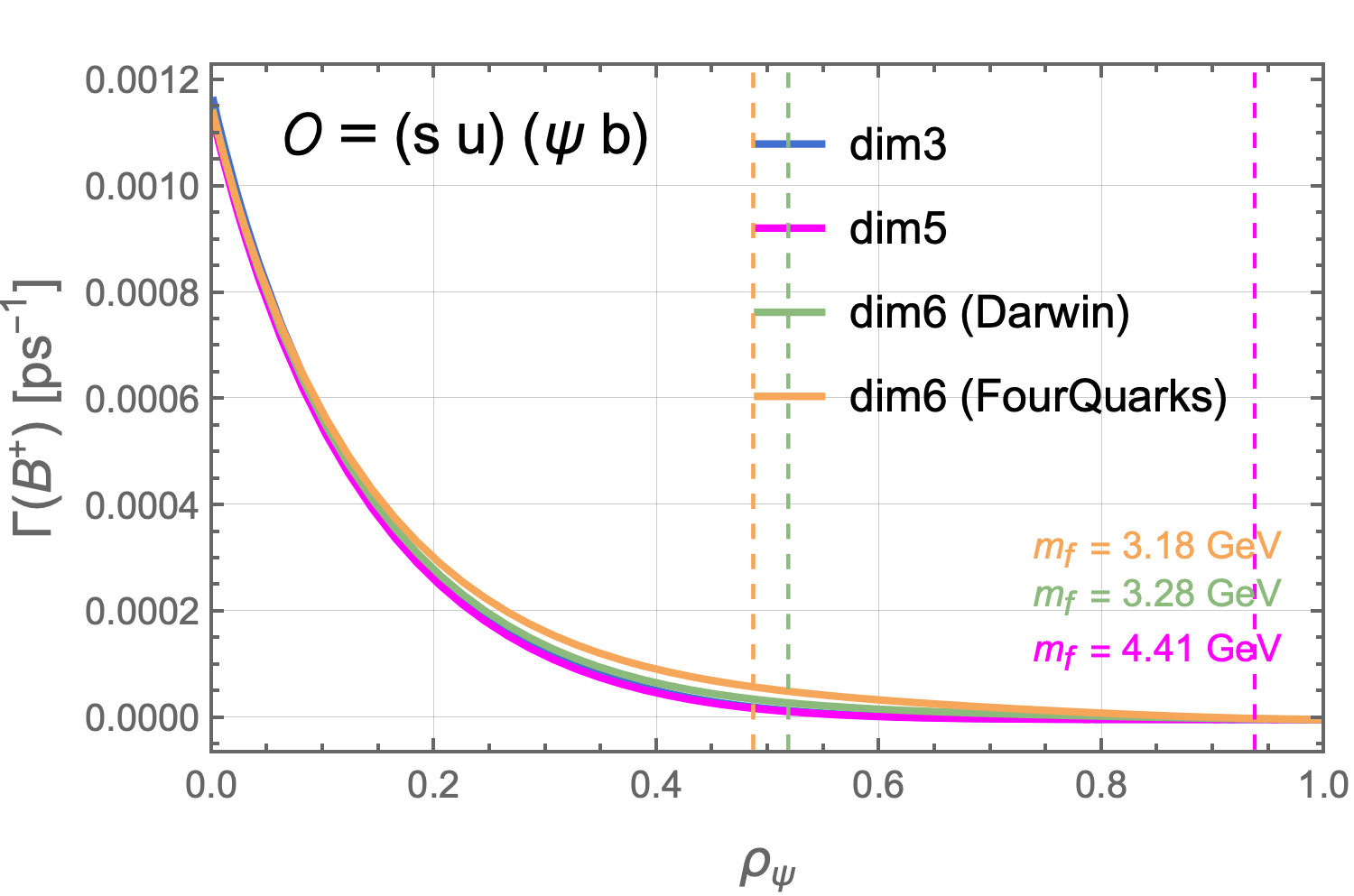}
    \end{subfigure}
    \hfill
    \begin{subfigure}{0.40\textwidth}
        \centering
        \includegraphics[width=\textwidth]{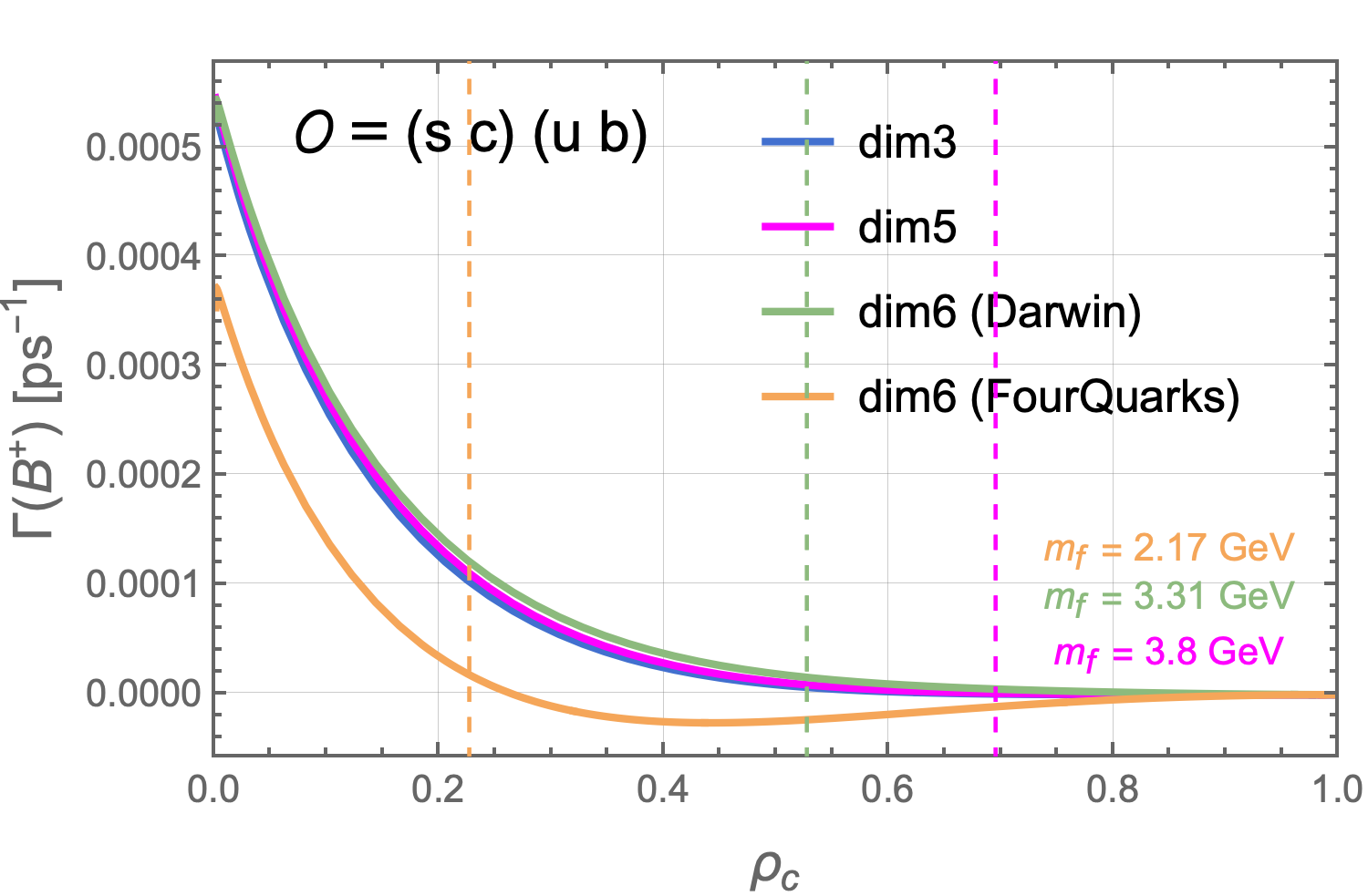}
    \end{subfigure}
        \caption{(Part 1): Total decay width of $\Gamma(B^+)$ arising purely from the Mesogenesis (BSM) operators as a function of $\rho_f= (m_f/m_b)^2$. The figure displays the cumulative sum of all terms up to the indicated dimension. The vertical lines indicate the value of $\rho_f$, where the contribution exceeds the leading dimension-three $\Gamma_3$ term (the corresponding $m_f$ values are also indicated).}\label{fig:DecayRatesTotal2}
\end{figure}
\begin{figure}[ht]
    \centering
    \hfill
    \begin{subfigure}{0.40\textwidth}
        \centering
        \includegraphics[width=\textwidth]{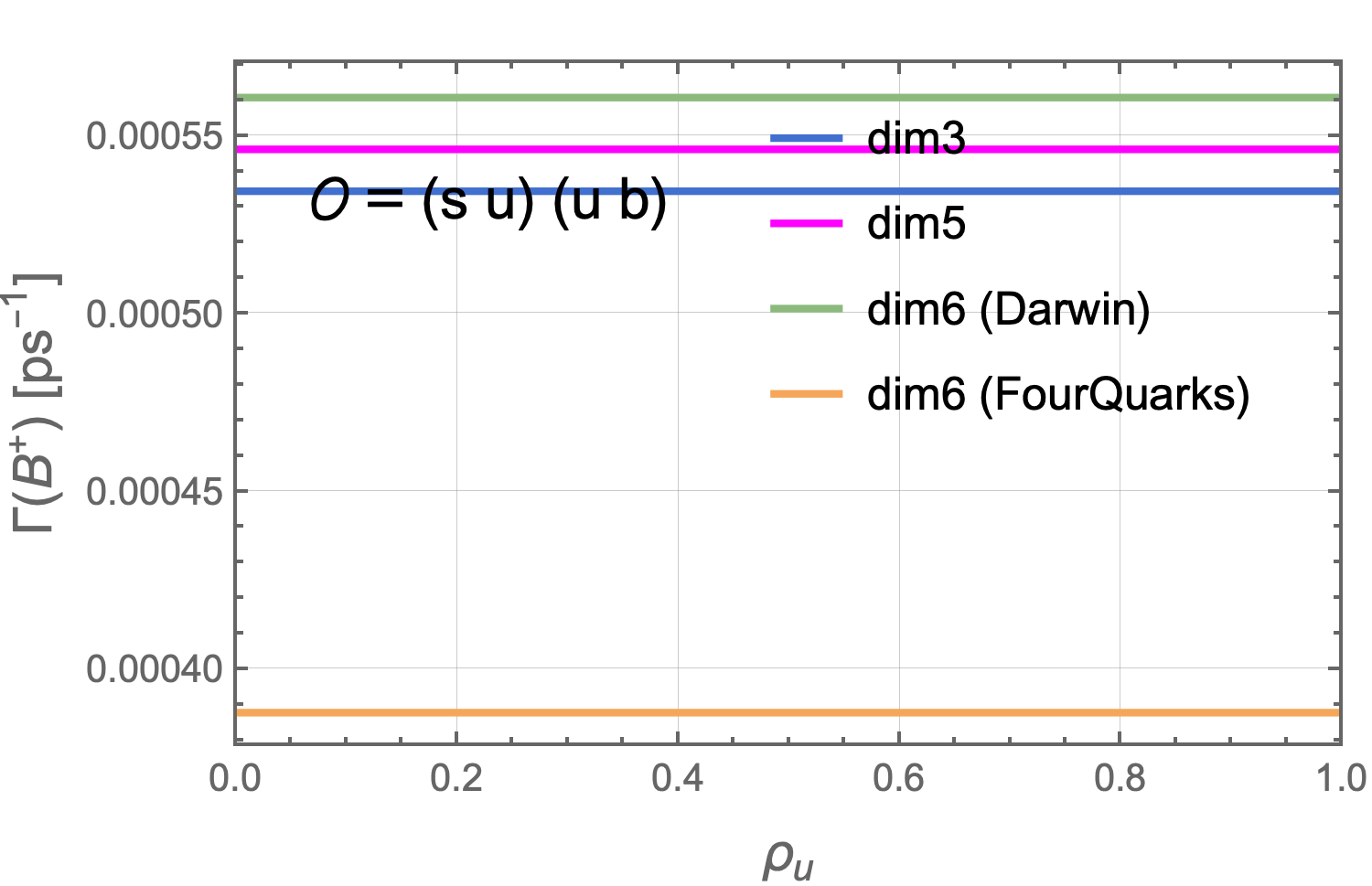}
    \end{subfigure}
     \hfill
    \begin{subfigure}{0.40\textwidth}
        \centering
        \includegraphics[width=\textwidth]{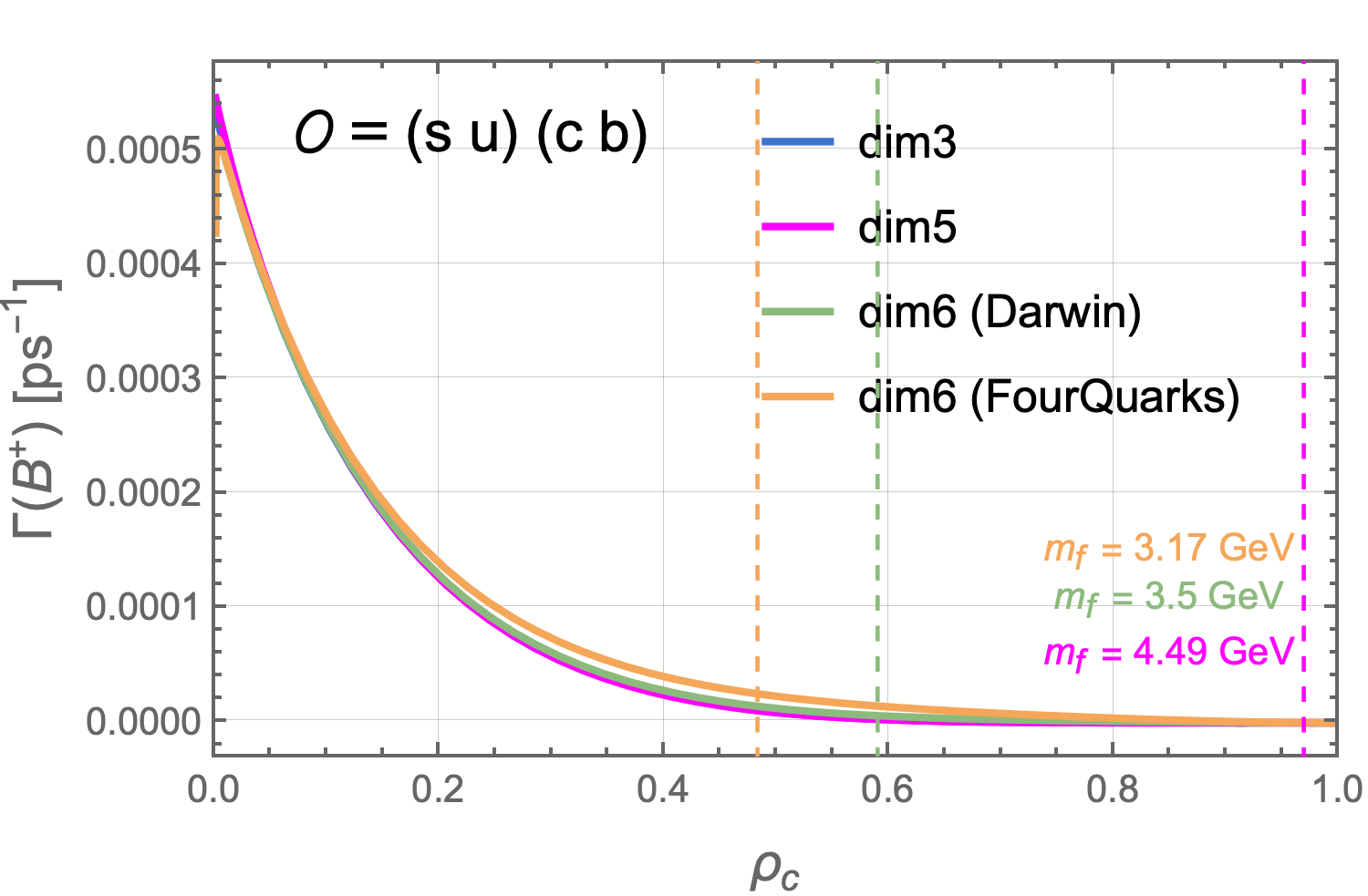}
    \end{subfigure}

  \medskip 
    
    \begin{subfigure}{0.43\textwidth}
        \includegraphics[width=\textwidth]{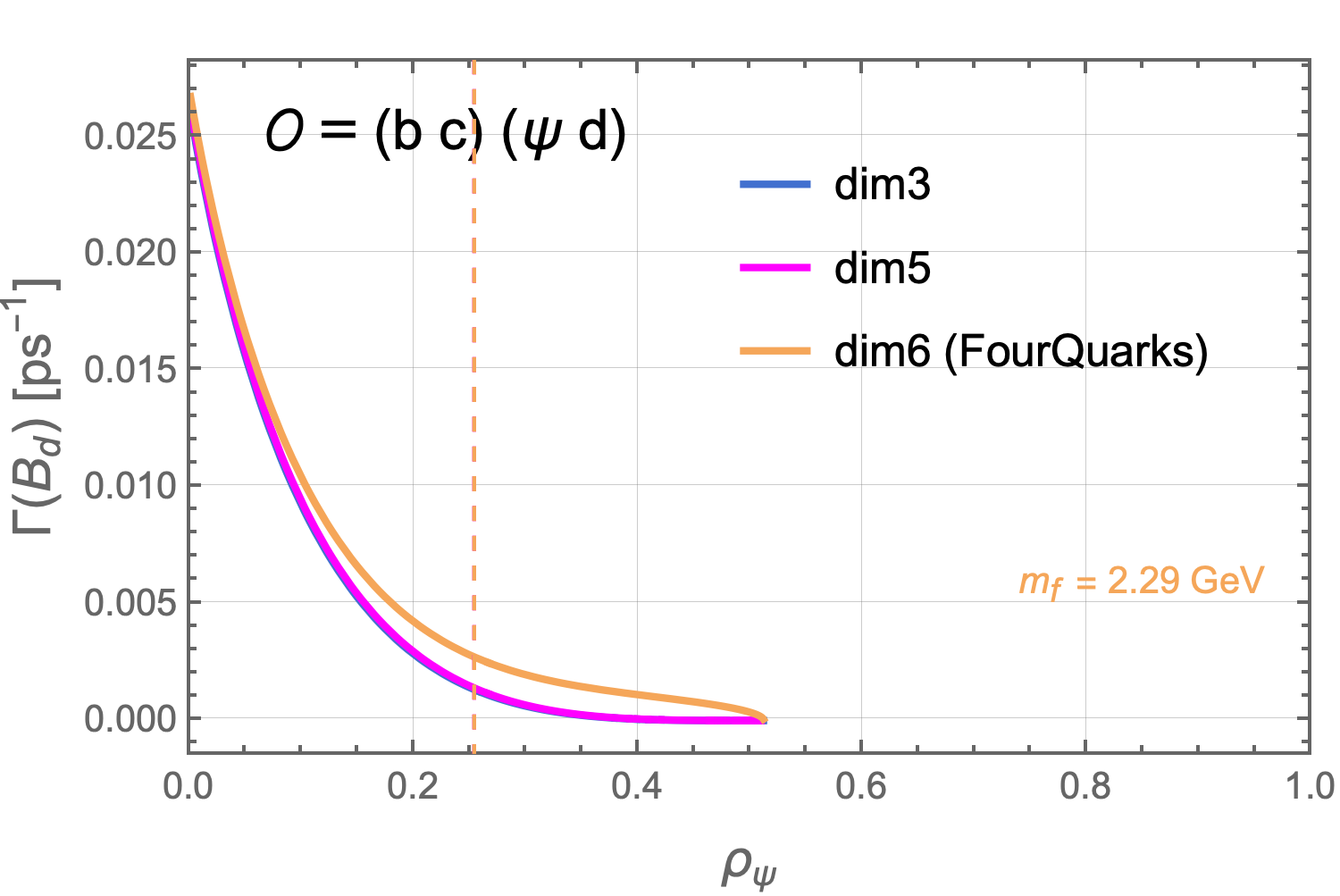}
    \end{subfigure}
        \caption{(Part 2): Total decay width of $\Gamma(B^+)$ arising purely from the Mesogenesis (BSM) operators as a function of $\rho_f= (m_f/m_b)^2$, with the same color coding as in the previous figure. }\label{fig:DecayRatesTotal2}
\end{figure}

\begin{figure}[ht]
    \centering
    \begin{subfigure}{0.40\textwidth}
        \centering
        \includegraphics[width=\textwidth]{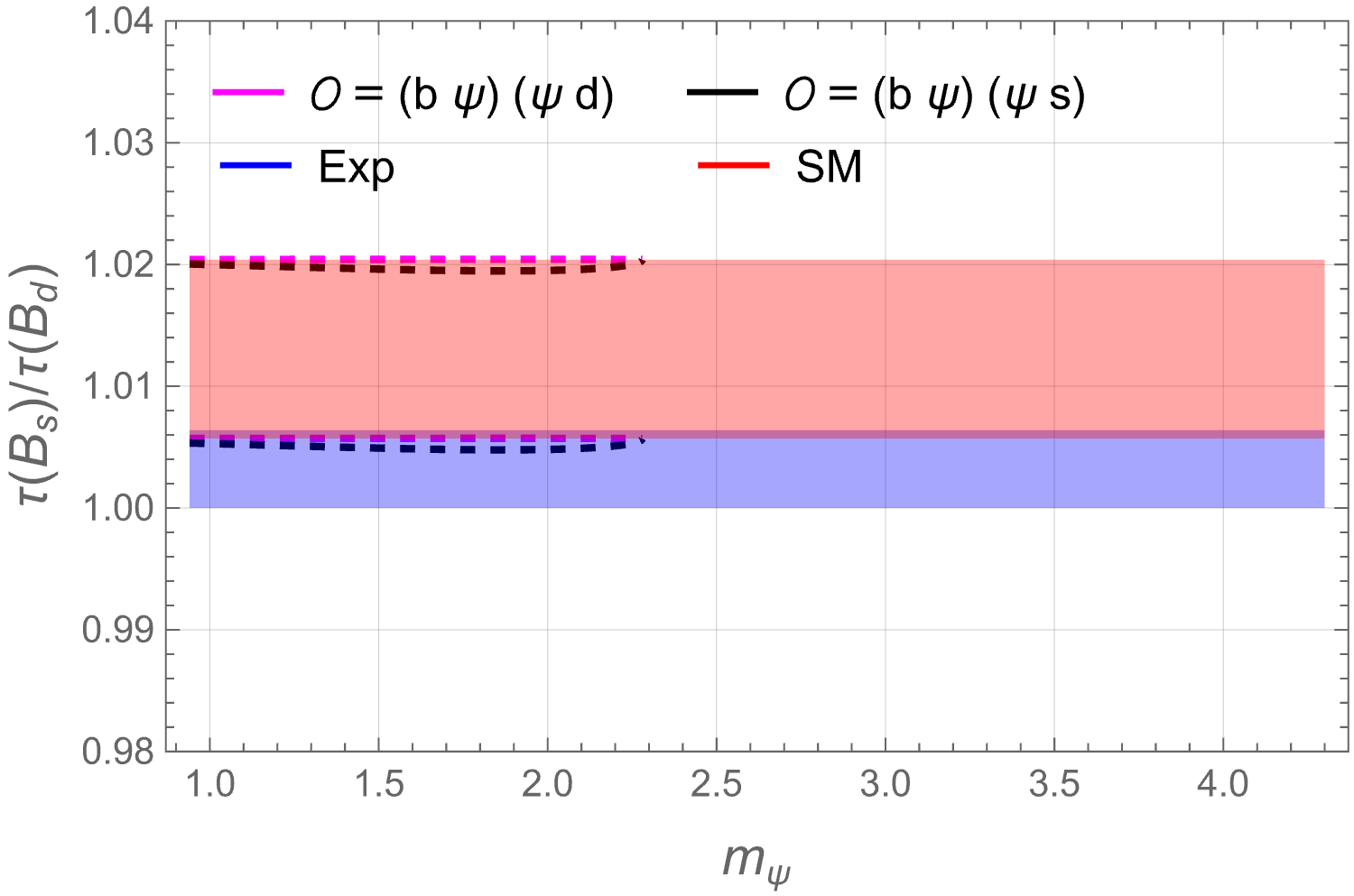}
    \end{subfigure}
    \hfill
    \begin{subfigure}{0.40\textwidth}
        \centering
        \includegraphics[width=\textwidth]{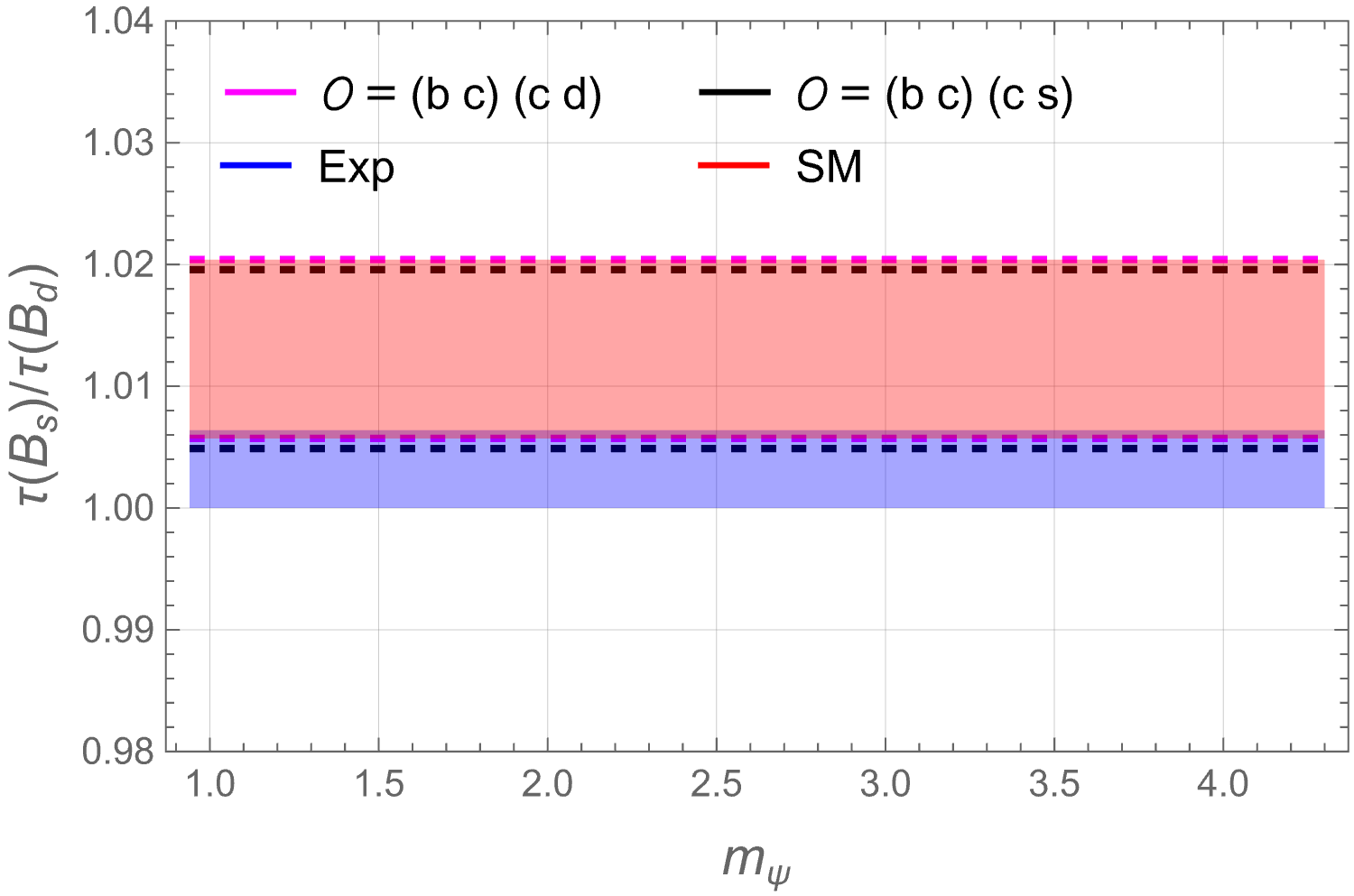}
    \end{subfigure}
     \hfill
    \begin{subfigure}{0.40\textwidth}
        \centering
        \includegraphics[width=\textwidth]{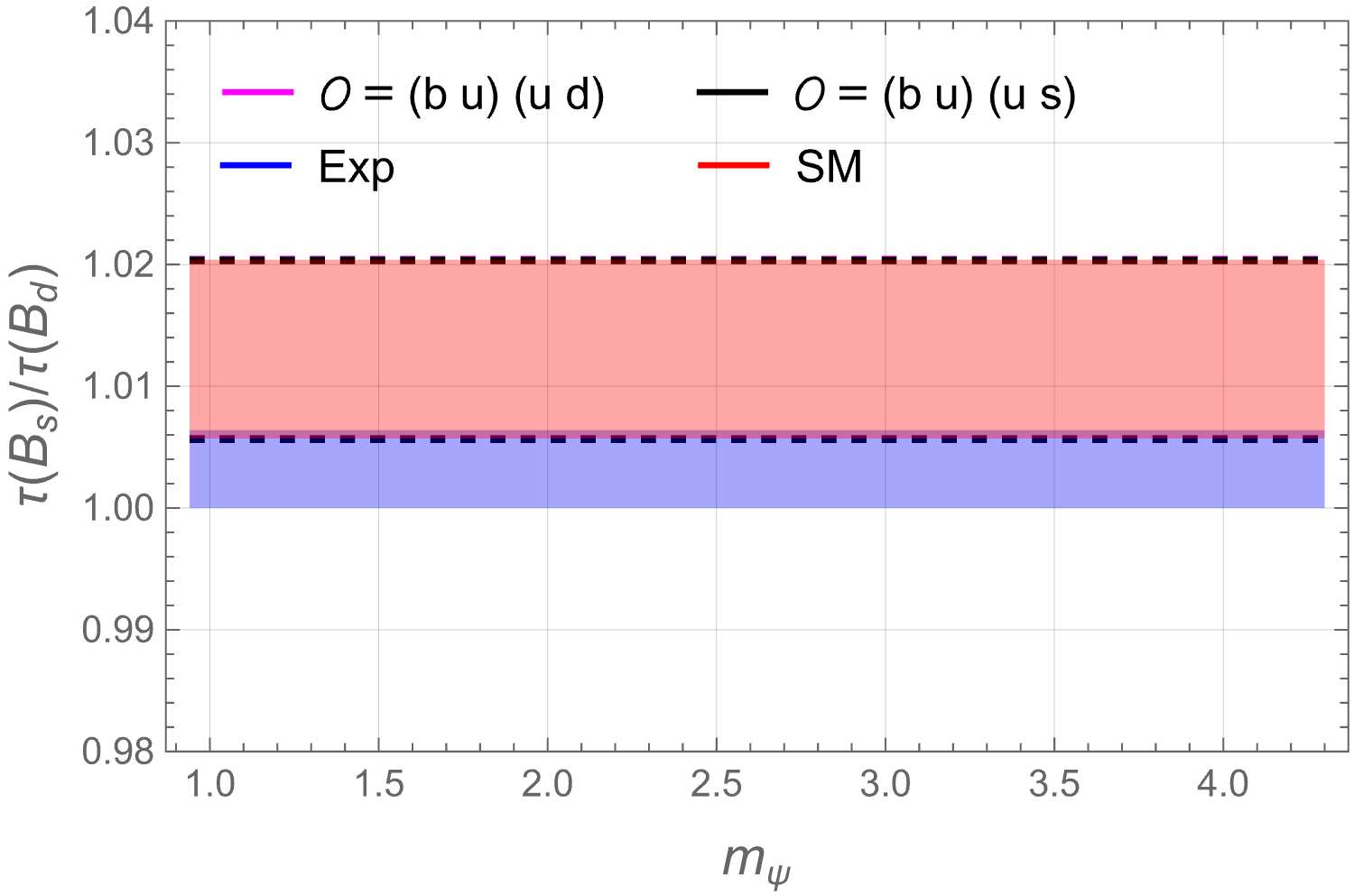}
    \end{subfigure}
     \hfill
    \begin{subfigure}{0.40\textwidth}
        \centering
        \includegraphics[width=\textwidth]{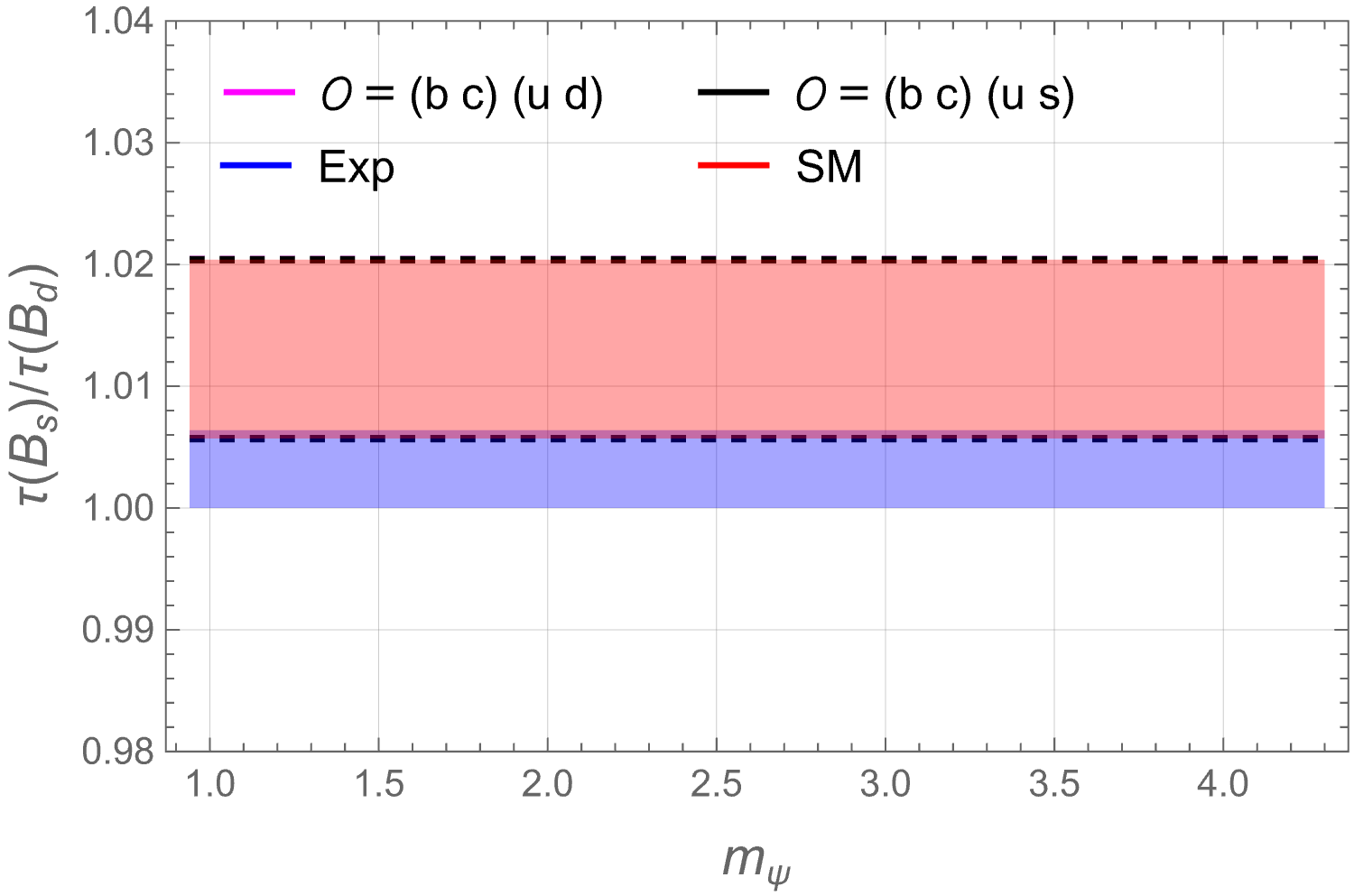}
    \end{subfigure}
     \hfill
    \begin{subfigure}{0.40\textwidth}
        \centering
        \includegraphics[width=\textwidth]{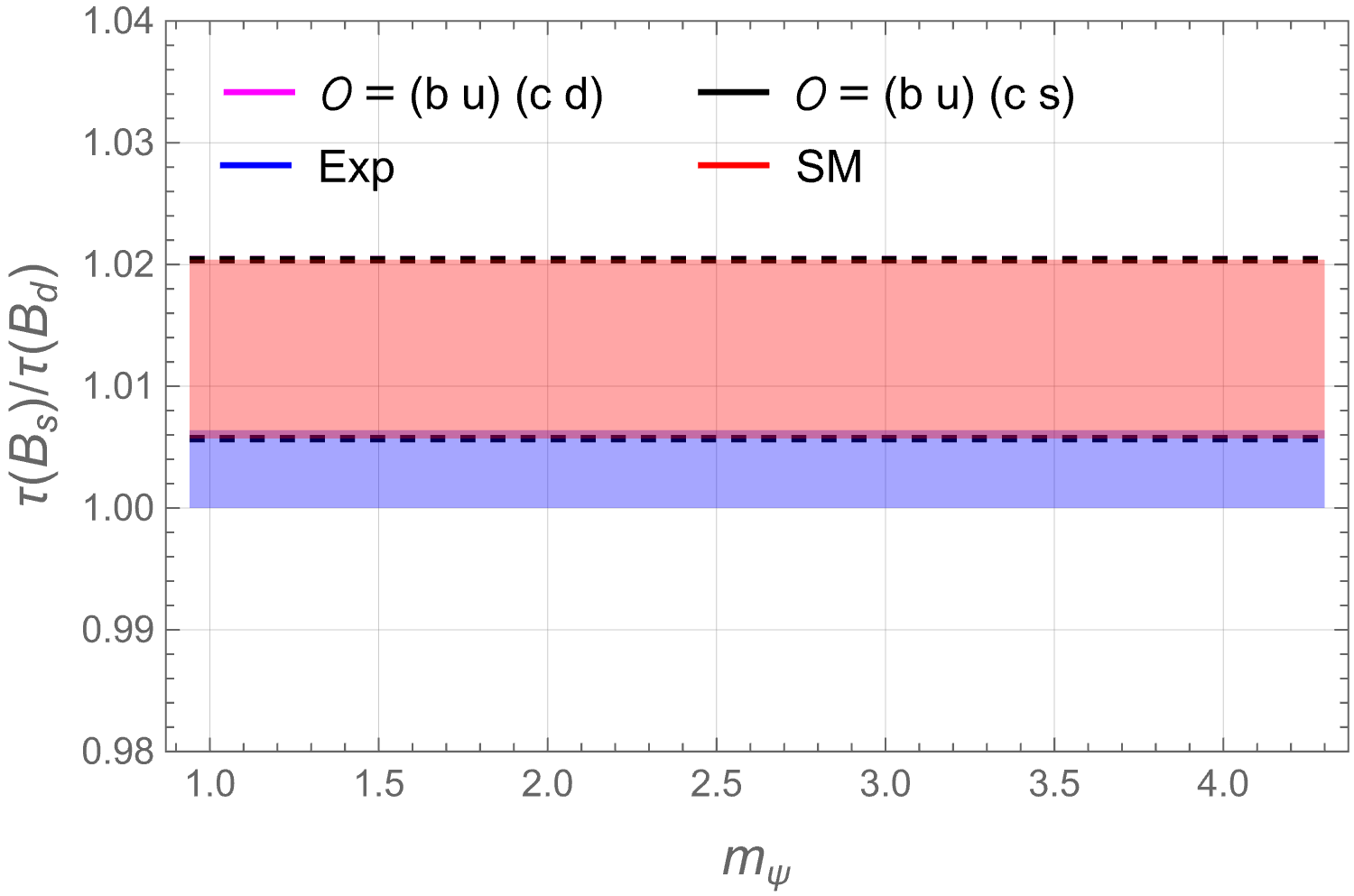}
    \end{subfigure}
       \caption{Lifetime ratio $\tau(B_s)/\tau(B_d)$ as a function of the dark fermion mass $m_\psi$, 
including both SM and SM+NP predictions for representative operators.  
The NP contribution is found to be negligible compared to present theoretical uncertainties. (Operators involving two distinct massive final states are not shown, as the corresponding Darwin-term contribution has not yet been calculated.)}
        \label{fig:BsBdAll}
\end{figure}

\begin{figure}[ht]
    \centering
    \begin{subfigure}{0.40\textwidth}
        \centering
        \includegraphics[width=\textwidth]{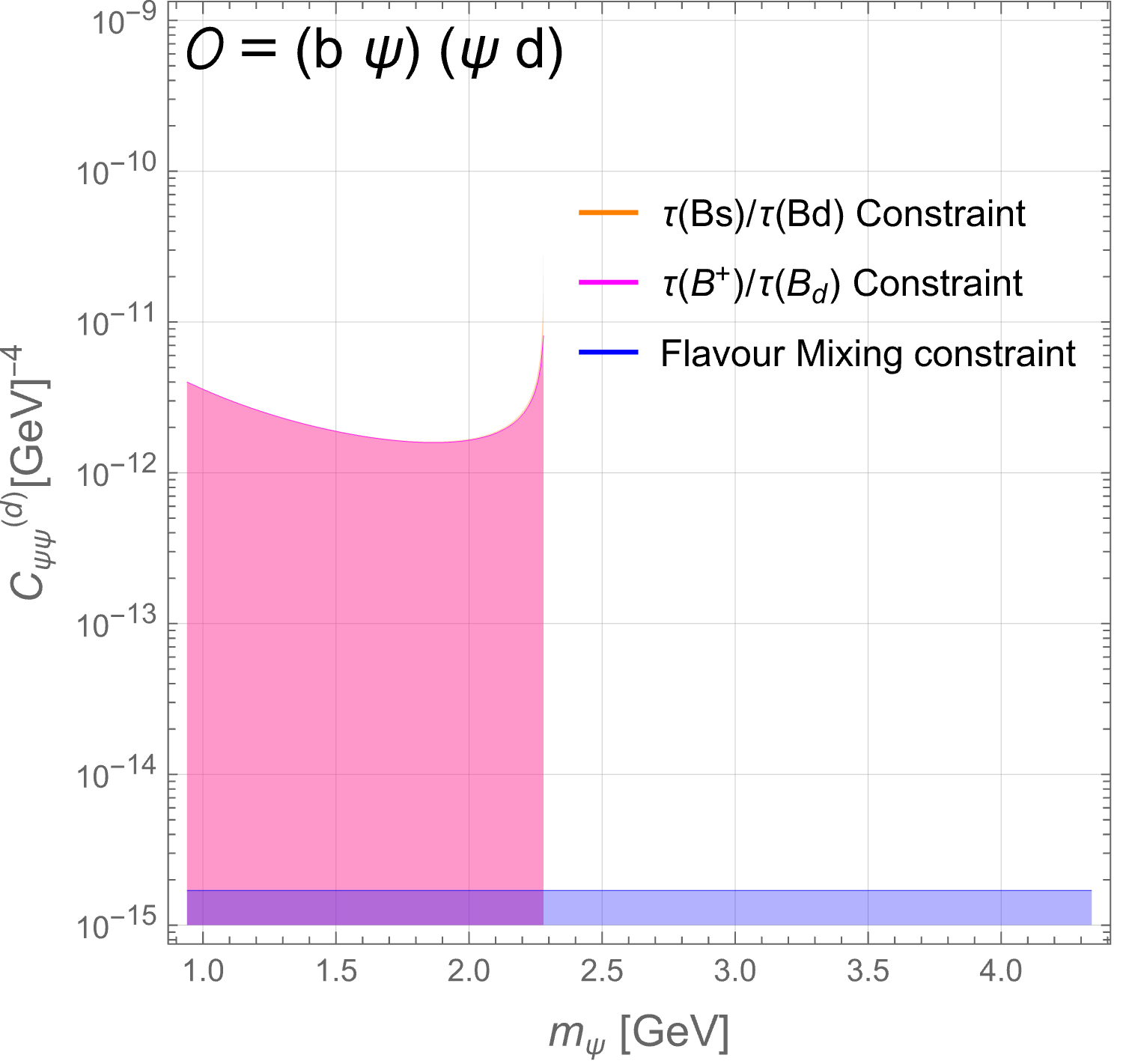}
    \end{subfigure}
    \hfill
    \begin{subfigure}{0.40\textwidth}
        \centering
        \includegraphics[width=\textwidth]{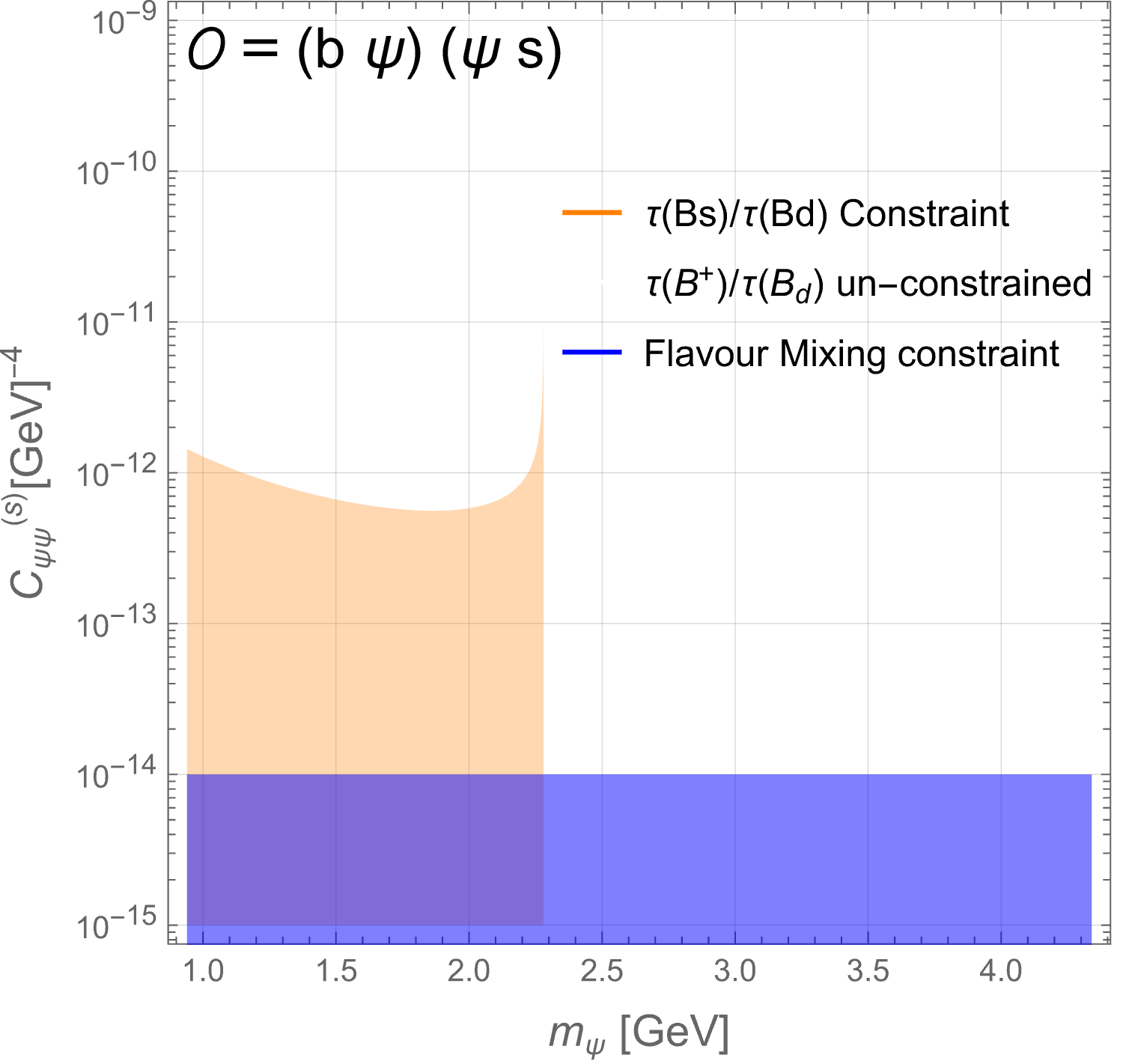}
    \end{subfigure}
    \begin{subfigure}{0.40\textwidth}
        \centering
        \includegraphics[width=\textwidth]{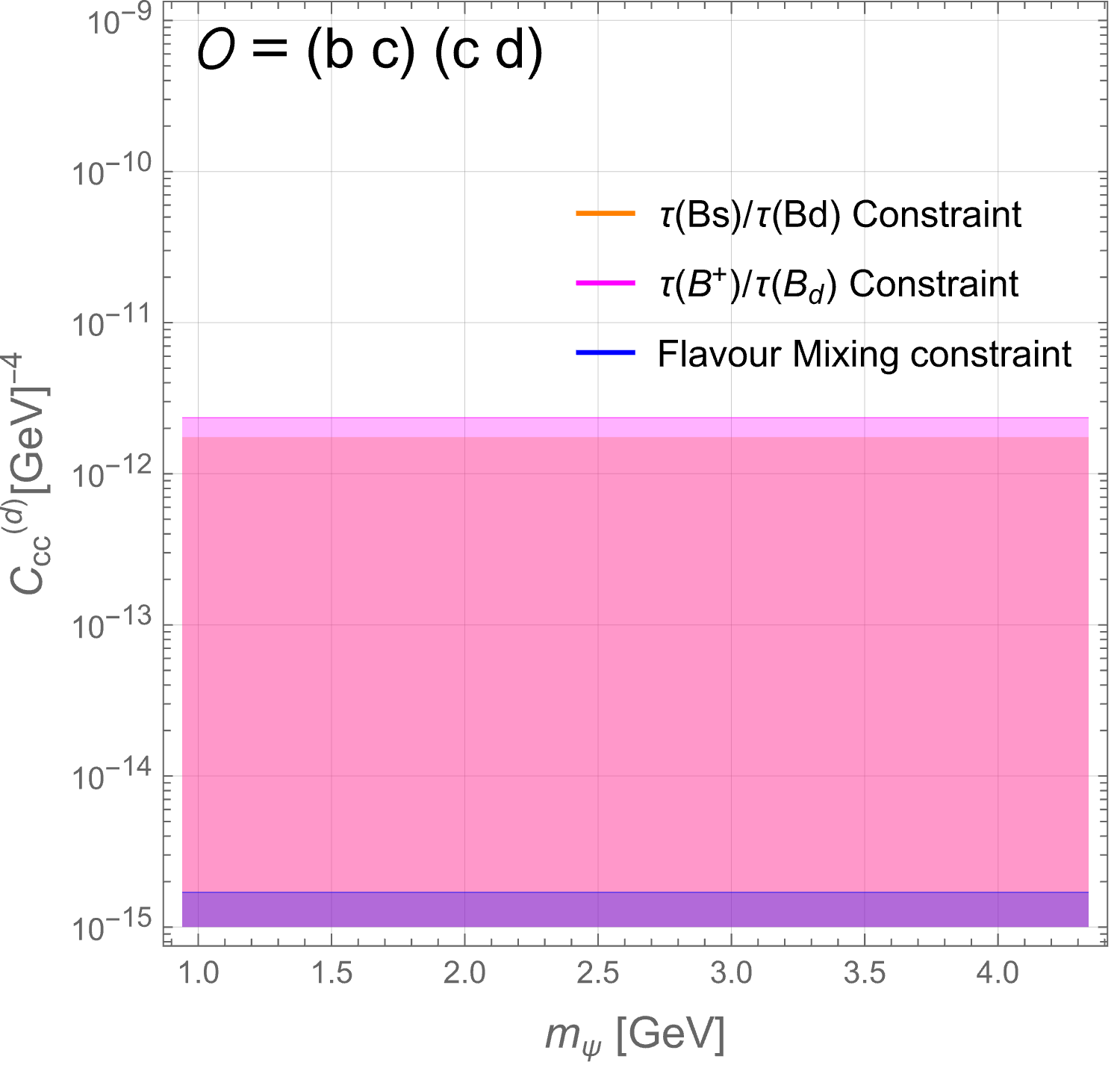}
    \end{subfigure}
      \hfill
    \begin{subfigure}{0.40\textwidth}
        \centering
        \includegraphics[width=\textwidth]{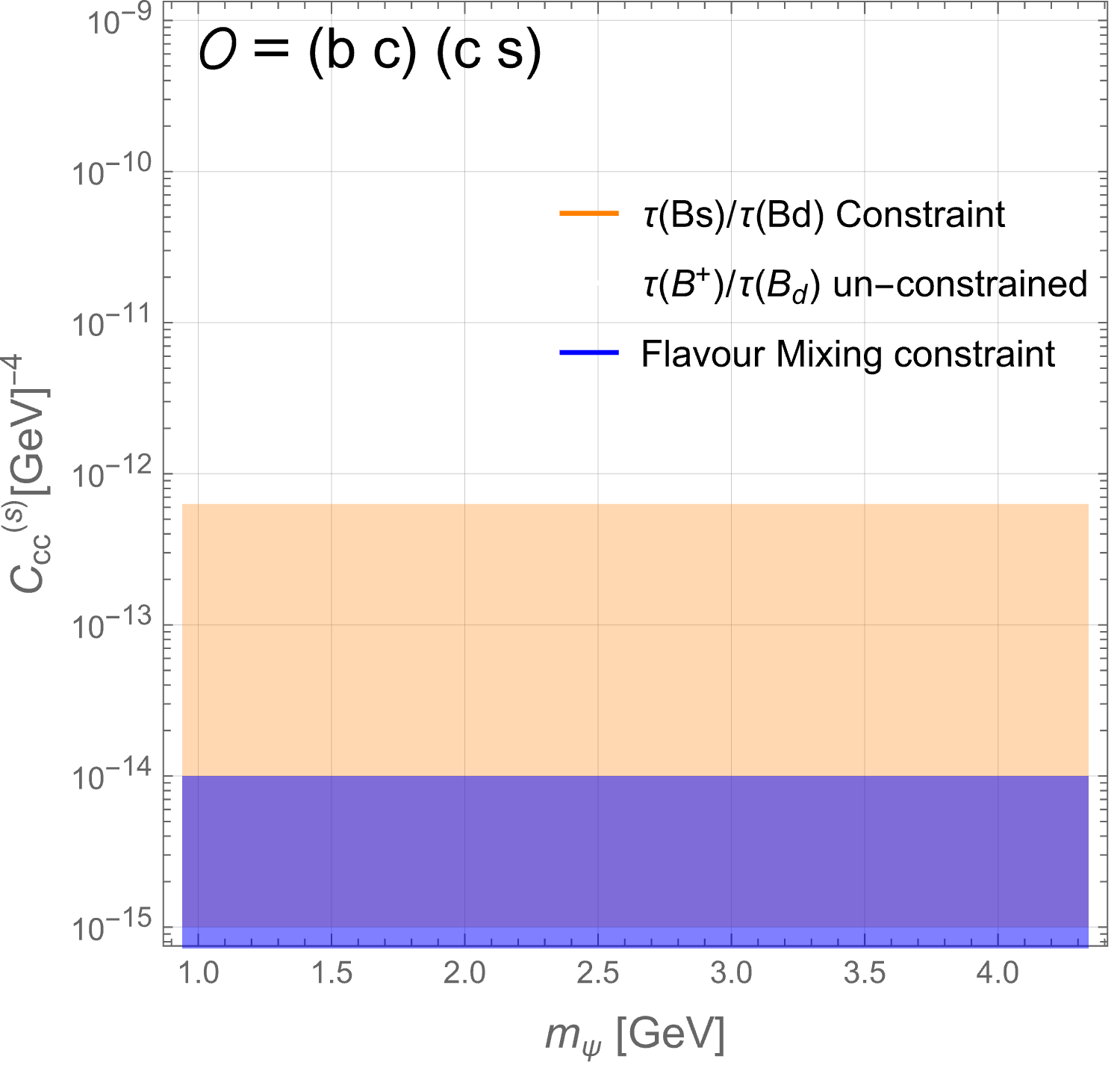}
    \end{subfigure}

     \begin{subfigure}{0.40\textwidth}
        \centering
        \includegraphics[width=\textwidth]{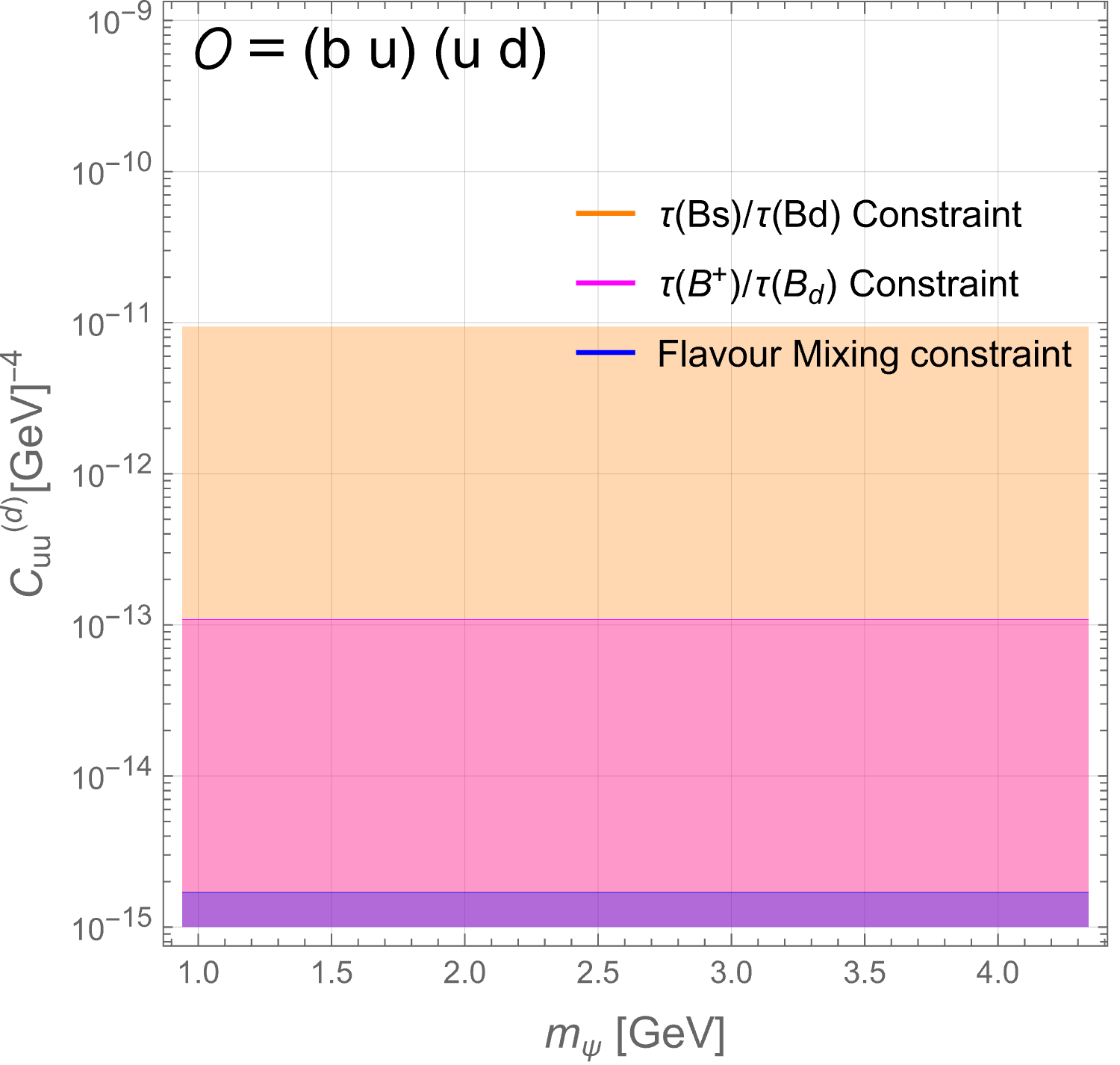}
    \end{subfigure}
      \hfill
    \begin{subfigure}{0.40\textwidth}
        \centering
        \includegraphics[width=\textwidth]{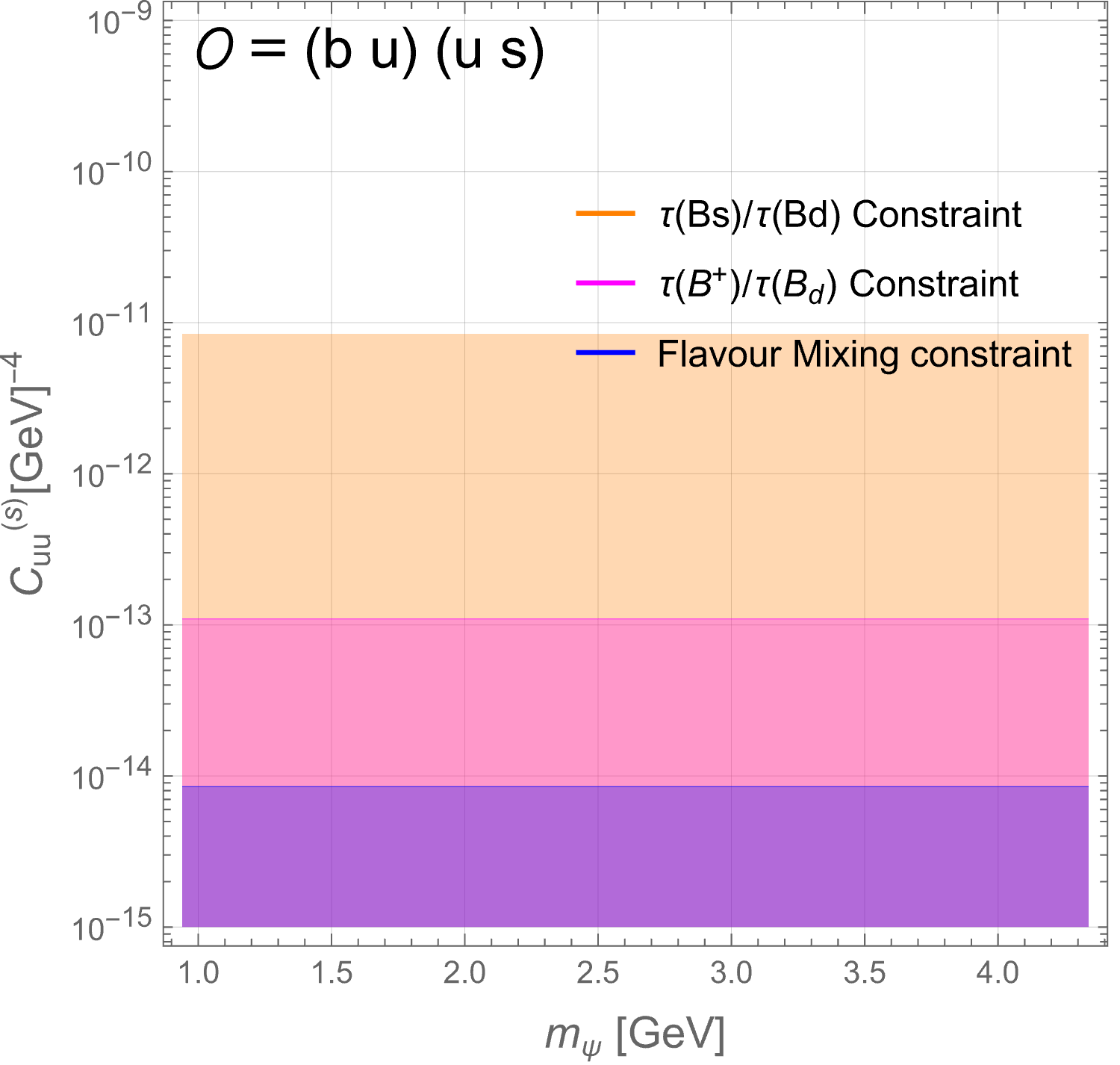}
    \end{subfigure}
       \caption{(Part 1): Constraints on the Wilson coefficients as a function of the dark fermion mass $m_\psi$, derived from the lifetime ratios $\tau(B_s)/\tau(B_d)$ (orange) and $\tau(B^+)/\tau(B_d)$ (magenta), compared to collider constraints (blue).}
        \label{fig:BsBdConstraints2}
\end{figure}

\begin{figure}[ht]
    \centering
    \begin{subfigure}{0.40\textwidth}
        \centering
        \includegraphics[width=\textwidth]{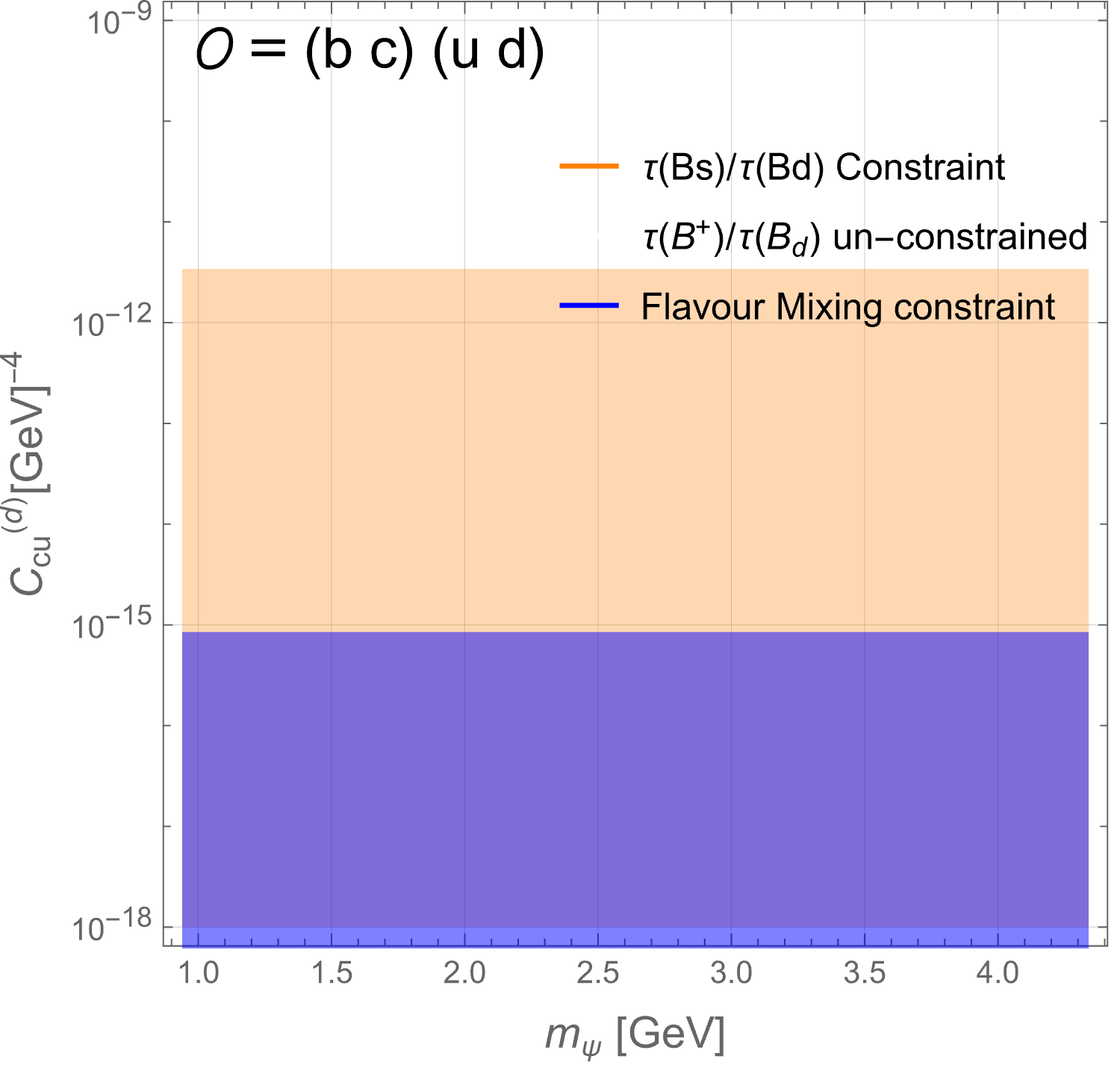}
    \end{subfigure}
    \hfill
    \begin{subfigure}{0.40\textwidth}
        \centering
        \includegraphics[width=\textwidth]{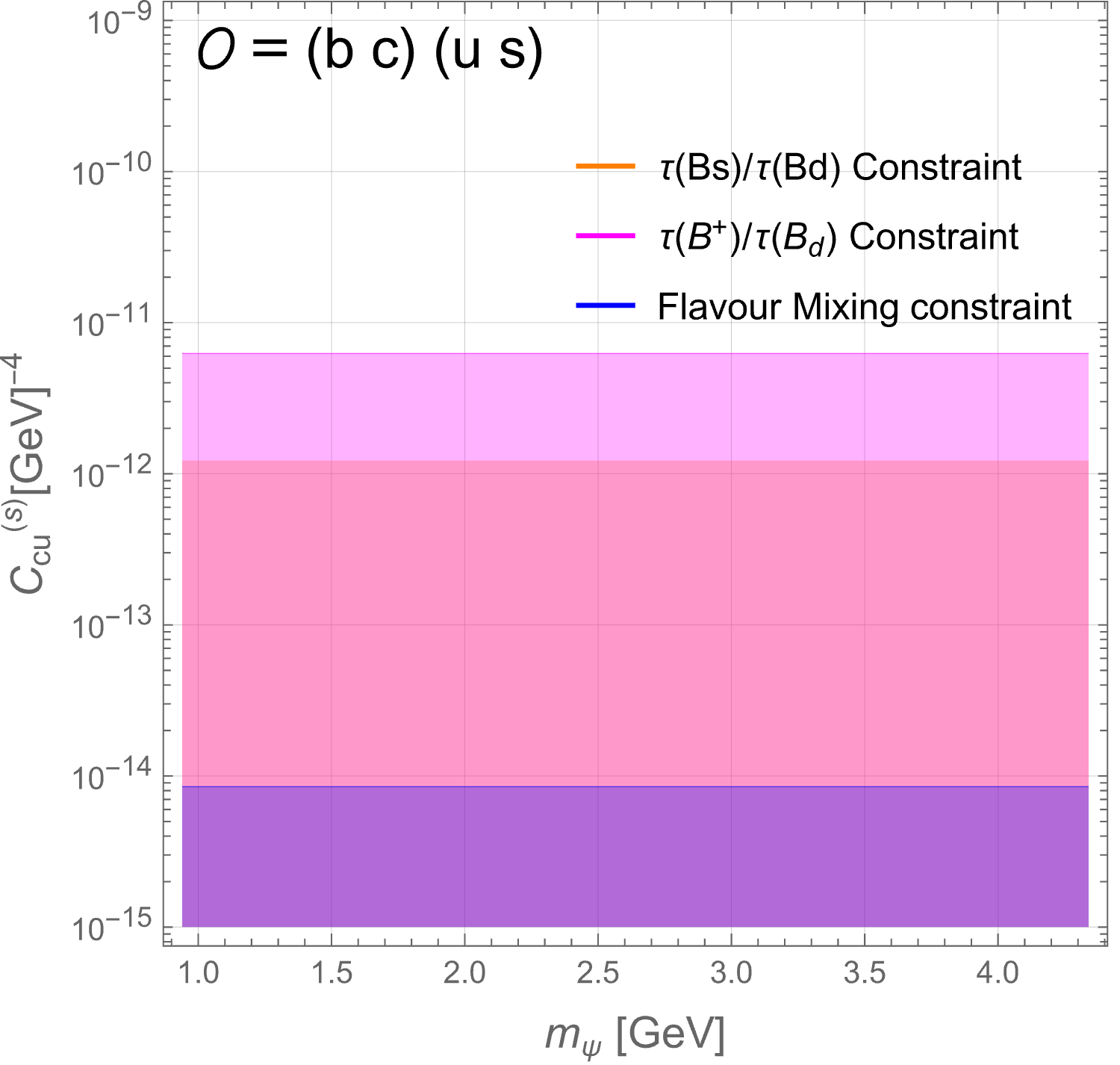}
    \end{subfigure}
     \begin{subfigure}{0.40\textwidth}
        \centering
        \includegraphics[width=\textwidth]{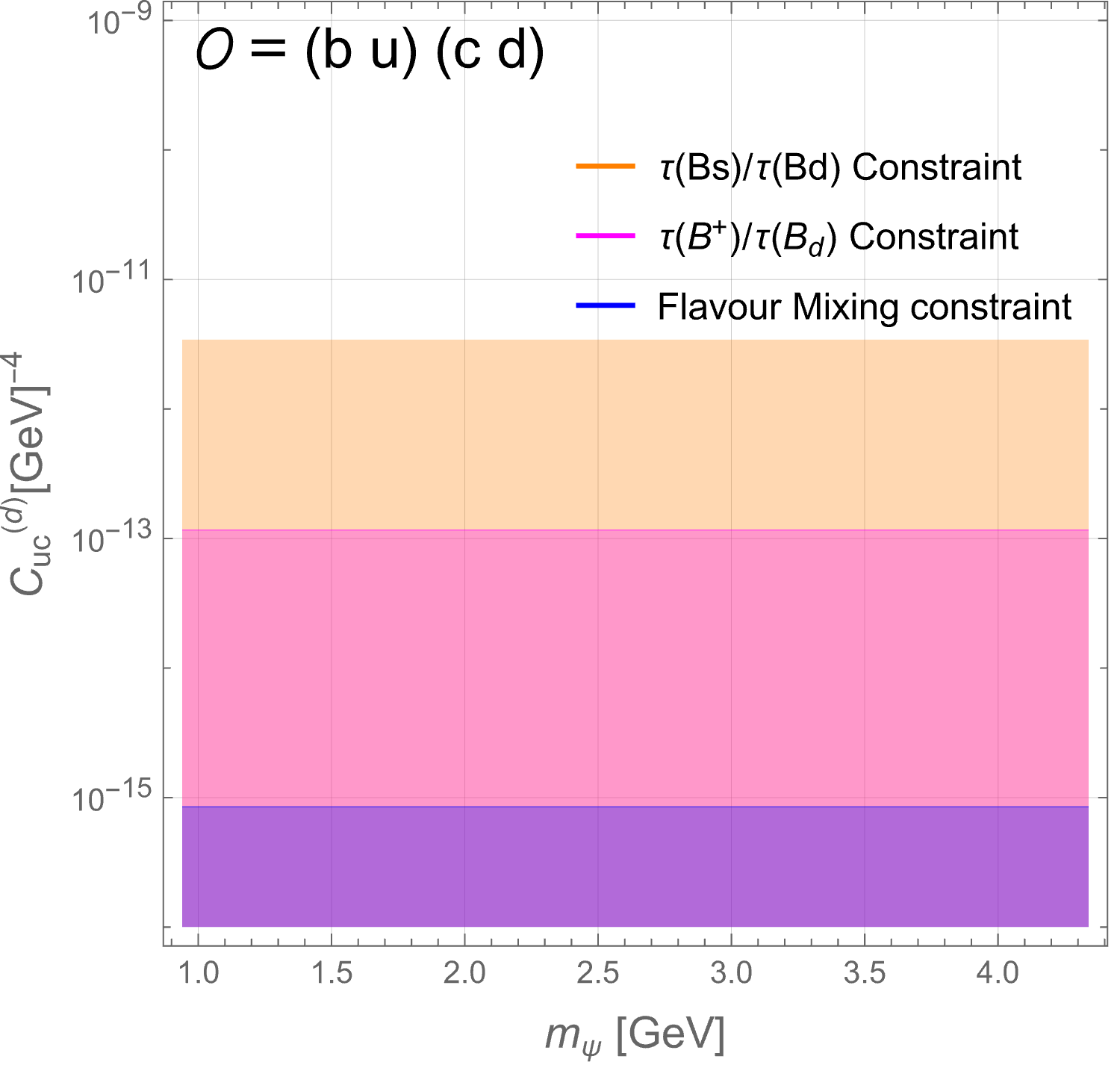}
    \end{subfigure}
    \hfill
    \begin{subfigure}{0.40\textwidth}
        \centering
        \includegraphics[width=\textwidth]{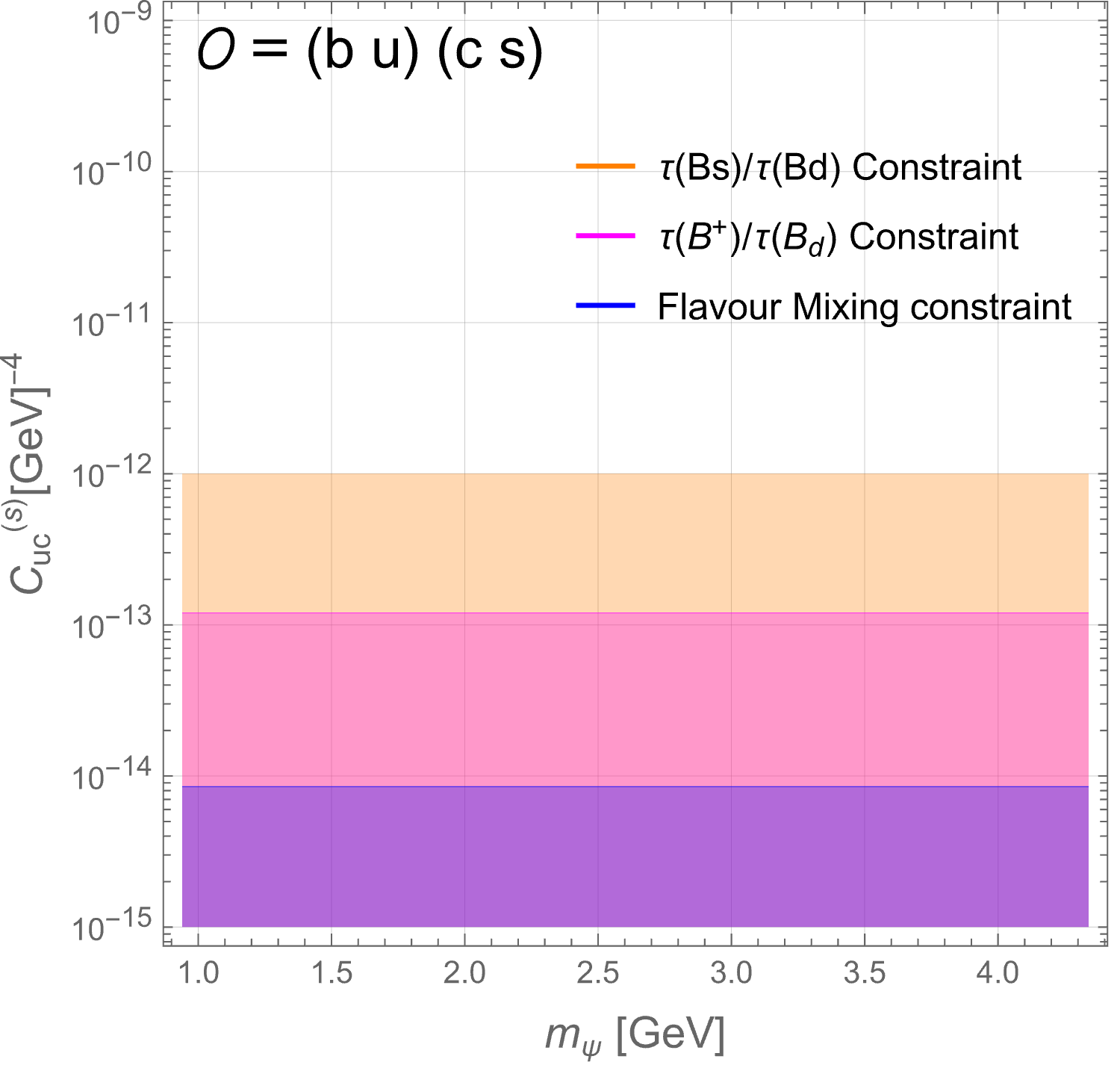}
    \end{subfigure}
       \caption{(Part 2): Constraints on the Wilson coefficients as a function of the dark fermion mass $m_\psi$, with the same color coding as in the previous figure. (Operators involving two distinct massive final states are not shown, as the corresponding Darwin-term contribution has not yet been calculated.)}
        \label{fig:BsBdConstraints3}
\end{figure}

\clearpage

\bibliographystyle{JHEP}
\bibliography{References}

\end{document}